\documentclass[10pt,english,prd,aps,showpacs,epsf,floats]{revtex4}
\usepackage[T1]{fontenc}
\usepackage[latin1]{inputenc}
\usepackage{amsmath}
\usepackage{amssymb}
\usepackage{amsfonts}
\usepackage{babel}

\makeatletter
\makeatletter
\input{tcilatex}
\makeatother
\makeatother

\begin{document}

\title{Information Geometry of Quantum Entangled Gaussian Wave-Packets}
\author{D.-H. Kim}
\email{ki1313@yahoo.com}
\affiliation{Institute for the Early Universe, Ewha Womans University, Seoul 120-750,
South Korea}
\affiliation{International Institute for Theoretical Physics and Mathematics
Einstein-Galilei, via Santa Gonda 14, 59100 Prato, Italy}
\author{S. A. Ali}
\affiliation{International Institute for Theoretical Physics and Mathematics
Einstein-Galilei, via Santa Gonda 14, 59100 Prato, Italy}
\affiliation{Department of Physics, State University of New York at Albany, 1400
Washington Avenue, Albany, NY 12222, USA}
\affiliation{Department of Arts and Sciences, Albany College of Pharmacy and Health
Sciences, 106 New Scotland Avenue, Albany, NY 12208, USA}
\author{C. Cafaro}
\affiliation{School of Science and Technology, Physics Division, University of Camerino,
I-62032 Camerino, Italy}
\author{S. Mancini}
\affiliation{School of Science and Technology, Physics Division, University of Camerino,
I-62032 Camerino, Italy}

\begin{abstract}
Describing and understanding the essence of quantum entanglement and its
connection to dynamical chaos is of great scientific interest. In this work,
using information geometric (IG) techniques, we investigate the effects of
micro-correlations on the evolution of maximal probability paths on
statistical manifolds induced by systems whose microscopic degrees of
freedom are Gaussian distributed. We use the statistical manifolds
associated with correlated and non-correlated Gaussians to model the
scattering induced quantum entanglement of two spinless, structureless,
non-relativistic particles, the latter represented by minimum uncertainty
Gaussian wave-packets. Knowing that the degree of entanglement is quantified
by the purity $\mathcal{P}$\ of the system, we express the purity for $s$%
-wave scattering in terms of the micro-correlation coefficient $r$ - a
quantity that parameterizes the correlated microscopic degrees of freedom of
the system; thus establishing a connection between entanglement and
micro-correlations. Moreover, the correlation coefficient $r$ is readily
expressed in terms of physical quantities involved in the scattering, the
precise form of which is obtained via our IG approach. It is found that the
entanglement duration can be controlled by the initial momentum $p_{\mathrm{o%
}}$, momentum spread $\sigma _{\mathrm{o}}$ and $r$. Furthermore, we obtain
exact expressions for the IG analogue of standard indicators of chaos such
as the sectional curvatures, Jacobi field intensities and the Lyapunov
exponents. We then present an analytical estimate of the information
geometric entropy (IGE); a suitable measure that quantifies the complexity
of geodesic paths on curved manifolds. Finally, we present concluding
remarks addressing the usefulness of an IG characterization of both
entanglement and complexity in quantum physics.
\end{abstract}

\maketitle

\section{\textbf{Introduction}}

One of the most debated features of composite quantum mechanical systems is
their ability to become entangled \cite{epr, Schrodinger}. By quantum
entanglement we mean quantum correlations among the distinct subsystems of
the entire composite quantum system. For such correlated quantum systems, it
is not possible to specify the quantum state of any subsystem independently
of the remaining subsystems.

The generation of quantum entanglement among spatially separated particles
requires non-local interactions through which quantum correlations are
dynamically created \cite{Law, Kubler, Dur}.

Quantum entanglement is an indispensable resource for quantum information
processes \cite{nielsen-book}. Continuous Variable Quantum Systems (CVQS)
are also an interesting topic in quantum information theory \cite{braunstein}%
. By CVQS we refer to quantum mechanical systems on which one can - in
principle - perform measurements of certain observables whose eigenvalue
spectrum is continuous. Examples of CVQS are the quantized motion of massive
particles with the corresponding position and momentum observables and the
quantized mode of the electromagnetic field with its quadrature observables
among others. Most examples of possible applications of entanglement in
continuous variable quantum information is based on EPR states \cite{epr} or
in the optical case, approximations of EPR states using squeezed states \cite%
{Furusawa, Silberhorn, Ou}. Continuous variable entanglement has been
investigated in the context of photon-atom scattering \cite{Chan},
photoionization processes \cite{Grobe, Liu, Chandra, Fedorov}, trapped atoms 
\cite{mancini}, and classically chaotic systems \cite{Miller, Jacquod}.
Correlated CVQS can be used as an invaluable non-classical resource for
quantum computation and quantum communication \cite{braunstein}.

The most realistic approach to the generation of entangled continuous
variable systems is via dynamical interaction, of which local scattering
events (collisions) are a natural, ubiquitous type \cite{Bubhardt}.
Scattering can result in a decomposition of the wave function into
transmission and reflection modes. Due to the mutual interactions present in
scattering processes (such as interference between incoming and reflected
parts of the wave function of the composite system), quantum particles can
become entangled. Moreover, scattering may result in a distortion (due to
rapid fluctuation of scattering amplitude with relative momentum for
instance \cite{Law, tal}) of the shape of the wave function. In cases with
constant amplitudes the wave function of the system may be rendered
inseparable as a consequence of reflection induced distortion. Interference
between incoming and reflected parts of the wave function of the system or
the distortion effect can result in a non-separable post collision
two-particle state. Entanglement generation in non-relativistic scattering
of distinguishable particles has been investigated by a number of
researchers \cite{Schulman, Mack, Schulman1, Law, tal, Wang, Wang1,
Schmuser, Bubhardt}. Most treatments consider interactions among similar
particle types \cite{Schulman, Law, tal, Wang, Wang1, Bubhardt}. It is
however, unclear as to how the interaction (scattering) potentials and
incident particle energies control the strength of entanglement \cite{Law}.
As will be seen in what follows, the information geometric approach employed
in the present work lends some degree of clarification on this issue.
Furthermore, describing and understanding the complexity of quantum
processes is still an open problem and our present knowledge on the
relations among complexity, chaoticity and quantum entanglement are not at
all satisfactory \cite{C09}. As we will see, our work sheds some lights on
this issue as well.

In this article, we explore the potential utility of the Information
Geometric Approach to Chaos (IGAC) \cite{ca-2, cafaroPD, cafaroPA,
cafaroPA2010, cafaro-mancini-1, CarAli} for analyzing quantum mechanical
systems. The IGAC is a theoretical framework developed to study chaos in
informational geodesic flows on statistical manifolds associated with
probabilistic descriptions of physical systems.

We seek to provide a quantitative estimate of the degree of entanglement of
CVQS in terms of information geometric quantities such as solutions to
geodesic equations (expected values of momentum and momentum spread in this
case) and micro-correlation coefficient $r$. The quantity $r$ parameterizes
the correlated microscopic degrees of freedom of the system. An important
question that arises is whether or not $r$ can be understood in terms of
physically measurable quantities.

As described above, when particles with no initial correlations collide,
they may emerge from the interaction entangled \cite{Harshman}. Hence, we
consider two CVQS with Gaussian continuous degrees of freedom that are
prepared independently, interact via a scattering process\ mediated by an
interaction (scattering) potential with finite range and separate again. We
investigate the entanglement of the two-particle wave function of the system
generated by such a scattering event. In this context we ask the question:
how much entanglement between the two particles is generated and on what
does it depend? Surprisingly, there are only a few studies of entanglement
production from the scattering of two particles \cite{Chen, Simon}. The
nature of quantum entanglement arising from $s$-wave scattering has yet to
be fully explored \cite{Gisin}. We choose to use Gaussian states \cite%
{Ferraro, Simon} since many important properties of these states can often
be obtained in an analytic fashion. Moreover, it is known that a good way to
describe naturally occurring quantum states is as spatially localized
Gaussian wave-packets, or as density matrices built from them \cite%
{Schulman1, Zurek, Tegmark, Tegmark1, Giulini}.

For a system of two spinless, structureless, non-relativistic particles with
no internal degrees of freedom, a complete set of commuting observables is
furnished by the momentum operators of each particle \cite{Harshman}. The
continuous variables in our case are therefore taken to be the momentum of
each particle. We investigate how the initial conditions and the magnitude
of $r$ of the system affect entanglement; specifically, the duration of
entanglement. By duration of entanglement we mean the temporal behavior of
the magnitude of entanglement. The IGAC is also used to estimate the extent
to which the complexity of the geodesic information flow (continuous
spectrum of expected values of some relevant observable) of the quantum
system is affected by $r$.

The layout of this article is as follows. In Section \ref{sec-entang-wave},
we reexamine the $s$-wave scattering induced quantum entanglement of two
spinless, structureless, non-relativistic particles, which are represented
by two-particle Gaussian wave-packets \cite{Wang}. We exploit the fact that
the three-dimensional scattering problem can be effectively reduced to a
one-dimensional problem as far as the three-dimensional representations of
wave-packets are isotropic. In Section \ref{sec-info-geo-pers}, we outline
the main ideas behind the IGAC and present the information geometry of
correlated and uncorrelated Gaussian statistical manifolds, which is to be
employed in our investigation of quantum entanglement and complexity in the
following sections. In Section \ref{sec-application}, we use information
geometric techniques in conjunction with standard partial wave quantum
scattering theory to provide an information geometric characterization of
quantum entanglement. In Section \ref{sec-complexity}, we obtain exact
expressions for the information geometric analogue of standard indicators of
chaos such as sectional curvatures, Jacobi field intensities and Lyapunov
exponents. Finally, we present an analytical estimate of the information
geometric entropy (IGE) and this allows us to connect quantum entanglement
to the complexity of informational geodesic flows in a quantitative manner.
Our concluding remarks are presented in Section \ref{sec-final}.

\section{Entanglement and Gaussian wave-packet scattering processes\label%
{sec-entang-wave}}

In this Section, we reexamine the $s$-wave scattering induced quantum
entanglement of two spinless, structureless, non-relativistic particles,
represented by two-particle Gaussian wave-packets as presented in \cite{Wang}%
. For ease of analysis, we exploit the fact that the three-dimensional
scattering problem can be effectively reduced to a one-dimensional problem
as the three-dimensional representations of wave-packets are isotropic.

\subsection{The Pre-collisional Scenario and the Effective Dimensional
Reduction}

For the purpose of modeling a head-on collision we consider two identical
(but distinguishable), spinless particles in momentum space, each
represented by minimum uncertainty Gaussian wave-packets. Before collision,
particles $1$\ and $2$\ are initially located far from each other - a linear
distance\ $\mathbf{R}_{\mathrm{o}}$ - each having the initial average
momentum $\left\langle \mathbf{p}_{1}\right\rangle _{\mathrm{o}}=\mathbf{p}_{%
\mathrm{o}}$ and $\left\langle \mathbf{p}_{2}\right\rangle _{\mathrm{o}}=-%
\mathbf{p}_{\mathrm{o}}$, respectively, with equal momentum dispersion $%
\sigma _{\mathrm{o}}$ (see Figure \ref{fig1}). The normalized, separable
(i.e., non-entangled) two-particle Gaussian wave function representing the
situation before collision is then given by \cite{Wang}%
\begin{equation}
\psi \left( \mathbf{k}_{1},\mathbf{k}_{2}\right) =\psi _{1}\left( \mathbf{k}%
_{1}\right) \otimes \psi _{2}\left( \mathbf{k}_{2}\right) ,  \label{dim1}
\end{equation}%
with respective single particle wave functions%
\begin{equation}
\psi _{1/2}\left( \mathbf{k}_{1/2}\right) =a\left( \mathbf{k}%
_{1/2},\left\langle \mathbf{k}_{1/2}\right\rangle _{\mathrm{o}};\sigma _{k%
\mathrm{o}}\right) e^{i\left( \mathbf{k}_{1/2}-\left\langle \mathbf{k}%
_{1/2}\right\rangle _{\mathrm{o}}\right) \cdot \mathbf{q}_{1/2}},
\label{dim2}
\end{equation}%
where 
\begin{equation}
a\left( \mathbf{k}_{1/2},\left\langle \mathbf{k}_{1/2}\right\rangle _{%
\mathrm{o}};\sigma _{k\mathrm{o}}\right) \equiv \left( \frac{1}{2\pi \sigma
_{k\mathrm{o}}^{2}}\right) ^{3/4}\exp \left[ -\frac{\left( \mathbf{k}%
_{1/2}-\left\langle \mathbf{k}_{1/2}\right\rangle _{\mathrm{o}}\right) ^{2}}{%
4\sigma _{k\mathrm{o}}^{2}}\right] ,  \label{dim3}
\end{equation}%
with $\mathbf{k}_{1/2}=\frac{\mathbf{p}_{1/2}}{\hbar }$, $\left\langle 
\mathbf{k}_{1/2}\right\rangle _{\mathrm{o}}=\frac{\left\langle \mathbf{p}%
_{1/2}\right\rangle _{\mathrm{o}}}{\hbar }=\pm \frac{\mathbf{p}_{\mathrm{o}}%
}{\hbar }=\pm \mathbf{k}_{\mathrm{o}}$, $\sigma _{k\mathrm{o}}=\frac{\sigma
_{\mathrm{o}}}{\hbar }$ and $\mathbf{q}_{1/2}=\mp \frac{1}{2}\mathbf{R}_{%
\mathrm{o}}$. Observe that wave functions in (\ref{dim2}) satisfy the
normalization conditions%
\begin{equation}
\int \psi _{1}^{\ast }\left( \mathbf{k}_{1}\right) \psi _{1}\left( \mathbf{k}%
_{1}\right) d^{3}\mathbf{k}_{1}=\int a^{2}\left( \mathbf{k}_{1},\left\langle 
\mathbf{k}_{1}\right\rangle _{\mathrm{o}};\sigma _{k\mathrm{o}}\right) d^{3}%
\mathbf{k}_{1}=1  \label{dim4}
\end{equation}%
and%
\begin{equation}
\int \psi _{2}^{\ast }\left( \mathbf{k}_{2}\right) \psi _{2}\left( \mathbf{k}%
_{2}\right) d^{3}\mathbf{k}_{2}=\int a^{2}\left( \mathbf{k}_{2},\left\langle 
\mathbf{k}_{2}\right\rangle _{\mathrm{o}};\sigma _{k\mathrm{o}}\right) d^{3}%
\mathbf{k}_{2}=1.  \label{dim5}
\end{equation}%
The type of state described by (\ref{dim1}) is ubiquitous when describing
quantum systems of continuous variables.\FRAME{fhFU}{9.4366cm}{4.7742cm}{0pt%
}{\Qcb{An illustration of the two-particle system before and after a head-on
collision. Before collision the two particles are initially the distance $%
\mathbf{R}_{\mathrm{o}}$ away from each other and move toward each other
with the momenta $\mathbf{p}_{\mathrm{o}}$ and $-\mathbf{p}_{\mathrm{o}}$,
respectively; both particles have the identical momentum spread $\protect%
\sigma _{\mathrm{o}}$ ($\hbar /\protect\sigma _{\mathrm{o}}$ in
configuration space). After collision a large spherical shell represents the
scattered part of the single particle density (i.e. either particle $1$ or
particle $2$) under the $s$-wave approximation. The arrows indicate that
after collision the two particles move away from each other with the
momenta, $-\mathbf{p}_{\mathrm{o}}$ and $\mathbf{p}_{\mathrm{o}}$,
respectively~\protect\cite{Wang}.}}{\Qlb{fig1}}{fig1.png}{\special{language
"Scientific Word";type "GRAPHIC";maintain-aspect-ratio TRUE;display
"USEDEF";valid_file "F";width 9.4366cm;height 4.7742cm;depth
0pt;original-width 23.0401in;original-height 11.5707in;cropleft "0";croptop
"1";cropright "1";cropbottom "0";filename '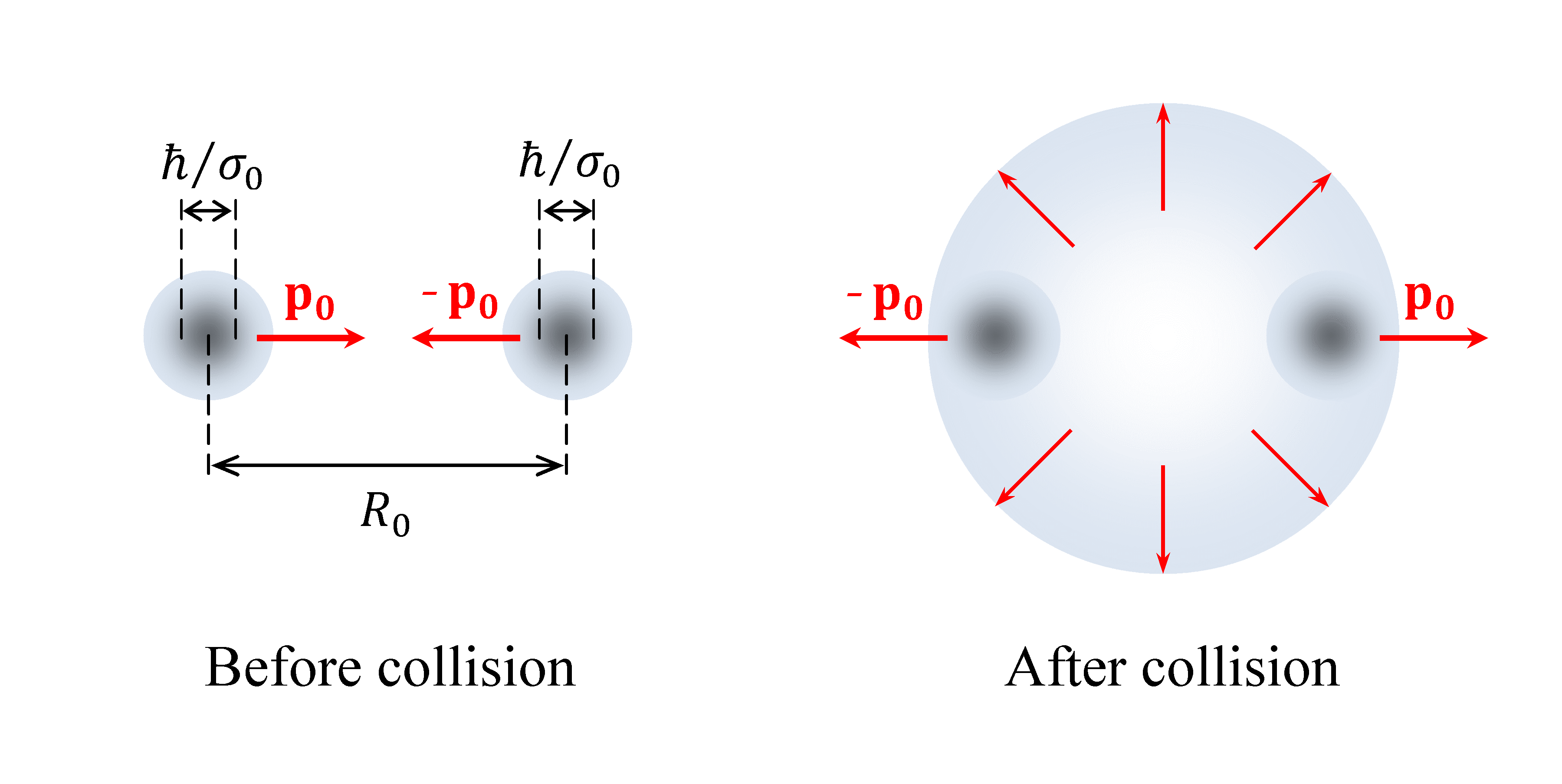';file-properties
"XNPEU";}}

One should note that the three-dimensional Gaussian wave-packet (\ref{dim1})
is isotropic. That is to say, in polar coordinates the representation of the
separable two-particle state exhibits a functional dependence on the radial
variable only. For this reason the three-dimensional vectorial
representation (\ref{dim1}) may be effectively reduced to a one-dimensional
representation. This may be demonstrated as follows. First, using (\ref{dim1}%
), (\ref{dim2}) and (\ref{dim3}), we express the two-particle wave-packet in
Cartesian coordinates as%
\begin{equation}
\psi \left( \left( k_{1}^{x},k_{1}^{y},k_{1}^{z}\right) ,\left(
k_{2}^{x},k_{2}^{y},k_{2}^{z}\right) \right) =\psi _{1}\left(
k_{1}^{x},k_{1}^{y},k_{1}^{z}\right) \otimes \psi _{2}\left(
k_{2}^{x},k_{2}^{y},k_{2}^{z}\right) ,  \label{dim6}
\end{equation}%
where 
\begin{equation}
\psi _{1/2}\left( k_{1/2}^{x},k_{1/2}^{y},k_{1/2}^{z}\right) =\left( \frac{1%
}{2\pi \sigma _{k\mathrm{o}}^{2}}\right) ^{3/4}\dprod\limits_{j=x,y,z}\exp %
\left[ -\frac{\left( k_{1/2}^{j}-\left\langle k_{1/2}^{j}\right\rangle _{%
\mathrm{o}}\right) ^{2}}{4\sigma _{k\mathrm{o}}^{2}}\right] e^{i\left(
k_{1/2}^{j}-\left\langle k_{1/2}^{j}\right\rangle _{\mathrm{o}}\right)
q_{1/2}^{j}},  \label{dim7}
\end{equation}%
with $k_{1/2}^{j}=\mathbf{k}_{1/2}\cdot \mathbf{e}_{j}$, $\left\langle
k_{1/2}^{j}\right\rangle _{\mathrm{o}}=\left\langle \mathbf{k}%
_{1/2}\right\rangle _{\mathrm{o}}\cdot \mathbf{e}_{j}$, $q_{1/2}^{j}=$ $%
\mathbf{q}_{1/2}\cdot \mathbf{e}_{j}$, the components in a Cartesian basis $%
\left\{ \mathbf{e}_{j}\right\} $. Then the probability density for this
wave-packet becomes 
\begin{eqnarray}
\left\vert \psi \left( \mathbf{k}_{1},\mathbf{k}_{2}\right) \right\vert ^{2}
&=&\psi ^{\ast }\left( \left( k_{1}^{x},k_{1}^{y},k_{1}^{z}\right) ,\left(
k_{2}^{x},k_{2}^{y},k_{2}^{z}\right) \right) \psi \left( \left(
k_{1}^{x},k_{1}^{y},k_{1}^{z}\right) ,\left(
k_{2}^{x},k_{2}^{y},k_{2}^{z}\right) \right)  \notag \\
&=&\left( \frac{1}{2\pi \sigma _{k\mathrm{o}}^{2}}\right)
^{3}\dprod\limits_{j=x,y,z}\exp \left[ -\frac{\left( k_{1}^{j}-\left\langle
k_{1}^{j}\right\rangle _{\mathrm{o}}\right) ^{2}}{2\sigma _{k\mathrm{o}}^{2}}%
\right] \exp \left[ -\frac{\left( k_{2}^{j}-\left\langle
k_{2}^{j}\right\rangle _{\mathrm{o}}\right) ^{2}}{2\sigma _{k\mathrm{o}}^{2}}%
\right] .  \label{dim8}
\end{eqnarray}%
Upon integrating $\left\vert \psi \left( \mathbf{k}_{1},\mathbf{k}%
_{2}\right) \right\vert ^{2}$ over $d^{3}\mathbf{k}_{1}d^{3}\mathbf{k}_{2}$,
one obtains 
\begin{equation}
\int\limits_{-\infty }^{+\infty }\int\limits_{-\infty }^{+\infty
}\int\limits_{-\infty }^{+\infty }\int\limits_{-\infty }^{+\infty
}\int\limits_{-\infty }^{+\infty }\int\limits_{-\infty }^{+\infty
}\dprod\limits_{j=x,y,z}dk_{1}^{j}dk_{2}^{j}\left\vert \psi \left( \mathbf{k}%
_{1},\mathbf{k}_{2}\right) \right\vert ^{2}=I_{1}I_{2},  \label{dim9}
\end{equation}%
where 
\begin{equation}
I_{l}\equiv \left( \frac{1}{2\pi \sigma _{k\mathrm{o}}^{2}}\right)
^{3/2}\int\limits_{-\infty }^{+\infty }\int\limits_{-\infty }^{+\infty
}\int\limits_{-\infty }^{+\infty }\dprod\limits_{j=x,y,z}dk_{l}^{j}\exp %
\left[ -\frac{\left( k_{l}^{j}-\left\langle k_{l}^{j}\right\rangle _{\mathrm{%
o}}\right) ^{2}}{2\sigma _{k\mathrm{o}}^{2}}\right] \text{ with }l=1,2.
\label{dim10}
\end{equation}%
Converting the integrals $I_{l}$ ($l=1,2$) into polar-coordinate
representation yields%
\begin{eqnarray}
I_{l} &=&\left( \frac{1}{2\pi \sigma _{k\mathrm{o}}^{2}}\right)
^{3/2}\int\limits_{-\infty }^{+\infty }\int\limits_{-\infty }^{+\infty
}\int\limits_{-\infty }^{+\infty }\dprod\limits_{j=x,y,z}dk_{l}^{j}\exp %
\left[ -\frac{\left( k_{l}^{j}\right) ^{2}}{2\sigma _{k\mathrm{o}}^{2}}%
\right]  \notag \\
&=&\left( \frac{1}{2\pi \sigma _{k\mathrm{o}}^{2}}\right) ^{3/2}4\pi
\dint\limits_{0}^{+\infty }dk_{l}k_{l}^{2}\exp \left( -\frac{k_{l}^{2}}{%
2\sigma _{k\mathrm{o}}^{2}}\right) ,  \label{dim11}
\end{eqnarray}%
where $k_{l}^{2}\equiv \mathbf{k}_{l}^{2}=\left( k_{l}^{x}\right)
^{2}+\left( k_{l}^{y}\right) ^{2}+\left( k_{l}^{z}\right) ^{2}$ ($l=1,2$).
The remaining integral in (\ref{dim11}) can be replaced with 
\begin{eqnarray}
\dint\limits_{0}^{+\infty }dk_{l}k_{l}^{2}\exp \left( -\frac{k_{l}^{2}}{%
2\sigma _{k\mathrm{o}}^{2}}\right) &=&\sigma _{k\mathrm{o}%
}^{2}\dint\limits_{0}^{+\infty }dk_{l}\exp \left( -\frac{k_{l}^{2}}{2\sigma
_{k\mathrm{o}}^{2}}\right)  \label{dim12} \\
&=&\frac{\sigma _{k\mathrm{o}}^{2}}{2}\dint\limits_{-\infty }^{+\infty
}dk_{l}\exp \left( -\frac{k_{l}^{2}}{2\sigma _{k\mathrm{o}}^{2}}\right) =%
\frac{\sigma _{k\mathrm{o}}^{2}}{2}\dint\limits_{-\infty }^{+\infty
}dk_{l}\exp \left[ -\frac{\left( k_{l}-\left\langle k_{l}\right\rangle _{%
\mathrm{o}}\right) ^{2}}{2\sigma _{k\mathrm{o}}^{2}}\right] ,  \notag
\end{eqnarray}%
where we have changed the domain of $k_{l}$ from $\left[ 0,+\infty \right) $
to $\left( -\infty ,+\infty \right) $ in order to obtain the last equality.
By substituting (\ref{dim12}) into (\ref{dim11}), followed by substituting (%
\ref{dim11}) into the right-hand side of (\ref{dim9}) and finally inserting (%
\ref{dim8}) into the left-hand side of (\ref{dim9}), we establish%
\begin{eqnarray}
&&\left( \frac{1}{2\pi \sigma _{k\mathrm{o}}^{2}}\right)
^{3}\int\limits_{-\infty }^{+\infty }\int\limits_{-\infty }^{+\infty
}\int\limits_{-\infty }^{+\infty }\int\limits_{-\infty }^{+\infty
}\int\limits_{-\infty }^{+\infty }\int\limits_{-\infty }^{+\infty
}\dprod\limits_{l=1,2}\dprod\limits_{j=x,y,z}dk_{l}^{j}\exp \left[ -\frac{%
\left( k_{l}^{j}-\left\langle k_{l}^{j}\right\rangle _{\mathrm{o}}\right)
^{2}}{2\sigma _{k\mathrm{o}}^{2}}\right]  \notag \\
&=&\frac{1}{2\pi \sigma _{k\mathrm{o}}^{2}}\dint\limits_{-\infty }^{+\infty
}\dint\limits_{-\infty }^{+\infty }\dprod\limits_{l=1,2}dk_{l}\exp \left[ -%
\frac{\left( k_{l}-\left\langle k_{l}\right\rangle _{\mathrm{o}}\right) ^{2}%
}{2\sigma _{k\mathrm{o}}^{2}}\right] .  \label{dim13}
\end{eqnarray}%
The integration of $\left\vert \psi \left( \mathbf{k}_{1},\mathbf{k}%
_{2}\right) \right\vert ^{2}$ over $d^{3}\mathbf{k}_{1}d^{3}\mathbf{k}_{2}$
now reads 
\begin{eqnarray}
\diint \left\vert \psi \left( \mathbf{k}_{1},\mathbf{k}_{2}\right)
\right\vert ^{2}d^{3}\mathbf{k}_{1}d^{3}\mathbf{k}_{2} &=&\left( \frac{1}{%
2\pi \sigma _{k\mathrm{o}}^{2}}\right) ^{3}\diint \exp \left[ -\frac{\left( 
\mathbf{k}_{1}-\left\langle \mathbf{k}_{1}\right\rangle _{\mathrm{o}}\right)
^{2}+\left( \mathbf{k}_{2}-\left\langle \mathbf{k}_{2}\right\rangle _{%
\mathrm{o}}\right) ^{2}}{2\sigma _{k\mathrm{o}}^{2}}\right] d^{3}\mathbf{k}%
_{1}d^{3}\mathbf{k}_{2}  \notag \\
&=&\frac{1}{2\pi \sigma _{k\mathrm{o}}^{2}}\dint\limits_{-\infty }^{+\infty
}\dint\limits_{-\infty }^{+\infty }\exp \left[ -\frac{\left(
k_{1}-\left\langle k_{1}\right\rangle _{\mathrm{o}}\right) ^{2}+\left(
k_{2}-\left\langle k_{2}\right\rangle _{\mathrm{o}}\right) ^{2}}{2\sigma _{k%
\mathrm{o}}^{2}}\right] dk_{1}dk_{2}.  \label{dim14}
\end{eqnarray}%
Thus, we may effectively reduce the three dimensional two-particle wave
function $\psi \left( \mathbf{k}_{1},\mathbf{k}_{2}\right) $ expressed via (%
\ref{dim1}), (\ref{dim2}) and (\ref{dim3}) to the two-particle
one-dimensional wave function $\psi \left( k_{1},k_{2}\right) $ given by%
\begin{equation}
\psi \left( k_{1},k_{2}\right) =\psi _{1}\left( k_{1}\right) \otimes \psi
_{2}\left( k_{2}\right) ,  \label{eq:int0}
\end{equation}%
with respective single-particle wave functions,%
\begin{equation}
\psi _{1/2}\left( k_{1/2}\right) =a\left( k_{1/2},\left\langle
k_{1/2}\right\rangle _{\mathrm{o}};\sigma _{k\mathrm{o}}\right) e^{i\left(
k_{1/2}-\left\langle k_{1/2}\right\rangle _{\mathrm{o}}\right) q_{1/2}},
\label{int1}
\end{equation}%
where 
\begin{equation}
a\left( k_{1/2},\left\langle k_{1/2}\right\rangle _{\mathrm{o}};\sigma _{k%
\mathrm{o}}\right) \equiv \left( \frac{1}{2\pi \sigma _{k\mathrm{o}}^{2}}%
\right) ^{1/4}\exp \left[ -\frac{\left( k_{1/2}-\left\langle
k_{1/2}\right\rangle _{\mathrm{o}}\right) ^{2}}{4\sigma _{k\mathrm{o}}^{2}}%
\right] ,  \label{int2}
\end{equation}%
and $k_{1/2}=\frac{p_{1/2}}{\hbar }\in \left( -\infty ,+\infty \right) $, $%
\left\langle k_{1/2}\right\rangle _{\mathrm{o}}=\frac{\left\langle
p_{1/2}\right\rangle _{\mathrm{o}}}{\hbar }=\pm \frac{p_{\mathrm{o}}}{\hbar }%
=\pm k_{\mathrm{o}}$, $\sigma _{k\mathrm{o}}=\frac{\sigma _{\mathrm{o}}}{%
\hbar }$, $q_{1/2}=\mp \frac{1}{2}R_{\mathrm{o}}$. The wave functions (\ref%
{int1}) satisfy the normalization conditions 
\begin{equation}
\int\limits_{-\infty }^{+\infty }\psi _{1/2}^{\ast }\left( k_{1/2}\right)
\psi _{1/2}\left( k_{1/2}\right) dk_{1/2}=\int\limits_{-\infty }^{+\infty
}a^{2}\left( k_{1/2},\left\langle k_{1/2}\right\rangle _{\mathrm{o}};\sigma
_{k\mathrm{o}}\right) dk_{1/2}=1.
\end{equation}

\subsection{The Post-collisional Scenario}

After collision, the wave function for the two-particle system in the long
time limit takes the form \cite{Wang}:%
\begin{equation}
\psi \left( \mathbf{k}_{1},\mathbf{k}_{2},t\right) =\left( N\right) ^{-1/2}%
\left[ \psi _{1}\left( \mathbf{k}_{1}\right) \psi _{2}\left( \mathbf{k}%
_{2}\right) e^{-i\hbar \left( k_{1}^{2}+k_{2}^{2}\right) t/\left( 2m\right)
}+\varepsilon \psi _{\mathrm{scat}}\left( \mathbf{k}_{1},\mathbf{k}%
_{2},t\right) \right] ,  \label{eq:psi_aft_coll}
\end{equation}%
where $N$ and $\varepsilon $ are normalization constants such that $\psi $
and $\psi _{\mathrm{scat}}$ are both normalized to unity, $m$ denotes the
mass of each particle. Here $\psi _{1/2}\left( \mathbf{k}_{1/2}\right) $ is
given by (\ref{dim2}) and (\ref{dim3}) and thus we have 
\begin{eqnarray}
\psi _{1}\left( \mathbf{k}_{1}\right) \psi _{2}\left( \mathbf{k}_{2}\right)
e^{-i\hbar \left( k_{1}^{2}+k_{2}^{2}\right) t/\left( 2m\right) } &=&\left( 
\frac{1}{2\pi \sigma _{k\mathrm{o}}^{2}}\right) ^{3/2}\exp \left[ -\frac{%
\left( \mathbf{k}_{1}-\mathbf{k}_{\mathrm{o}}\right) ^{2}+\left( \mathbf{k}%
_{2}+\mathbf{k}_{\mathrm{o}}\right) ^{2}}{4\sigma _{k\mathrm{o}}^{2}}\right]
\notag \\
&&\times e^{-i\left( k_{1}-k_{\mathrm{o}}\right) R_{\mathrm{o}}/2+i\left(
k_{2}+k_{\mathrm{o}}\right) R_{\mathrm{o}}/2-i\hbar \left(
k_{1}^{2}+k_{2}^{2}\right) t/\left( 2m\right) }.  \label{psi12b}
\end{eqnarray}%
We treat $\left\vert \varepsilon \right\vert \ll 1$ as a small number.
Following \cite{Wang}, one can write%
\begin{equation}
\varepsilon \psi _{\mathrm{scat}}\left( \mathbf{k}_{1},\mathbf{k}%
_{2},t\right) =\varepsilon \psi _{\mathrm{c.m.}}\left( \mathbf{K},t\right)
\eta \left( \mathbf{k},t\right) ,  \label{sc1}
\end{equation}%
where 
\begin{equation}
\psi _{\mathrm{c.m.}}\left( \mathbf{K},t\right) =\left( \frac{1}{4\pi \sigma
_{k\mathrm{o}}^{2}}\right) ^{3/4}\exp \left( -\frac{\mathbf{K}^{2}}{8\sigma
_{k\mathrm{o}}^{2}}\right) e^{-i\hbar K^{2}t/\left( 2M\right) },  \label{sc2}
\end{equation}%
and the scattering part $\varepsilon \eta \left( \mathbf{k},t\right) $ is
given by 
\begin{equation}
\varepsilon \eta \left( \mathbf{k},t\right) =\frac{1}{\left( 2\pi \right)
^{3}}\diint \psi _{\mathrm{rel}}\left( \mathbf{k}^{\prime },0\right) f\left(
k^{\prime }\right) \frac{e^{ik^{\prime }r}}{r}e^{-i\mathbf{k\cdot r}-i\hbar
k^{\prime 2}t/\left( 2\mu \right) }d^{3}\mathbf{k}^{\prime }d^{3}\mathbf{r}
\label{sc3}
\end{equation}%
with 
\begin{equation}
\psi _{\mathrm{rel}}\left( \mathbf{k}^{\prime },0\right) =\left( \frac{1}{%
\pi \sigma _{k\mathrm{o}}^{2}}\right) ^{3/4}\exp \left[ -\frac{\left( 
\mathbf{k}^{\prime }-k_{\mathrm{o}}\mathbf{\hat{k}}\right) ^{2}}{2\sigma _{k%
\mathrm{o}}^{2}}\right] e^{-i\left( k-k_{\mathrm{o}}\right) R_{\mathrm{o}}}.
\label{sc4}
\end{equation}%
Here we have adopted the center of mass and relative coordinates such that
the conjugate momenta $\mathbf{K}\equiv \mathbf{k}_{1}+\mathbf{k}_{2}$ and $%
\mathbf{k}\equiv \frac{1}{2}\left( \mathbf{k}_{1}-\mathbf{k}_{2}\right) $
are used along with the total mass, $M=2m$ and the reduced mass $\mu =m/2$.
The quantity $f\left( k\right) \equiv \frac{e^{i2\theta \left( k\right) }-1}{%
2ik}$ is the $s$-wave scattering amplitude due to the $s$-wave scattering
phase shift $\theta \left( k\right) $. Inserting (\ref{sc4}) into (\ref{sc3}%
) and performing the integral, we obtain%
\begin{equation}
\varepsilon \eta \left( \mathbf{k},t\right) \approx \left( \frac{1}{\pi
\sigma _{k\mathrm{o}}^{2}}\right) ^{3/4}\exp \left[ -\frac{\left( \mathbf{k}%
-k_{\mathrm{o}}\mathbf{\hat{k}}\right) ^{2}}{2\sigma _{k\mathrm{o}}^{2}}%
\right] \varrho \left( k\right) e^{-i\left( k-k_{\mathrm{o}}\right) R_{%
\mathrm{o}}-i\hbar k^{2}t/\left( 2\mu \right) },  \label{sc5}
\end{equation}%
where 
\begin{equation}
\varrho \left( k\right) \equiv \frac{4i\left( k_{\mathrm{o}}-i\sigma _{k%
\mathrm{o}}^{2}R_{\mathrm{o}}\right) k^{2}f\left( k\right) }{\sigma _{k%
\mathrm{o}}^{2}},  \label{sc6}
\end{equation}%
and the approximation has been made under the assumption of low energy $s$%
-wave scattering. Then by (\ref{sc1}), (\ref{sc2}) and (\ref{sc5}) we find 
\begin{eqnarray}
\varepsilon \psi _{\mathrm{scat}}\left( \mathbf{k}_{1},\mathbf{k}%
_{2},t\right) &\approx &\left( \frac{1}{2\pi \sigma _{k\mathrm{o}}^{2}}%
\right) ^{3/2}\exp \left[ -\frac{\mathbf{K}^{2}+4\left( \mathbf{k}-k_{%
\mathrm{o}}\mathbf{\hat{k}}\right) ^{2}}{8\sigma _{k\mathrm{o}}^{2}}\right] 
\notag \\
&&\times \varrho \left( k\right) e^{-i\left( k-k_{\mathrm{o}}\right) R_{%
\mathrm{o}}-i\hbar K^{2}t/\left( 2M\right) -i\hbar k^{2}t/\left( 2\mu
\right) }.  \label{eq:epsilon_psi}
\end{eqnarray}%
Due to the fact that%
\begin{eqnarray}
&&\exp \left[ -\frac{\left( \mathbf{k}_{1}-\mathbf{k}_{\mathrm{o}}\right)
^{2}+\left( \mathbf{k}_{2}+\mathbf{k}_{\mathrm{o}}\right) ^{2}}{4\sigma _{k%
\mathrm{o}}^{2}}\right] e^{-i\left( k_{1}-k_{\mathrm{o}}\right) R_{\mathrm{o}%
}/2+i\left( k_{2}+k_{\mathrm{o}}\right) R_{\mathrm{o}}/2-i\hbar \left(
k_{1}^{2}+k_{2}^{2}\right) t/\left( 2m\right) }  \notag \\
&=&\exp \left[ -\frac{\mathbf{K}^{2}+4\left( \mathbf{k}-k_{\mathrm{o}}%
\mathbf{\hat{k}}\right) ^{2}}{8\sigma _{k\mathrm{o}}^{2}}\right] e^{-i\left(
k-k_{\mathrm{o}}\right) R_{\mathrm{o}}-i\hbar K^{2}t/\left( 2M\right)
-i\hbar k^{2}t/\left( 2\mu \right) },
\end{eqnarray}%
we may combine (\ref{psi12b}) and (\ref{eq:epsilon_psi}) to rewrite (\ref%
{eq:psi_aft_coll}) as%
\begin{eqnarray}
\psi \left( \mathbf{k}_{1},\mathbf{k}_{2},t\right) &=&\left( N\right)
^{-1/2}\left( \frac{1}{2\pi \sigma _{k\mathrm{o}}^{2}}\right) ^{3/2}\exp %
\left[ -\frac{\mathbf{K}^{2}+4\left( \mathbf{k}-k_{\mathrm{o}}\mathbf{\hat{k}%
}\right) ^{2}}{8\sigma _{k\mathrm{o}}^{2}}\right]  \notag \\
&&\times \left[ 1+\varrho \left( k\right) \right] e^{-i\left( k-k_{\mathrm{o}%
}\right) R_{\mathrm{o}}-i\hbar K^{2}t/\left( 2M\right) -i\hbar k^{2}t/\left(
2\mu \right) }.  \label{sc7}
\end{eqnarray}%
Our dimensional analysis carried out in the separable case (see the previous
Subsection) applies equally well in the entangled case. Hence, we may reduce
the three-dimensional wave-packet $\psi \left( \mathbf{k}_{1},\mathbf{k}%
_{2},t\right) $ expressed via (\ref{eq:psi_aft_coll}) to the one-dimensional
one,%
\begin{equation}
\psi \left( k_{1},k_{2},t\right) =\left( N\right) ^{-1/2}\left[ \psi
_{1}\left( k_{1}\right) \psi _{2}\left( k_{2}\right) e^{-i\hbar \left(
k_{1}^{2}+k_{2}^{2}\right) t/\left( 2m\right) }+\varepsilon \psi _{\mathrm{%
scat}}\left( k_{1},k_{2},t\right) \right] ,  \label{psi0}
\end{equation}%
where the single-particle wave function $\psi _{1/2}\left( k_{1/2}\right) $
is specified via (\ref{int1}) and (\ref{int2}), together with $k_{1/2}=\frac{%
p_{1/2}}{\hbar }\in \left( -\infty ,+\infty \right) $, $\left\langle
k_{1/2}\right\rangle _{\mathrm{o}}=\frac{\left\langle p_{1/2}\right\rangle _{%
\mathrm{o}}}{\hbar }=\pm \frac{p_{\mathrm{o}}}{\hbar }=\pm k_{\mathrm{o}}$, $%
\sigma _{k\mathrm{o}}=\frac{\sigma _{\mathrm{o}}}{\hbar }$, $q_{1/2}=\mp 
\frac{1}{2}R_{\mathrm{o}}$. In analogy to (\ref{sc7}), $\psi \left(
k_{1},k_{2},t\right) $ can be rewritten as 
\begin{eqnarray}
\psi \left( k_{1},k_{2},t\right) &=&\left( N\right) ^{-1/2}\left( \frac{1}{%
2\pi \sigma _{k\mathrm{o}}^{2}}\right) ^{1/2}\exp \left[ -\frac{%
K^{2}+4\left( k-k_{\mathrm{o}}\right) ^{2}}{8\sigma _{k\mathrm{o}}^{2}}%
\right]  \notag \\
&&\times \left[ 1+\varrho \left( k\right) \right] e^{-i\left( k-k_{\mathrm{o}%
}\right) R_{\mathrm{o}}-i\hbar K^{2}t/\left( 2M\right) -i\hbar k^{2}t/\left(
2\mu \right) },  \label{psi1}
\end{eqnarray}%
where we adopt the one-dimensional center of mass and relative coordinates,\
whose conjugate momenta are defined as $K\equiv k_{1}+k_{2}\in \left(
-\infty ,+\infty \right) $ and $k\equiv \frac{1}{2}\left( k_{1}-k_{2}\right)
\in \left( -\infty ,+\infty \right) $, and $\varrho \left( k\right) $ is
given by (\ref{sc6}). Separating variables in (\ref{psi1}), we may write%
\begin{eqnarray}
\left\vert \psi \left( k_{1},k_{2},t\right) \right\vert ^{2} &=&\psi \left(
k_{1},k_{2},t\right) \psi ^{\ast }\left( k_{1},k_{2},t\right)  \notag \\
&=&\frac{N^{-1}}{2\pi \sigma _{k\mathrm{o}}^{2}}\exp \left( -\frac{K^{2}}{%
4\sigma _{k\mathrm{o}}^{2}}\right) \exp \left( -\frac{\tilde{k}^{2}}{\sigma
_{k\mathrm{o}}^{2}}\right) \left[ 1+2\Re \left( \varrho \left( k\right)
\right) +\left\vert \varrho \left( k\right) \right\vert ^{2}\right] ,
\label{eq:psi_sq}
\end{eqnarray}%
where $\tilde{k}\equiv k-k_{\mathrm{o}}$ and $\Re $ denotes the real part of 
$\varrho \left( k\right) $. However, we find that the complex-valued
scattering amplitude $f\left( k\right) $ can be approximated as real since $%
f\left( k\right) =\frac{\theta \left( k\right) }{k}+\mathcal{O}\left( \theta
^{2}\right) $ for $\theta \left( k\right) \ll 1$. In view of this fact and (%
\ref{sc6}) we employ the following approximations:%
\begin{equation}
\Re \left( \varrho \left( k\right) \right) \approx 4R_{\mathrm{o}%
}k^{2}f\left( k\right) ,  \label{eq:Re_varrho}
\end{equation}%
\begin{equation}
\left\vert \varrho \left( k\right) \right\vert ^{2}\approx \frac{16\left( k_{%
\mathrm{o}}^{2}+\sigma _{k\mathrm{o}}^{4}R_{\mathrm{o}}^{2}\right)
k^{4}\left( f\left( k\right) \right) ^{2}}{\sigma _{k\mathrm{o}}^{4}}.
\label{rho1}
\end{equation}%
Upon Taylor expanding $f\left( k\right) $ and $\left( f\left( k\right)
\right) ^{2}$ around $k=k_{\mathrm{o}}$, we re-express $\Re \left( \varrho
\left( k\right) \right) $ in (\ref{eq:Re_varrho}) and $\left\vert \varrho
\left( k\right) \right\vert ^{2}$ in (\ref{rho1}) as%
\begin{equation}
\Re \left( \varrho \left( k\right) \right) \approx 4R_{\mathrm{o}%
}k^{2}\dsum\limits_{n=0}^{\infty }A_{n}\left( k-k_{\mathrm{o}}\right) ^{n}
\label{rho2}
\end{equation}%
and%
\begin{equation}
\left\vert \varrho \left( k\right) \right\vert ^{2}\approx \frac{16\left( k_{%
\mathrm{o}}^{2}+\sigma _{k\mathrm{o}}^{4}R_{\mathrm{o}}^{2}\right) k^{4}}{%
\sigma _{k\mathrm{o}}^{4}}\dsum\limits_{n=0}^{\infty }B_{n}\left( k-k_{%
\mathrm{o}}\right) ^{n},  \label{rho3}
\end{equation}%
respectively. Here the quantities $A_{n}$ and $B_{n}$ are appropriate
coefficients determined from the expansions $f\left( k\right)
=\sum_{n=0}^{\infty }A_{n}\left( k-k_{\mathrm{o}}\right) ^{n}$ and $\left(
f\left( k\right) \right) ^{2}=\sum_{n=0}^{\infty }B_{n}\left( k-k_{\mathrm{o}%
}\right) ^{n}$, respectively. By inserting (\ref{rho2}) and (\ref{rho3})
into (\ref{eq:psi_sq}) we can integrate $\left\vert \psi \right\vert ^{2}$
as follows:%
\begin{eqnarray}
2\pi \sigma _{k\mathrm{o}}^{2}N\dint\limits_{-\infty }^{+\infty
}\dint\limits_{-\infty }^{+\infty }\left\vert \psi \right\vert
^{2}dk_{1}dk_{2} &=&2\pi \sigma _{k\mathrm{o}}^{2}N\dint\limits_{-\infty
}^{+\infty }\dint\limits_{-\infty }^{+\infty }\left\vert \psi \right\vert
^{2}dKd\tilde{k}  \notag \\
&\approx &\dint\limits_{-\infty }^{+\infty }dK\exp \left( -\frac{K^{2}}{%
4\sigma _{k\mathrm{o}}^{2}}\right) \dint\limits_{-\infty }^{+\infty }d\tilde{%
k}\exp \left( -\frac{\tilde{k}^{2}}{\sigma _{k\mathrm{o}}^{2}}\right) \left[
1+8R_{\mathrm{o}}\left( \tilde{k}^{2}+2k_{\mathrm{o}}\tilde{k}+k_{\mathrm{o}%
}^{2}\right) \dsum\limits_{n=0}^{\infty }A_{n}\tilde{k}^{n}+\right.  \notag
\\
&&\left. +\frac{16\left( k_{\mathrm{o}}^{2}+\sigma _{k\mathrm{o}}^{4}R_{%
\mathrm{o}}^{2}\right) }{\sigma _{k\mathrm{o}}^{4}}\left( \tilde{k}^{4}+4k_{%
\mathrm{o}}\tilde{k}^{3}+6k_{\mathrm{o}}^{2}\tilde{k}^{2}+4k_{\mathrm{o}}^{3}%
\tilde{k}+k_{\mathrm{o}}^{4}\right) \dsum\limits_{n=0}^{\infty }B_{n}\tilde{k%
}^{n}\right] ,  \label{rho4}
\end{eqnarray}%
where $\tilde{k}\equiv k-k_{\mathrm{o}}$. From \cite{Gradshtyne} we find%
\begin{equation}
\dint\limits_{-\infty }^{+\infty }d\tilde{k}\exp \left( -\frac{\tilde{k}^{2}%
}{\sigma _{k\mathrm{o}}^{2}}\right) \tilde{k}^{n}=\delta _{n,2m}\left(
2m-1\right) !!\sqrt{\pi }\sigma _{k\mathrm{o}}\left( \frac{\sigma _{k\mathrm{%
o}}^{2}}{2}\right) ^{m},  \label{rho7}
\end{equation}%
where $m=0,1,2,\ldots $. It should be noted however, that we have
approximated $f\left( k\right) $ as a real-valued function $f\left( k\right)
\approx \frac{\theta \left( k\right) }{k}$, assuming $\theta \left( k\right)
\ll 1$. For low energy $s$-wave scattering, which is the case presently
under consideration, we have $k\ll 1$ and $\theta \left( k\right) =-ka_{%
\mathrm{s}}+\mathcal{O}\left( k^{2}\right) $, where the parameter $a_{%
\mathrm{s}}$ of dimension length is defined as the $s$-wave scattering
length \cite{Landau}. This leads to $f\left( k\right) \approx -a_{\mathrm{s}%
} $ and $f^{[p]}\left( k\right) \approx 0$, where the superscript $^{[p]}$
denotes any $p$-th order derivative ($p=1,2,\ldots $). Hence, we have $%
A_{0}= $ $-a_{\mathrm{s}}$, $A_{1}=A_{2}=\cdots =0$ and $B_{0}=a_{\mathrm{s}%
}^{2}$, $B_{1}=B_{2}=\cdots =0$. Making use of these coefficients as well as
(\ref{rho7}), one may compute the integral in (\ref{rho4}) to obtain 
\begin{eqnarray}
2\pi \sigma _{k\mathrm{o}}^{2}N\dint\limits_{-\infty }^{+\infty
}\dint\limits_{-\infty }^{+\infty }\left\vert \psi \right\vert
^{2}dk_{1}dk_{2} &=&2\pi \sigma _{k\mathrm{o}}^{2}N\dint\limits_{-\infty
}^{+\infty }\dint\limits_{-\infty }^{+\infty }\left\vert \psi \right\vert
^{2}dKd\tilde{k}  \notag \\
&\approx &2\pi \sigma _{k\mathrm{o}}^{2}\left[ 1-4\left( 2k_{\mathrm{o}%
}^{2}+\sigma _{k\mathrm{o}}^{2}\right) R_{\mathrm{o}}a_{\mathrm{s}}+\frac{%
4\left( k_{\mathrm{o}}^{2}+\sigma _{k\mathrm{o}}^{4}R_{\mathrm{o}%
}^{2}\right) \left( 4k_{\mathrm{o}}^{4}+12k_{\mathrm{o}}^{2}\sigma _{k%
\mathrm{o}}^{2}+3\sigma _{k\mathrm{o}}^{4}\right) }{\sigma _{k\mathrm{o}}^{4}%
}a_{\mathrm{s}}^{2}\right] .  \label{rho8}
\end{eqnarray}%
Indeed, one would obtain the same result as (\ref{rho8}) to the leading
order by evaluating the following integral:%
\begin{eqnarray}
&&\dint\limits_{-\infty }^{+\infty }\dint\limits_{-\infty }^{+\infty
}dk_{1}dk_{2}\exp \left\{ -\frac{1}{2\left( 1-r_{\mathrm{QM}}^{2}\right) }%
\left[ \frac{\left( k_{1}-k_{\mathrm{o}}\right) ^{2}}{\sigma _{k\mathrm{o}%
}^{2}}-2r_{\mathrm{QM}}\frac{\left( k_{1}-k_{\mathrm{o}}\right) \left(
k_{2}+k_{\mathrm{o}}\right) }{\sigma _{k\mathrm{o}}^{2}}+\frac{\left(
k_{2}+k_{\mathrm{o}}\right) ^{2}}{\sigma _{k\mathrm{o}}^{2}}\right] \right\}
\notag \\
&=&2\pi \sigma _{k\mathrm{o}}^{2}\sqrt{1-r_{\mathrm{QM}}^{2}}=2\pi \sigma _{k%
\mathrm{o}}^{2}\left[ 1-\frac{1}{2}r_{\mathrm{QM}}^{2}+\mathcal{O}\left( r_{%
\mathrm{QM}}^{4}\right) \right] ,  \label{rho11}
\end{eqnarray}%
where%
\begin{equation}
r_{\mathrm{QM}}\equiv \sqrt{8\left( 2k_{\mathrm{o}}^{2}+\sigma _{k\mathrm{o}%
}^{2}\right) R_{\mathrm{o}}a_{\mathrm{s}}}\ll 1.  \label{rho12}
\end{equation}%
Therefore, to a good approximation, we may replace the probability density
in (\ref{eq:psi_sq}) with 
\begin{equation}
P_{\text{\textrm{QM}}}^{\text{after}}\equiv \left\vert \psi ^{\text{after}%
}\left( k_{1},k_{2},t\right) \right\vert ^{2}\simeq \frac{\exp \left\{ -%
\frac{1}{2\left( 1-r_{\mathrm{QM}}^{2}\right) }\left[ \frac{\left( k_{1}-k_{%
\mathrm{o}}\right) ^{2}}{\sigma _{k\mathrm{o}}^{2}}-2r_{\mathrm{QM}}\frac{%
\left( k_{1}-k_{\mathrm{o}}\right) \left( k_{2}+k_{\mathrm{o}}\right) }{%
\sigma _{k\mathrm{o}}^{2}}+\frac{\left( k_{2}+k_{\mathrm{o}}\right) ^{2}}{%
\sigma _{k\mathrm{o}}^{2}}\right] \right\} }{2\pi \sigma _{k\mathrm{o}}^{2}%
\sqrt{1-r_{\mathrm{QM}}^{2}}},  \label{eq:psi_sq2}
\end{equation}%
where the integral in (\ref{rho11}) has been normalized. In case $r_{\mathrm{%
QM}}=0$, (\ref{eq:psi_sq2}) reduces to 
\begin{equation}
P_{\text{\textrm{QM}}}^{\text{after}}=P_{\text{\textrm{QM}}}^{\text{before}%
}=\left\vert \psi ^{\text{before}}\left( k_{1},k_{2}\right) \right\vert ^{2}=%
\frac{1}{2\pi \sigma _{k\mathrm{o}}^{2}}\exp \left[ -\frac{\left( k_{1}-k_{%
\mathrm{o}}\right) ^{2}+\left( k_{2}+k_{\mathrm{o}}\right) ^{2}}{2\sigma _{k%
\mathrm{o}}^{2}}\right] ,  \label{Pbef}
\end{equation}%
which is verified via (\ref{eq:int0}), (\ref{int1}) and (\ref{int2}).

The obtained expressions for the probability densities $P_{\text{\textrm{QM}}%
}^{\text{before}}$ and $P_{\text{\textrm{QM}}}^{\text{after}}$ motivate our
information geometric investigation as will be explained in the next Section.

\section{The information geometric perspective\label{sec-info-geo-pers}}

In this Section, we outline the main ideas behind the IGAC and present the
information geometry of correlated and uncorrelated Gaussian statistical
manifolds employed in our investigation of scattering induced quantum
entanglement.

\subsection{On the IGAC}

\textbf{\ }IGAC \cite{carlo-tesi, carlo-CSF} is a theoretical framework
developed to study the complexity of informational geodesic flows describing
physical. It is the information geometric analogue of conventional
geometrodynamical approaches to chaos \cite{casetti, di bari, cafaroPD,
cafaroPA, fox} where the classical configuration space\ is replaced by a
curved statistical manifold\textbf{\ }with the additional possibility of
considering chaotic dynamics arising from non-conformally flat metrics.
Additionally, it is an information geometric extension of the Jacobi
geometrodynamics (the geometrization of a Hamiltonian system by transforming
it to a geodesic flow \cite{jacobi}).

More specifically, IGAC \ is the application of entropic dynamics (ED) \cite%
{caticha1} to complex systems of arbitrary nature. ED is a theoretical
framework that arises from the combination of inductive inference (Maximum
Relative Entropy methods, \cite{caticha2, adom}) and Information Geometry
(IG), that is, Riemannian geometry applied to probability theory \cite{amari}%
. IGAC\ extends the applicability of ED to temporally-complex (chaotic)
dynamical systems on curved statistical manifolds and relevant measures of
chaoticity of such an IGAC have been identified \cite{carlo-tesi}.

The essential ideas underlying the IGAC and the construction of statistical
manifolds are presented in what follows.\textbf{\ }Let the probability
distribution function (PDF) $P\left( X|\Theta \right) $ represent the
maximally probable description of the system being considered. The quantity $%
X$ is a random variable that represents a microstate of the system, while $%
\Theta $ represents a macrostate. The sets $\left\{ X\right\} $ and $\left\{
\Theta \right\} $ form the microspace $\mathcal{X}$ and the parameter space $%
\mathcal{D}_{\Theta }$, respectively. The set of probability distributions
forms the statistical manifold $\mathcal{M}$. A geodesic curve on a curved
statistical manifold $\mathcal{M}$ represents the maximum probability path a
complex dynamical system explores in its evolution between initial and final
macrostates $\Theta _{\text{I}}$ and $\Theta _{\text{F}}$, respectively.
Each point of the geodesic on an $n$-dimensional statistical manifold $%
\mathcal{M}$ represents a macrostate parametrized by the macroscopic
dynamical variables $\left\{ \Theta \right\} $. Furthermore, each macrostate
is in a one-to-one correspondence with the probability distribution $P\left(
X|\Theta \right) $ representing the maximally probable description of the
system being considered. The main goal of an ED model is that of inferring 
\emph{\textquotedblleft macroscopic predictions\textquotedblright }\ in the
absence of detailed knowledge of the microscopic nature of the arbitrary
complex systems being considered. More explicitly, by \textquotedblleft
macroscopic prediction\textquotedblright\ we mean knowledge of the
statistical parameters (expectation values) of the probability distribution
function that best reflects what is known about the system. This is an
important conceptual point. The probability distribution reflects the system
in general, not the microstates. Once the microstates have been defined, we
then select the relevant information about the system. In other words, we
have to select the macrospace of the system. We emphasize that knowledge of
both initial and final macrostates is not necessary to carry out macroscopic
predictions. For instance, one may only have knowledge of the initial state
and assume that the system evolves to other states, without actually knowing
what the final state is. In such a case, it can be shown that the system
moves continuously and irreversibly along the entropy gradient \cite%
{catichaIED}. We note that in its present form the IGAC\ can only be applied
to CVQS. This restricted applicability is due to the fact that the IGAC is
used to understand the evolution of continuous trajectories on $\mathcal{M}$%
. In the context of quantum mechanical systems, the set $\left\{ \Theta
\right\} $ would correspond to the continuous eigenvalue spectrum of some
observable (expected values). The IGAC\ must be reformulated in order to be
applicable to general quantum systems. Such a reformulation is currently in
progress.

For a brief overview of some of the latest applications of the IGAC to both
classical and quantum scenarios, we refer to \cite{carlo-stefano}.

\subsection{Gaussian Statistical Models and Micro-correlations}

Here, we introduce the notion of Gaussian statistical models (manifolds) in
either absence or presence of correlations between the microscopic degrees
of freedom of the system (i.e. micro-correlations).

\subsubsection{Statistical Models in Absence of Micro-correlations}

Consider a statistical model whose microstates span a $n$-dimensional space
labeled by the variables $\left\{ X\right\} =\left\{ x_{1},x_{2},\ldots
,x_{n}\right\} $ with $x_{j}\in \mathbb{R}$, $\forall j=1,\ldots ,n$. We
assume the only testable information pertaining to the quantities $x_{j}$
consists of the expectation values $\left\langle x_{j}\right\rangle $ and
the variance $\Delta x_{j}$. The set of these expected values define the $2n$%
-dimensional space of macrostates of the system. Our $2n$-dimensional
statistical model represents a macroscopic (i.e. probabilistic) description
of a microscopic, $n$-dimensional physical system evolving over\textbf{\ }a $%
n$-dimensional (micro) space.\ We assume that all information relevant to
the dynamical evolution of the system is contained in the probability
distributions. For this reason, no other information is required. Each
macrostate may be thought as a point of a $2n$-dimensional statistical
manifold with coordinates given by the numerical values of the expectations $%
^{\left( 1\right) }\vartheta _{j}$ and $^{\left( 2\right) }\vartheta _{j}$.
The available \textit{relevant information} can be written in the form of
the following $2n$ information constraint equations:%
\begin{equation}
\left\langle x_{j}\right\rangle =\dint\limits_{-\infty }^{+\infty
}dx_{j}x_{j}P_{j}\left( x_{j}|^{\left( 1\right) }\vartheta _{j},^{\left(
2\right) }\vartheta _{j}\right) ,\text{ \ }\Delta x_{j}=\left[
\dint\limits_{-\infty }^{+\infty }dx_{j}\left( x_{j}-\left\langle
x_{j}\right\rangle \right) ^{2}P_{j}\left( x_{j}|^{\left( 1\right)
}\vartheta _{j},^{\left( 2\right) }\vartheta _{j}\right) \right] ^{\frac{1}{2%
}}.  \label{C1}
\end{equation}%
The probability distributions $P_{j}\left( x_{j}|^{\left( 1\right)
}\vartheta _{j},^{\left( 2\right) }\vartheta _{j}\right) $ in (\ref{C1}) are
constrained by the conditions of normalization,%
\begin{equation}
\dint\limits_{-\infty }^{+\infty }dx_{j}P_{j}\left( x_{j}|^{\left( 1\right)
}\vartheta _{j},^{\left( 2\right) }\vartheta _{j}\right) =1.  \label{C2}
\end{equation}%
Maximum Relative Entropy methods \cite{caticha(REII), caticha-giffin,
caticha2, adom} allow us to associate a probability distribution $P\left(
X|\Theta \right) $ to each point in the space of states $\left\{ \Theta
\right\} $. The\textbf{\ }distribution that best reflects the information
contained in the prior distribution\textbf{\ }$m\left( X\right) $\textbf{\ }%
updated by the information\textbf{\ }$\left( \left\langle x_{j}\right\rangle
,\Delta x_{j}\right) $\textbf{\ }is obtained by maximizing the relative
entropy 
\begin{equation}
S\left( \Theta \right) =-\int dXP\left( X|\Theta \right) \ln \left( \frac{%
P\left( X|\Theta \right) }{m\left( X\right) }\right) .  \label{RE}
\end{equation}%
As a working hypothesis, the prior $m\left( X\right) $ is set to be uniform
since we assume the lack of prior available information about the system
(postulate of equal \textit{a priori} probabilities). Information theory
identifies the Gaussian distribution as the maximum entropy distribution if
only the expectation value and the variance are known \cite{jaynes2}.
Indeed, upon maximizing (\ref{RE}) given the constraints (\ref{C1}) and (\ref%
{C2}), we obtain%
\begin{equation}
P\left( X|\Theta \right) =\dprod\limits_{j=1}^{n}P_{j}\left( x_{j}|^{\left(
1\right) }\vartheta _{j},^{\left( 2\right) }\vartheta _{j}\right) ,
\label{PDG}
\end{equation}%
where%
\begin{equation}
P_{j}\left( x_{j}|^{\left( 1\right) }\vartheta _{j},^{\left( 2\right)
}\vartheta _{j}\right) =\left( 2\pi \sigma _{j}^{2}\right) ^{-\frac{1}{2}%
}\exp \left[ -\frac{\left( x_{j}-\mu _{j}\right) ^{2}}{2\sigma _{j}^{2}}%
\right] ,
\end{equation}%
and\textbf{\ }in standard notation for Gaussians, $^{\left( 1\right)
}\vartheta _{j}\overset{\text{def}}{=}\left\langle x_{j}\right\rangle \equiv
\mu _{j}$, $^{\left( 2\right) }\vartheta _{j}\overset{\text{def}}{=}\Delta
x_{j}\equiv \sigma _{j}$. The probability distribution (\ref{PDG}) encodes
the available information concerning the system.

The statistical manifold $\mathcal{M}$ associated to (\ref{PDG}) is formally
defined as follows:%
\begin{equation}
\mathcal{M}=\left\{ \left. P\left( X|\Theta \right) =\underset{j=1}{\overset{%
n}{\dprod }}P_{j}\left( x_{j}|\mu _{j},\sigma _{j}\right) :\Theta =\left(
\vartheta ^{1},\ldots ,\vartheta ^{2n}\right) \in \mathcal{D}_{\Theta
}^{\left( \text{total}\right) }\right\vert P\left( X|\Theta \right) \geq
0\right\} .  \label{manifold}
\end{equation}%
The parameter space $\mathcal{D}_{\Theta }^{\left( \text{total}\right) }$
(homeomorphic to\textbf{\ }$\mathcal{M}$\textbf{)} is defined as%
\begin{equation}
\mathcal{D}_{\Theta }^{\left( \text{total}\right) }\overset{\text{def}}{=}%
\dbigotimes\limits_{k=1}^{2n}\mathcal{I}_{\vartheta ^{k}}=\left( \mathcal{I}%
_{\vartheta ^{1}}\otimes \mathcal{I}_{\vartheta ^{2}}\cdots \otimes \mathcal{%
I}_{\vartheta ^{2n}}\right) \subseteq \mathbb{R}^{2n},  \label{is}
\end{equation}%
where $\mathcal{I}_{\vartheta ^{k}}$ is a subset of $\mathbb{R}$ and
represents the entire range of accessible values for the macrovariable $%
\vartheta ^{k}$.

The line element $ds^{2}$ arising from (\ref{PDG}) is \cite{cafaroIJTP}%
\begin{equation}
ds_{\mathcal{M}}^{2}=g_{ab}\left( \Theta \right) d\vartheta ^{a}d\vartheta
^{b}=\dsum\limits_{j=1}^{n}\left( \frac{1}{\sigma _{j}^{2}}d\mu _{j}^{2}+%
\frac{2}{\sigma _{j}^{2}}d\sigma _{j}^{2}\right) \text{ with }a,b=1,\ldots
,2n.
\end{equation}%
Note that we have assumed uncoupled constraints among microvariables $x_{j}$%
. In other words, we assumed that information about correlations between the
microvariables need not to be tracked. This assumption leads to the
simplified product rule (\ref{PDG}).

A measure of distinguishability among the macrostates of the Gaussian model
is achieved by assigning a probability distribution $P\left( X|\Theta
\right) $ to each $2n$-dimensional macrostate $\Theta \overset{\text{def}}{=}%
\left\{ \left( ^{\left( 1\right) }\vartheta _{j},^{\left( 2\right)
}\vartheta _{j}\right) \right\} _{n\text{-pairs}}$ $=\left\{ \left(
\left\langle x_{j}\right\rangle ,\Delta x_{j}\right) \right\} _{n\text{-pairs%
}}$. The process of assigning a probability distribution to each state
provides $\mathcal{M}$ with a metric structure. Specifically, the Fisher-Rao
information metric $g_{ab}\left( \Theta \right) $ \cite{amari} is a measure
of distinguishability among macrostates on the statistical manifold $%
\mathcal{M}$, 
\begin{equation}
g_{ab}\left( \Theta \right) =\int dXP\left( X|\Theta \right) \partial
_{a}\ln P\left( X|\Theta \right) \partial _{b}\ln P\left( X|\Theta \right)
=4\int dX\partial _{a}\sqrt{P\left( X|\Theta \right) }\partial _{b}\sqrt{%
P\left( X|\Theta \right) },  \label{FRM}
\end{equation}%
with $a,b=1,\ldots ,2n$ and $\partial _{a}=\frac{\partial }{\partial
\vartheta ^{a}}$. It assigns an information geometry to the space of states.
The information metric $g_{ab}\left( \Theta \right) $ is a symmetric and
positive definite Riemannian metric. For the sake of completeness and in
view of its potential relevance in the study of correlations, we point out
that the Fisher-Rao metric satisfies the following two properties: 1)
invariance under (invertible) transformations of microvariables $\left\{
X\right\} \in \mathcal{X}$; 2) covariance under reparametrization of the
statistical macrospace $\left\{ \Theta \right\} \in \mathcal{D}_{\Theta }$.
The invariance of $g_{ab}\left( \Theta \right) $ under reparametrization of
the microspace $\mathcal{X}$ implies that \cite{amari},%
\begin{equation}
\mathcal{X}\subseteq \mathbb{R}^{n}\ni x\longmapsto y\overset{\text{def}}{=}%
f\left( x\right) \in \mathcal{Y}\subseteq \mathbb{R}^{n}\Longrightarrow
p\left( x|\vartheta \right) \longmapsto p^{\prime }\left( y|\vartheta
\right) =\left[ \frac{1}{\left\vert \frac{\partial f}{\partial x}\right\vert 
}p\left( x|\vartheta \right) \right] _{x=f^{-1}\left( y\right) }.
\end{equation}%
The covariance under reparametrization of the parameter space\textbf{\ }$%
\mathcal{D}_{\Theta }$\textbf{\ }(homeomorphic to\textbf{\ }$\mathcal{M}$%
\textbf{)} implies that \cite{amari},%
\begin{equation}
\mathcal{D}_{\Theta }\ni \vartheta \longmapsto \vartheta ^{\prime }\overset{%
\text{def}}{=}f\left( \vartheta \right) \in \mathcal{D}_{\Theta ^{\prime
}}\Longrightarrow g_{ab}\left( \vartheta \right) \longmapsto g_{ab}^{\prime
}\left( \vartheta ^{\prime }\right) =\left[ \frac{\partial \vartheta ^{c}}{%
\partial \vartheta ^{\prime a}}\frac{\partial \vartheta ^{d}}{\partial
\vartheta ^{\prime b}}g_{cd}\left( \vartheta \right) \right] _{\vartheta
=f^{-1}\left( \vartheta ^{\prime }\right) },
\end{equation}%
where%
\begin{equation}
g_{ab}^{\prime }\left( \vartheta ^{\prime }\right) =\int dxp^{\prime }\left(
x|\vartheta ^{\prime }\right) \partial _{a}^{\prime }\ln p^{\prime }\left(
x|\vartheta ^{\prime }\right) \partial _{b}^{\prime }\ln p^{\prime }\left(
x|\vartheta ^{\prime }\right) ,
\end{equation}%
with $\partial _{a}^{\prime }=\frac{\partial }{\partial \vartheta ^{\prime a}%
}$ and $p^{\prime }\left( x|\vartheta ^{\prime }\right) =p\left( x|\vartheta
=f^{-1}\left( \vartheta ^{\prime }\right) \right) $.

\subsubsection{Statistical Models in Presence of Micro-correlations}

Coupled constraints would lead to a {}\textquotedblleft
generalized\textquotedblright\ product rule in (\ref{PDG}) and to a metric
tensor (\ref{FRM}) with non-trivial off-diagonal elements (covariance
terms). In presence of correlated degrees of freedom $\left\{ x_{j}\right\} $%
, the \textquotedblleft generalized\textquotedblright\ product rule becomes%
\begin{equation}
P_{\text{total}}\left( x_{1},\ldots ,x_{n}\right)
=\dprod\limits_{j=1}^{n}P_{j}\left( x_{j}\right) \overset{\text{correlations}%
}{\longrightarrow }P_{\text{total}}^{\prime }\left( x_{1},\ldots
,x_{n}\right) \neq \dprod\limits_{j=1}^{n}P_{j}\left( x_{j}\right) ,
\end{equation}%
where%
\begin{equation}
P_{\text{total}}^{\prime }\left( x_{1},\ldots ,x_{n}\right) =P_{n}\left(
x_{n}|x_{1},\ldots ,x_{n-1}\right) P_{n-1}\left( x_{n-1}|x_{1},\ldots
,x_{n-2}\right) \cdots P_{2}\left( x_{2}|x_{1}\right) P_{1}\left(
x_{1}\right) .
\end{equation}%
For instance, correlations in the degrees of freedom may be introduced in
terms of the following information-constraints,%
\begin{equation}
x_{j}=f_{j}\left( x_{1},\ldots ,x_{j-1}\right) ,\text{ }\forall j=2,\ldots
,n.
\end{equation}%
In such a case, we obtain%
\begin{equation}
P_{\text{total}}^{\prime }\left( x_{1},\ldots ,x_{n}\right) =\delta \left(
x_{n}-f_{n}\left( x_{1},\ldots ,x_{n-1}\right) \right) \delta \left(
x_{n-1}-f_{n-1}\left( x_{1},\ldots ,x_{n-2}\right) \right) \cdots \delta
\left( x_{2}-f_{2}\left( x_{1}\right) \right) P_{1}\left( x_{1}\right) ,
\end{equation}%
where the $j$-th probability distribution $P_{j}\left( x_{j}\right) $ is
given by%
\begin{equation}
P_{j}\left( x_{j}\right) =\int \cdots \int dx_{1}\cdots
dx_{j-1}dx_{j+1}\cdots dx_{n}P_{\text{total}}^{\prime }\left( x_{1},\ldots
,x_{n}\right) .
\end{equation}%
Correlations between the microscopic degrees of freedom of the system $%
\left\{ x_{j}\right\} $ (micro-correlations) are conventionally introduced
by means of the correlation coefficients $r_{ij}^{\left( \text{micro}\right)
}$ \cite{roz}, 
\begin{equation}
r_{ij}^{\left( \text{micro}\right) }=r\left( x_{i},x_{j}\right) \overset{%
\text{def}}{=}\frac{\left\langle x_{i}x_{j}\right\rangle -\left\langle
x_{i}\right\rangle \left\langle x_{j}\right\rangle }{\sigma _{i}\sigma _{j}}%
\text{ with }\sigma _{i}=\sqrt{\left\langle \left( x_{i}-\left\langle
x_{i}\right\rangle \right) ^{2}\right\rangle },  \label{corr-coeff}
\end{equation}%
with $r_{ij}^{\left( \text{micro}\right) }\in \left( -1,1\right) $ and $%
i,j=1,\ldots ,n$. For the $2n$-dimensional Gaussian statistical model in
presence of micro-correlations, the system is described by the following
probability distribution $P\left( X|\Theta \right) $:%
\begin{equation}
P\left( X|\Theta \right) =\frac{1}{\left[ \left( 2\pi \right) ^{n}\det
C\left( \Theta \right) \right] ^{\frac{1}{2}}}\exp \left[ -\frac{1}{2}\left(
X-M\right) ^{t}\cdot C^{-1}\left( \Theta \right) \cdot \left( X-M\right) %
\right] \neq \dprod\limits_{j=1}^{n}\left( 2\pi \sigma _{j}^{2}\right) ^{-%
\frac{1}{2}}\exp \left[ -\frac{\left( x_{j}-\mu _{j}\right) ^{2}}{2\sigma
_{j}^{2}}\right] ,  \label{CG}
\end{equation}%
where $X=\left( x_{1},\ldots ,x_{n}\right) $, $M=\left( \mu _{1},\ldots ,\mu
_{n}\right) $ and $C\left( \Theta \right) $ is the $\left( 2n\times
2n\right) $-dimensional (non-singular) covariance matrix.

In what follows, we will introduce the three-dimensional micro-correlated
Gaussian statistical model being investigated.

\subsection{The Two-variable Micro-correlated Gaussian Statistical Model}

Consider micro-correlated Gaussian statistical models with $2n=4$. For $n=2$%
, (\ref{CG}) leads to the probability distribution $P\left( x,y|\mu _{x},\mu
_{y},\sigma _{x},\sigma _{y}\right) $ \cite{roz},%
\begin{equation}
P\left( x,y|\mu _{x},\mu _{y},\sigma _{x},\sigma _{y};r\right) =\frac{\exp
\left\{ -\frac{1}{2\left( 1-r^{2}\right) }\left[ \frac{\left( x-\mu
_{x}\right) ^{2}}{\sigma _{x}^{2}}-2r\frac{\left( x-\mu _{x}\right) \left(
y-\mu _{y}\right) }{\sigma _{x}\sigma _{y}}+\frac{\left( y-\mu _{y}\right)
^{2}}{\sigma _{y}^{2}}\right] \right\} }{2\pi \sigma _{x}\sigma _{y}\sqrt{%
1-r^{2}}},  \label{2g}
\end{equation}%
a bivariate normal distribution where $\sigma _{x}>0$, $\sigma _{y}>0$, $%
r\in \left( -1,1\right) $, $X=\left( x,y\right) $, $\Theta =\left( \mu
_{x},\mu _{y},\sigma _{x},\sigma _{y}\right) $ and $C\left( \Theta \right) $,%
\begin{equation}
C_{ij}=\left[ 
\begin{array}{cc}
\sigma _{x}^{2} & r\sigma _{x}\sigma _{y}+\mu _{x}\mu _{y} \\ 
r\sigma _{y}\sigma _{x}+\mu _{y}\mu _{x} & \sigma _{y}^{2}%
\end{array}%
\right] \text{ with }i,j=1,2.
\end{equation}%
Substituting (\ref{2g}) in (\ref{FRM}), the Fisher-Rao information metric $%
g_{ab}\left( \mu _{x},\mu _{y},\sigma _{x},\sigma _{y};r\right) $ becomes%
\begin{equation}
g_{ab}\left( \mu _{x},\mu _{y},\sigma _{x},\sigma _{y};r\right) =\left( 
\begin{array}{cccc}
-\frac{1}{\sigma _{x}^{2}\left( r^{2}-1\right) } & 0 & \frac{r}{\sigma
_{x}\sigma _{y}\left( r^{2}-1\right) } & 0 \\ 
0 & -\frac{2-r^{2}}{\sigma _{x}^{2}\left( r^{2}-1\right) } & 0 & \frac{r^{2}%
}{\sigma _{x}\sigma _{y}\left( r^{2}-1\right) } \\ 
\frac{r}{\sigma _{x}\sigma _{y}\left( r^{2}-1\right) } & 0 & -\frac{1}{%
\sigma _{y}^{2}\left( r^{2}-1\right) } & 0 \\ 
0 & \frac{r^{2}}{\sigma _{x}\sigma _{y}\left( r^{2}-1\right) } & 0 & -\frac{%
2-r^{2}}{\sigma _{y}^{2}\left( r^{2}-1\right) }%
\end{array}%
\right) .  \label{corr}
\end{equation}%
The infinitesimal line element $ds_{\mathcal{M}_{\text{corr.}}^{4\text{D}%
}}^{2}$ relative to $g_{ab}\left( \mu _{x},\mu _{y},\sigma _{x},\sigma
_{y};r\right) $ is given by%
\begin{eqnarray}
ds_{\mathcal{M}_{\text{corr.}}^{4\text{D}}}^{2} &=&g_{11}\left( \sigma
_{x};r\right) d\mu _{x}^{2}+g_{33}\left( \sigma _{y};r\right) d\mu
_{y}^{2}+g_{22}\left( \sigma _{x};r\right) d\sigma _{x}^{2}+g_{44}\left(
\sigma _{y};r\right) d\sigma _{y}^{2}+2g_{13}\left( \sigma _{x},\sigma
_{y};r\right) d\mu _{x}d\mu _{y}  \notag \\
&&+2g_{24}\left( \sigma _{x},\sigma _{y};r\right) d\sigma _{x}d\sigma _{y},
\label{le1}
\end{eqnarray}%
where%
\begin{eqnarray}
g_{11}\left( \sigma _{x};r\right) &=&-\frac{1}{\sigma _{x}^{2}\left(
r^{2}-1\right) },\text{ }g_{13}\left( \sigma _{x},\sigma _{y};r\right) =%
\frac{r}{\sigma _{x}\sigma _{y}\left( r^{2}-1\right) },\text{ }g_{22}\left(
\sigma _{x};r\right) =-\frac{2-r^{2}}{\sigma _{x}^{2}\left( r^{2}-1\right) },
\notag \\
g_{24}\left( \sigma _{x},\sigma _{y};r\right) &=&\frac{r^{2}}{\sigma
_{x}\sigma _{y}\left( r^{2}-1\right) },\text{ }g_{31}\left( \sigma
_{x},\sigma _{y};r\right) =\frac{r}{\sigma _{x}\sigma _{y}\left(
r^{2}-1\right) },\text{ }g_{33}\left( \sigma _{y};r\right) =-\frac{1}{\sigma
_{y}^{2}\left( r^{2}-1\right) },  \notag \\
g_{42}\left( \sigma _{x},\sigma _{y};r\right) &=&\frac{r^{2}}{\sigma
_{x}\sigma _{y}\left( r^{2}-1\right) },\text{ }g_{44}\left( \sigma
_{y};r\right) =-\frac{2-r^{2}}{\sigma _{y}^{2}\left( r^{2}-1\right) }.
\end{eqnarray}%
It is rather difficult to present an analytical study of the IGAC associated
with infinitesimal line element $ds_{\mathcal{M}_{\text{corr.}}^{4\text{D}%
}}^{2}$ in (\ref{le1}). Such a study will be the subject of forthcoming
investigations. In the present work we consider a special class of Gaussian
models, namely those in which $\sigma _{y}=\sigma _{x}=\sigma $. Then the
probability distribution $P\left( x,y|\mu _{x},\mu _{y},\sigma _{x},\sigma
_{y};r\right) $ in (\ref{2g}) can be reduced to a simpler form, 
\begin{equation}
P\left( x,y|\mu _{x},\mu _{y},\sigma ;r\right) =\frac{\exp \left\{ -\frac{1}{%
2\left( 1-r^{2}\right) }\left[ \frac{\left( x-\mu _{x}\right) ^{2}}{\sigma
^{2}}-2r\frac{\left( x-\mu _{x}\right) \left( y-\mu _{y}\right) }{\sigma ^{2}%
}+\frac{\left( y-\mu _{y}\right) ^{2}}{\sigma ^{2}}\right] \right\} }{2\pi
\sigma ^{2}\sqrt{1-r^{2}}},  \label{eq:new_P}
\end{equation}%
where $\sigma >0$, $X=\left( x,y\right) $, $\Theta =\left( \mu _{x},\mu
_{y},\sigma \right) $ and $C\left( \Theta \right) $,%
\begin{equation}
C_{ij}=\left[ 
\begin{array}{cc}
\sigma ^{2} & r\sigma ^{2}+\mu _{x}\mu _{y} \\ 
r\sigma ^{2}+\mu _{y}\mu _{x} & \sigma ^{2}%
\end{array}%
\right] \text{ with }i,j=1,2.
\end{equation}%
\textbf{\ }The Fisher-Rao matrix $g_{ab}\left( \mu _{x},\mu _{y},\sigma
;r\right) $ associated with $P\left( x,y|\mu _{x},\mu _{y},\sigma ;r\right) $
reads%
\begin{equation}
g_{ab}\left( \mu _{x},\mu _{y},\sigma ;r\right) =\frac{1}{\sigma ^{2}}\left( 
\begin{array}{ccc}
-\frac{1}{r^{2}-1} & \frac{r}{r^{2}-1} & 0 \\ 
\frac{r}{r^{2}-1} & -\frac{1}{r^{2}-1} & 0 \\ 
0 & 0 & 4%
\end{array}%
\right) .  \label{cim}
\end{equation}%
\textbf{\ }The line element associated with metric $g_{ab}\left( \mu
_{x},\mu _{y},\sigma ;r\right) $ is given by%
\begin{eqnarray}
ds_{\mathcal{M}_{\text{corr.}}^{3\text{D}}}^{2} &=&g_{11}\left( \sigma
_{x};r\right) d\mu _{x}^{2}+g_{33}\left( \sigma _{y};r\right) d\mu
_{y}^{2}+2g_{13}\left( \sigma ;r\right) d\mu _{x}d\mu _{y}+\left[
g_{22}\left( \sigma ;r\right) +g_{44}\left( \sigma ;r\right) +2g_{24}\left(
\sigma ;r\right) \right] d\sigma ^{2}  \label{le} \\
&=&\frac{1}{\sigma ^{2}}\left( \frac{1}{1-r^{2}}d\mu _{x}^{2}+\frac{1}{%
1-r^{2}}d\mu _{y}^{2}-\frac{2r}{1-r^{2}}d\mu _{x}d\mu _{y}+4d\sigma
^{2}\right) .  \notag
\end{eqnarray}%
Observe\textbf{\ }that in the absence of micro-correlations, the
two-variable probability distribution (\ref{eq:new_P}) reduces to 
\begin{equation}
P\left( x,y|\mu _{x},\mu _{y},\sigma \right) =\frac{1}{2\pi \sigma ^{2}}\exp %
\left[ -\frac{\left( x-\mu _{x}\right) ^{2}}{2\sigma ^{2}}\right] \exp \left[
-\frac{\left( y-\mu _{y}\right) ^{2}}{2\sigma ^{2}}\right] ,  \label{newp1}
\end{equation}%
while the metric and corresponding line element become%
\begin{equation}
g_{ab}\left( \mu _{x},\mu _{y},\sigma \right) =\frac{1}{\sigma ^{2}}\left( 
\begin{array}{ccc}
1 & 0 & 0 \\ 
0 & 1 & 0 \\ 
0 & 0 & 4%
\end{array}%
\right)  \label{no-corrg}
\end{equation}%
and%
\begin{equation}
ds_{\mathcal{M}_{\text{non-corr.}}^{3D}}^{2}=\frac{1}{\sigma ^{2}}\left(
d\mu _{x}^{2}+d\mu _{y}^{2}+4d\sigma ^{2}\right) ,
\end{equation}%
respectively. In what follows we limit our analysis to the study of
non-negative micro-correlations, that is, we will consider $r\in \lbrack
0,1) $.

\subsection{The Information Dynamics on the Statistical Manifold $\mathcal{M}%
_{\text{corr.}}^{3\text{D}}$}

The information dynamics on the manifold $\mathcal{M}_{\text{corr.}}^{3\text{%
D}}$ represented by (\ref{cim}) can be derived from a standard principle of
least action of Jacobi type \cite{caticha1}. The geodesic equations for the
macrovariables of the Gaussian ED model are given by\textit{\ nonlinear}
second order coupled ordinary differential equations, 
\begin{equation}
\frac{d^{2}\vartheta ^{a}}{d\tau ^{2}}+\Gamma _{bc}^{a}\frac{d\vartheta ^{b}%
}{d\tau }\frac{d\vartheta ^{c}}{d\tau }=0,  \label{GE}
\end{equation}%
where $a,b,c=1,2,3$ and we denote $\vartheta ^{1}=$ $\mu _{1}=\mu _{x}$, $%
\vartheta ^{2}=\mu _{2}=\mu _{y}$, $\vartheta ^{3}=\sigma $. The connection
coefficients $\Gamma _{bc}^{a}$ appearing in (\ref{GE}) are defined as \cite%
{felice}%
\begin{equation}
\Gamma _{bc}^{a}=\frac{1}{2}g^{ad}\left( \partial _{b}g_{dc}+\partial
_{c}g_{bd}-\partial _{d}g_{bc}\right) .  \label{c}
\end{equation}%
In our case, through (\ref{cim}) the non-vanishing connection coefficients
are given by%
\begin{equation}
\Gamma _{13}^{1}=-\frac{1}{\sigma },\text{ }\Gamma _{23}^{2}=-\frac{1}{%
\sigma }=\Gamma _{32}^{2},\text{ }\Gamma _{11}^{3}=-\frac{1}{4\sigma \left(
r^{2}-1\right) },\text{ }\Gamma _{12}^{3}=\frac{r}{4\sigma \left(
r^{2}-1\right) }=\Gamma _{21}^{3},\text{ }\Gamma _{22}^{3}=-\frac{1}{4\sigma
\left( r^{2}-1\right) },\text{ }\Gamma _{33}^{3}=-\frac{1}{\sigma }.
\label{c2}
\end{equation}%
The geodesic equations in (\ref{GE}) describe a \textit{reversible} dynamics
whose solution is the trajectory between an initial $\Theta _{\text{I}}$ and
a final macrostate $\Theta _{\text{F}}$. The trajectory can be equally well
traversed in both directions. In the case under consideration, substituting (%
\ref{c2}) in (\ref{GE}), the three geodesic equations become 
\begin{eqnarray}
0 &=&\frac{d^{2}\mu _{1}\left( \tau \right) }{d\tau ^{2}}-\frac{2}{\sigma
\left( \tau \right) }\frac{d\mu _{1}\left( \tau \right) }{d\tau }\frac{%
d\sigma \left( \tau \right) }{d\tau },  \label{eq:1} \\
0 &=&\frac{d^{2}\mu _{2}\left( \tau \right) }{d\tau ^{2}}-\frac{2}{\sigma
\left( \tau \right) }\frac{d\mu _{2}\left( \tau \right) }{d\tau }\frac{%
d\sigma \left( \tau \right) }{d\tau },  \label{eq:2} \\
0 &=&\frac{d^{2}\sigma \left( \tau \right) }{d\tau ^{2}}-\frac{1}{\sigma
\left( \tau \right) }\left( \frac{d\sigma \left( \tau \right) }{d\tau }%
\right) ^{2}-\frac{1}{4\sigma \left( \tau \right) \left( r^{2}-1\right) }%
\left[ \left( \frac{d\mu _{1}\left( \tau \right) }{d\tau }\right)
^{2}+\left( \frac{d\mu _{2}\left( \tau \right) }{d\tau }\right) ^{2}\right] +
\label{eq:3} \\
&&+\frac{r}{2\sigma \left( \tau \right) \left( r^{2}-1\right) }\frac{d\mu
_{1}\left( \tau \right) }{d\tau }\frac{d\mu _{2}\left( \tau \right) }{d\tau }%
.  \notag
\end{eqnarray}%
Integration of the above coupled system of nonlinear differential equations
is non-trivial. A detailed derivation of the geodesic paths is given in
Appendix \ref{app-geod}. After integration of (\ref{eq:1}), (\ref{eq:2}) and
(\ref{eq:3}), the geodesic trajectories for the non-correlated Gaussian
system become,%
\begin{eqnarray}
\mu _{1}\left( \tau ;0\right) &=&-\sqrt{p_{\mathrm{o}}^{2}+2\sigma _{\mathrm{%
o}}^{2}}\tanh \left( A_{\mathrm{o}}\tau \right) ,  \label{eq:MU01f} \\
\mu _{2}\left( \tau ;0\right) &=&\sqrt{p_{\mathrm{o}}^{2}+2\sigma _{\mathrm{o%
}}^{2}}\tanh \left( A_{\mathrm{o}}\tau \right) ,  \label{eq:MU02f} \\
\sigma \left( \tau ;0\right) &=&\sqrt{\frac{1}{2}p_{\mathrm{o}}^{2}+\sigma _{%
\mathrm{o}}^{2}}\frac{1}{\cosh \left( A_{\mathrm{o}}\tau \right) },
\label{eq:SIGMA0f}
\end{eqnarray}%
while for the correlated Gaussian system the geodesics read,%
\begin{eqnarray}
\mu _{1}(\tau ;r) &=&-\sqrt{\left( 1-r\right) \left( p_{\mathrm{o}%
}^{2}+2\sigma _{\mathrm{o}}^{2}\right) }\tanh \left( A_{\mathrm{o}}\tau
\right) ,  \label{eq:MUr1f} \\
\mu _{2}(\tau ;r) &=&\sqrt{\left( 1-r\right) \left( p_{\mathrm{o}%
}^{2}+2\sigma _{\mathrm{o}}^{2}\right) }\tanh \left( A_{\mathrm{o}}\tau
\right) ,  \label{eq:MUr2f} \\
\sigma (\tau ;r) &=&\sqrt{\frac{1}{2}p_{\mathrm{o}}^{2}+\sigma _{\mathrm{o}%
}^{2}}\frac{1}{\cosh \left( A_{\mathrm{o}}\tau \right) },  \label{eq:SIGMArf}
\end{eqnarray}%
where the subscript\textbf{\ }\textquotedblleft \textbf{$_{\mathrm{o}}$}%
\textquotedblright\ denotes the initial state, and 
\begin{eqnarray}
A_{\mathrm{o}} &\equiv &\frac{1}{\tau _{\mathrm{o}}}\sinh ^{-1}\left( \frac{%
p_{\mathrm{o}}}{\sqrt{2}\sigma _{\mathrm{o}}}\right)  \notag \\
&\overset{\frac{\sigma _{\mathrm{o}}}{p_{\mathrm{o}}}\ll 1}{=}&\frac{1}{\tau
_{\mathrm{o}}}\left\{ \ln \left( \frac{\sqrt{2}p_{\mathrm{o}}}{\sigma _{%
\mathrm{o}}}\right) +\frac{1}{2}\left( \frac{\sigma _{\mathrm{o}}}{p_{%
\mathrm{o}}}\right) ^{2}-\frac{3}{8}\left( \frac{\sigma _{\mathrm{o}}}{p_{%
\mathrm{o}}}\right) ^{4}+\mathcal{O}\left[ \left( \frac{\sigma _{\mathrm{o}}%
}{p_{\mathrm{o}}}\right) ^{6}\right] \right\} ,  \label{a0}
\end{eqnarray}%
whose derivation is presented in Appendix \ref{app-refn}.

\section{Application of information geometry to quantum physics - purity,
scattering and quantum entanglement\label{sec-application}}

In this Section we use information geometric techniques in conjunction with
standard partial wave quantum scattering theory to provide an information
geometric characterization of quantum entanglement.

\subsection{Association of Quantum Systems with Information Geometric Systems%
}

We now focus on applying IG methods to the quantum entanglement\ produced by
a head-on collision between two Gaussian wave-packets in momentum space. We
observe from (\ref{eq:int0}) and (\ref{int1}) that the two-particle
probability density before collision is given by $\left\vert \psi ^{\text{%
before}}\left( k_{1},k_{2}\right) \right\vert ^{2}=a^{2}\left(
k_{1},\left\langle k_{1}\right\rangle _{\mathrm{o}};\sigma _{k\mathrm{o}%
}\right) a^{2}\left( k_{2},\left\langle k_{2}\right\rangle _{\mathrm{o}%
};\sigma _{k\mathrm{o}}\right) $. By letting $x=p_{1}=\hbar k_{1}$, $%
y=p_{2}=\hbar k_{2}$, $\mu _{x}=\mu _{p_{1}}=\hbar \mu _{k_{1}}$, $\mu
_{x}=\mu _{p_{1}}=\hbar \mu _{k_{1}}$, $\sigma =\sigma _{p}=\hbar \sigma
_{k} $ in (\ref{newp1}) and assigning $\mu _{p_{1}}\rightarrow \left\langle
p_{1}\right\rangle _{\mathrm{o}}=\hbar \left\langle k_{1}\right\rangle _{%
\mathrm{o}}=\hbar k_{\mathrm{o}}$, $\mu _{p_{2}}\rightarrow \left\langle
p_{2}\right\rangle _{\mathrm{o}}=\hbar \left\langle k_{2}\right\rangle _{%
\mathrm{o}}=-\hbar k_{\mathrm{o}}$, $\sigma _{p}\rightarrow \sigma _{\mathrm{%
o}}=\hbar \sigma _{k\mathrm{o}}$, we can identify the non-correlated
probability distribution (\ref{newp1}) with $\left\vert \psi ^{\text{before}%
}\left( k_{1},k_{2}\right) \right\vert ^{2}$ due to (\ref{Pbef}). That is,%
\begin{equation}
P_{\text{\textrm{QM}}}^{\text{before}}=P_{\text{non-corr.}}.  \label{idbef}
\end{equation}%
The information geometry associated with the two-particle system before
collision is specified by metric (\ref{no-corrg}).

In a similar manner, the probability density $\left\vert \psi ^{\text{after}%
}\left( k_{1},k_{2},t\right) \right\vert ^{2}$ in (\ref{eq:psi_sq2}) is
approximated with the Gaussian probability distribution (\ref{eq:new_P}).
Comparison of (\ref{eq:psi_sq2}) and (\ref{eq:new_P}) implies that when $r_{%
\mathrm{QM}}\ll 1$,%
\begin{equation}
P_{\text{\textrm{QM}}}^{\text{after}}\simeq P_{\text{corr.}},
\label{eq:gauss_id}
\end{equation}%
where%
\begin{equation}
P_{\text{corr.}}\equiv \frac{\exp \left\{ -\frac{1}{2\left( 1-r^{2}\right) }%
\left[ \frac{\left( k_{1}-k_{\mathrm{o}}\right) ^{2}}{\sigma _{k\mathrm{o}%
}^{2}}-2r\frac{\left( k_{1}-k_{\mathrm{o}}\right) \left( k_{2}+k_{\mathrm{o}%
}\right) }{\sigma _{k\mathrm{o}}^{2}}+\frac{\left( k_{2}+k_{\mathrm{o}%
}\right) ^{2}}{\sigma _{k\mathrm{o}}^{2}}\right] \right\} }{2\pi \sigma _{k%
\mathrm{o}}^{2}\sqrt{1-r^{2}}},  \label{eq:PIG}
\end{equation}%
with%
\begin{equation}
r=r_{\mathrm{QM}}.  \label{rQM}
\end{equation}%
The expression of $r_{\mathrm{QM}}$ in terms of physical quantities is given
by (\ref{rho12}). The information geometry associated with the two-particle
system after collision is specified by metric (\ref{cim}).

\subsection{Purity as a Measure of Quantum Entanglement}

The {}subsystem purity of a composite system of two particles engaged in a
head-on collision was calculated in \cite{Wang} by deriving the two-particle
wave function modified by\textbf{\ }$s$-wave scattering amplitudes. They
utilized the purity function\textbf{\ }$\mathcal{P}$\textbf{\ }as a measure
of entanglement. Formally, the purity function is defined as%
\begin{equation}
\mathcal{P}\overset{\text{def}}{=}\mathrm{Tr}\left( \rho _{1}^{2}\right) ,
\end{equation}%
where\textbf{\ }$\rho _{1}=\mathrm{Tr}_{2}\left( \rho _{12}\right) $\textbf{%
\ }is the reduced density matrix of particle\textbf{\ }$1$\textbf{\ }and%
\textbf{\ }$\rho _{12}$\textbf{\ }is the two-particle density matrix
associated with the post-collisional two-particle wave function, given by%
\textbf{\ }(\ref{eq:psi_aft_coll}). For pure two-particle states, the
smaller the value of $\mathcal{P}$ the higher the entanglement. That is, the
loss of purity provides an indicator of the degree of entanglement. Hence, a
disentangled product state corresponds to $\mathcal{P}=1$. We remark that
although\textbf{\ }$\mathcal{P}$\textbf{\ }shares similar features to
entropy, it is more readily accessible to theoretical analysis \cite{Grobe,
Gemmer, Jacquod} than the latter. In atomic physics, $\mathcal{P}$ has also
been employed to quantify two-body correlations in a multitude of dynamical
processes \cite{Grobe, Gemmer, Jacquod, Liu}.

Given the system, it was found in \cite{Wang} that the purity is
specifically expressed as%
\begin{equation}
\mathcal{P=}\diiiint \psi \left( \mathbf{k}_{1},\mathbf{k}_{2},t\right) \psi
\left( \mathbf{k}_{3},\mathbf{k}_{4},t\right) \psi ^{\ast }\left( \mathbf{k}%
_{1},\mathbf{k}_{4},t\right) \psi ^{\ast }\left( \mathbf{k}_{3},\mathbf{k}%
_{2},t\right) d^{3}\mathbf{k}_{1}d^{3}\mathbf{k}_{2}d^{3}\mathbf{k}_{3}d^{3}%
\mathbf{k}_{4}.  \label{purity1}
\end{equation}%
By employing the same dimensional analysis as developed in Section \ref%
{sec-entang-wave}, we may effectively reduce (\ref{purity1}) to%
\begin{equation}
\mathcal{P=}\diiiint \psi \left( k_{1},k_{2},t\right) \psi \left(
k_{3},k_{4},t\right) \psi ^{\ast }\left( k_{1},k_{4},t\right) \psi ^{\ast
}\left( k_{3},k_{2},t\right) dk_{1}dk_{2}dk_{3}dk_{4}.  \label{purity2}
\end{equation}%
Now, specifying the wave-packets in (\ref{purity2}) by means of (\ref{sc6})
and (\ref{psi1}) and performing the integral, one obtains%
\begin{equation}
\mathcal{P}=1-8\left( 2k_{\mathrm{o}}^{2}+\sigma _{k\mathrm{o}}^{2}\right)
R_{\mathrm{o}}a_{\mathrm{s}}+\mathcal{O}\left( a_{\mathrm{s}}^{2}\right) ,
\label{eq:P1}
\end{equation}%
where the parameter $a_{\mathrm{s}}$ is the $s$-wave scattering length,
defined from $f\left( k\right) \approx -a_{\mathrm{s}}$ for $k\ll 1$.
Employing the scattering cross section $\Sigma =4\pi a_{\mathrm{s}}^{2}$, we
may express the purity in an alternative manner, namely, 
\begin{equation}
\mathcal{P}=1-\frac{4\left( 2k_{\mathrm{o}}^{2}+\sigma _{k\mathrm{o}%
}^{2}\right) R_{\mathrm{o}}\sqrt{\Sigma }}{\sqrt{\pi }}+\mathcal{O}\left(
\Sigma \right) .  \label{eq:P2}
\end{equation}%
The $s$-wave scattering can also be understood in terms of a scattering
(interaction) potential and the scattering phase shift. Consider a
scattering potential%
\begin{equation}
V(x)=\left\{ 
\begin{array}{ll}
V, & \;0\leq x\leq L \\ 
0, & \;x>L%
\end{array}%
\right. ,
\end{equation}%
where $V$ denotes the height (for $V>0$; repulsive potential) or depth (for $%
V<0$; attractive potential) of the potential and $L$ the range of the
potential. Then solving the Schrödinger equation with this potential for the
scattered wave, we are led to \cite{Hunt}%
\begin{equation}
k_{\mathrm{in}}\cot \left( k_{\mathrm{in}}L\right) =k_{\mathrm{out}}\cot
\left( k_{\mathrm{out}}L+\theta \right) ,  \label{eq:cot}
\end{equation}%
with%
\begin{eqnarray}
k_{\mathrm{in}} &=&\frac{\sqrt{2\mu \left( \mathcal{E}-V\right) }}{\hbar }%
,\;\;\;0<x<L,  \label{eq:ka} \\
k_{\mathrm{out}} &=&\frac{\sqrt{2\mu \mathcal{E}}}{\hbar },\;\;\;x>L,
\label{eq:kb}
\end{eqnarray}%
where $\mu $ and $\mathcal{E}$ are the reduced mass and kinetic energy of
the two-particle system in the relative coordinates, respectively, and $k_{%
\mathrm{in}}$ and $k_{\mathrm{out}}$ represent\textbf{\ }the
conjugate-coordinate wave vectors inside and outside the potential region,
respectively. Equation (\ref{eq:cot}) together with (\ref{eq:ka}) and (\ref%
{eq:kb}) indicates that the scattering potential shifts the phase of the
scattered wave by $\theta $ at points beyond the scattering region. In the
next Subsection we will make use of this idea to determine the scattering
phase shift which is linked with the micro-correlation coefficient $r$ in
our statistical model.

\subsection{Information Geometric\ Interpretation\ of\ Quantum\ Entanglement}

We\ utilize\ the\ results\ of\ our\ information\ dynamics\ given\ by\
equations\ (\ref{eq:MU01f}),\ (\ref{eq:MU02f}),\ (\ref{eq:SIGMA0f}),\ (\ref%
{eq:MUr1f}),\ (\ref{eq:MUr2f})\ and\ (\ref{eq:SIGMArf})\ to\ furnish\ an\
information geometric\ interpretation\ of\ \textquotedblleft quantum\
entanglement\textquotedblright ,\ which\ is\ characterized\ by\ the\ purity\
given\ by\ (\ref{eq:P2}).

\subsubsection{Momentum-space Gaussian Statistical Models}

To achieve the above task, one joins two different charts of Gaussian
statistical manifolds, one without correlation (before collision) and the
other with correlation (after collision). The two models can be represented
by means of (\ref{Pbef}) and (\ref{eq:PIG}) with associated statistical
manifolds (\ref{no-corrg}) and (\ref{cim}), respectively.

The set of geodesic curves for each model is represented by equations (\ref%
{eq:MU01f}), (\ref{eq:MU02f}), (\ref{eq:SIGMA0f}) (for the non-correlated
model) and by equations (\ref{eq:MUr1f}), (\ref{eq:MUr2f}), (\ref{eq:SIGMArf}%
) (for the correlated model). The two sets are joined at the junction, $%
\tau=0$: $\tau<0$ (before collision) for the non-correlated model and $%
\tau\geq0$ (after collision) for the correlated model.

The set of curves given by equations (\ref{eq:MU01f}), (\ref{eq:MU02f}), (%
\ref{eq:SIGMA0f}) may be assigned to $\left\{ \left\langle p_{1\mathrm{b}%
}(\tau )\right\rangle ,\,\left\langle p_{2\mathrm{b}}(\tau )\right\rangle
,\,\left\langle \sigma _{\mathrm{b}}(\tau )\right\rangle \right\} $ while
the set given by equations (\ref{eq:MUr1f}), (\ref{eq:MUr2f}), (\ref%
{eq:SIGMArf}) may be assigned to $\left\{ \left\langle p_{1\mathrm{a}}(\tau
)\right\rangle ,\,\left\langle p_{2\mathrm{a}}(\tau )\right\rangle
,\,\left\langle \sigma _{\mathrm{a}}(\tau )\right\rangle \right\} $\textbf{.}
The subscripts \textquotedblleft $_{1}$\textquotedblright\ and
\textquotedblleft $_{2}$\textquotedblright ${}$ denote particle $1$ and
particle $2$, respectively; subscripts \textquotedblleft $_{\mathrm{b}}$%
\textquotedblright ${}$ and \textquotedblleft $_{\mathrm{a}}$%
\textquotedblright ${}$ denote `before' and `after' collision, respectively.
Then we may write the following two sets of equations: for $\tau <0$ (before
collision), 
\begin{eqnarray}
\left\langle p_{1\mathrm{b}}(\tau )\right\rangle &=&\mu _{1}\left( \tau
;0\right) =-\sqrt{p_{\mathrm{o}}^{2}+2\sigma _{\mathrm{o}}^{2}}\tanh \left(
A_{\mathrm{o}}\tau \right) ,  \label{eq:p1b} \\
\left\langle p_{2\mathrm{b}}(\tau )\right\rangle &=&\mu _{2}\left( \tau
;0\right) =\sqrt{p_{\mathrm{o}}^{2}+2\sigma _{\mathrm{o}}^{2}}\tanh \left(
A_{\mathrm{o}}\tau \right) ,  \label{eq:p2b} \\
\left\langle \sigma _{\mathrm{b}}(\tau )\right\rangle &=&\sigma \left( \tau
;0\right) =\sqrt{\frac{1}{2}p_{\mathrm{o}}^{2}+\sigma _{\mathrm{o}}^{2}}%
\frac{1}{\cosh \left( A_{\mathrm{o}}\tau \right) },  \label{eq:sigmab}
\end{eqnarray}%
while for $\tau \geq 0$ (after collision), 
\begin{eqnarray}
\left\langle p_{1\mathrm{a}}(\tau )\right\rangle &=&\mu _{1}\left( \tau
;r\right) =-\sqrt{\left( 1-r\right) \left( p_{\mathrm{o}}^{2}+2\sigma _{%
\mathrm{o}}^{2}\right) }\tanh \left( A_{\mathrm{o}}\tau \right) ,
\label{eq:p1a} \\
\left\langle p_{2\mathrm{a}}(\tau )\right\rangle &=&\mu _{2}\left( \tau
;r\right) =\sqrt{\left( 1-r\right) \left( p_{\mathrm{o}}^{2}+2\sigma _{%
\mathrm{o}}^{2}\right) }\tanh \left( A_{\mathrm{o}}\tau \right) ,
\label{eq:p2a} \\
\left\langle \sigma _{\mathrm{a}}(\tau )\right\rangle &=&\sigma \left( \tau
;r\right) =\sqrt{\frac{1}{2}p_{\mathrm{o}}^{2}+\sigma _{\mathrm{o}}^{2}}%
\frac{1}{\cosh \left( A_{\mathrm{o}}\tau \right) },  \label{eq:sigmaa}
\end{eqnarray}%
where $A_{\mathrm{o}}$ is given by (\ref{a0}). Here we recognize that the
momenta\ $\left\langle p_{1\mathrm{b}}(\tau )\right\rangle $ and\ $%
\left\langle p_{1\mathrm{a}}(\tau )\right\rangle $ asymptotically converge
to\ $\sqrt{p_{\mathrm{o}}^{2}+2\sigma _{\mathrm{o}}^{2}}$\ and $-\sqrt{%
\left( 1-r\right) \left( p_{\mathrm{o}}^{2}+2\sigma _{\mathrm{o}}^{2}\right) 
}$\ toward $\tau =-\infty $ and $\tau =+\infty $, respectively (the same is
true for\ $-\left\langle p_{2\mathrm{b}}(\tau )\right\rangle $ and\ $%
-\left\langle p_{2\mathrm{a}}(\tau )\right\rangle $) while $\left\langle
\sigma _{\mathrm{b}}(\tau )\right\rangle $ and $\left\langle \sigma _{%
\mathrm{a}}(\tau )\right\rangle $ are identical and vanishingly small toward 
$\tau =\pm \infty $. Furthermore, we observe that there is continuity
between $\left\langle p_{1/2\mathrm{b}}(\tau )\right\rangle $ and $%
\left\langle p_{1/2\mathrm{a}}(\tau )\right\rangle $ and between $%
\left\langle \sigma _{\mathrm{b}}(\tau )\right\rangle $ and $\left\langle
\sigma _{\mathrm{a}}(\tau )\right\rangle $ at the junction, $\tau =0$ (see
Figure \ref{fig2}). \FRAME{fhFU}{11.6319cm}{6.9336cm}{0pt}{\Qcb{Plots of $%
\left\langle p_{1}(\protect\tau )\right\rangle $, $\left\langle p_{2}(%
\protect\tau )\right\rangle $ and $\left\langle \protect\sigma (\protect\tau %
)\right\rangle $ before and after collision}}{\Qlb{fig2}}{fig2.png}{\special%
{language "Scientific Word";type "GRAPHIC";maintain-aspect-ratio
TRUE;display "USEDEF";valid_file "F";width 11.6319cm;height 6.9336cm;depth
0pt;original-width 14.2208in;original-height 8.4416in;cropleft "0";croptop
"1";cropright "1";cropbottom "0";filename '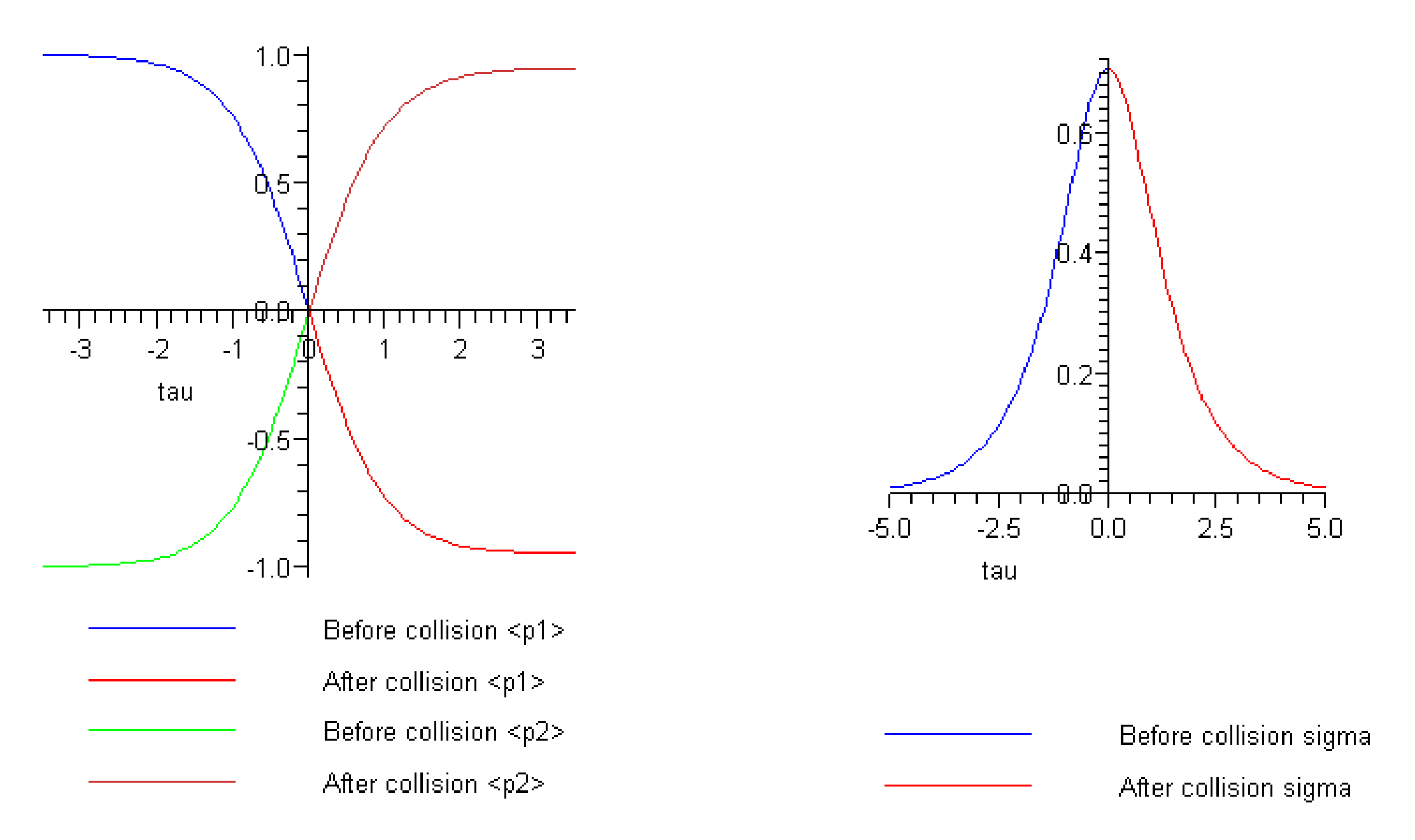';file-properties
"XNPEU";}}

\subsubsection{Correlation vs. Entanglement: Connection Established via
Scattering and Purity}

Intuitively, if the particles are not correlated (i.e. $r=0$)\ after
collision, then no entanglement should be present. In this scenario, the two
particle system would not experience any loss of purity so that $\mathcal{P}%
=1$. Indeed, this is verified by (\ref{eq:P1}). From (\ref{eq:P2}) the case $%
\mathcal{P}=1$ requires $a_{\mathrm{s}}=0$ or $\Sigma =0$, that is to say,
no scattering. For low energy $s$-wave scattering, $f\left( k\right) \approx
-a_{\mathrm{s}}$ and $\theta \left( k\right) =-ka_{\mathrm{s}}+\mathcal{O}%
\left( k^{2}\right) $. Thus one can readily determine the requirement
necessary to satisfy $a_{\mathrm{s}}=0$ or $\Sigma =0$, namely%
\begin{equation}
\theta =0.  \label{eq:theta_no-corr}
\end{equation}%
Equation (\ref{eq:theta_no-corr}) implies the $s$-wave scattering phase
shift must vanish if our system is non-correlated after collision.

A question that now arises is how to determine the scattering phase shift in
view of the fact that our statistical model is correlated after collision.
Initially, we need to examine how correlations affect the momentum geodesic
curve $\left\langle p_{1/2}(\tau )\right\rangle $. For this purpose we
define the\textbf{\ }momentum-difference curve $\left\langle p(\tau
)\right\rangle \equiv \frac{1}{2}\left[ \left\langle p_{2}(\tau
)\right\rangle -\left\langle p_{1}(\tau )\right\rangle \right] $. Comparison
of the following two equations, which follow from (\ref{eq:p1b}), (\ref%
{eq:p2b}) and (\ref{eq:p1a}), (\ref{eq:p2a}), 
\begin{eqnarray}
\left\langle p\left( \tau ;0\right) \right\rangle &\equiv &\frac{1}{2}\left[
\left\langle p_{2\mathrm{b}}(\tau )\right\rangle -\left\langle p_{1\mathrm{b}%
}(\tau )\right\rangle \right] =\sqrt{p_{\mathrm{o}}^{2}+2\sigma _{\mathrm{o}%
}^{2}}\tanh \left( A_{\mathrm{o}}\tau \right) ,  \label{eq:p_0} \\
\left\langle p\left( \tau ;r\right) \right\rangle &\equiv &\frac{1}{2}\left[
\left\langle p_{2\mathrm{a}}(\tau )\right\rangle -\left\langle p_{1\mathrm{a}%
}(\tau )\right\rangle \right] =\sqrt{\left( 1-r\right) \left( p_{\mathrm{o}%
}^{2}+2\sigma _{\mathrm{o}}^{2}\right) }\tanh \left( A_{\mathrm{o}}\tau
\right) ,  \label{eq:p_r}
\end{eqnarray}%
indicates that at any arbitrary time $\tau \geq 0$%
\begin{equation}
\left\langle p\left( \tau ;0\right) \right\rangle \geq \left\langle p\left(
\tau ;r\right) \right\rangle ,  \label{eq:p_compare}
\end{equation}%
while both (\ref{eq:p_0}) and (\ref{eq:p_r}) share the functional argument $%
A_{\mathrm{o}}\tau $. Condition (\ref{eq:p_compare}) implies that the
correlation causes the momentum to reduce for any $\tau \geq 0$ (relative to
the non-correlated situation)\textbf{.} This situation is analogous to the
change in momentum caused by a repulsive scattering potential (see (\ref%
{eq:ka}) and (\ref{eq:kb})). It is then reasonable to assume there exists
some connection between the scattering potential and the correlation.
Provided this connection is established, one should be able to determine the
scattering phase shift in terms of the correlation via equations (\ref%
{eq:cot}), (\ref{eq:ka}) and (\ref{eq:kb}). In this way, one can ultimately
establish a connection between quantum entanglement and the statistical
micro-correlation.

Recall that before collision (at the affine time $-\tau _{\mathrm{o}}$)
particles\ $1$\ and\ $2$\ are separated by\ a linear distance\ $R_{\mathrm{o}%
}$. Each particle has momenta\ $p_{\mathrm{o}}$ and\ $-p_{\mathrm{o}}$,
respectively and the same momentum spread\ $\sigma _{\mathrm{o}}$. Then from
(\ref{eq:p1b}), (\ref{eq:p2b}) and (\ref{eq:sigmab}) we have%
\begin{eqnarray}
p_{\mathrm{o}} &=&\left\langle p_{1\mathrm{b}}\left( -\tau _{\mathrm{o}%
}\right) \right\rangle =-\left\langle p_{2\mathrm{b}}\left( -\tau _{\mathrm{o%
}}\right) \right\rangle =\sqrt{p_{\mathrm{o}}^{2}+2\sigma _{\mathrm{o}}^{2}}%
\tanh \left( A_{\mathrm{o}}\tau _{\mathrm{o}}\right) ,  \label{eq:pb0} \\
\sigma _{\mathrm{o}} &=&\left\langle \sigma _{\mathrm{b}}\left( -\tau _{%
\mathrm{o}}\right) \right\rangle =\sqrt{\frac{1}{2}p_{\mathrm{o}}^{2}+\sigma
_{\mathrm{o}}^{2}}\frac{1}{\cosh \left( A_{\mathrm{o}}\tau _{\mathrm{o}%
}\right) }.  \label{eq:sigmab0}
\end{eqnarray}%
To give an estimate of how large $A_{\mathrm{o}}\tau _{\mathrm{o}}$ is, we
assume our momentum-space wave-packets initially have very narrow widths
compared to their momenta such that $\sigma _{\mathrm{o}}/p_{\mathrm{o}}\sim
10^{-3}$, for example.\textbf{\ }Then by (\ref{a0}) we find $A_{\mathrm{o}%
}\tau _{\mathrm{o}}\sim 7.254329369$. Using (\ref{eq:pb0}), we find $p_{%
\mathrm{o}}\sim 0.999999\times \sqrt{p_{\mathrm{o}}^{2}+2\sigma _{\mathrm{o}%
}^{2}}$, which is equivalent to $\sigma _{\mathrm{o}}/p_{\mathrm{o}}\sim
10^{-3}$.

For arbitrary $\tau \geq 0$ after collision, the system of particles $1$ and 
$2$, which initially carried momenta $p_{\mathrm{o}}$ and $-p_{\mathrm{o}}$,
respectively at $\tau =-\tau _{\mathrm{o}}$ before collision, now carries
the relative conjugate-momentum $\left\langle p\left( \tau ;r\right)
\right\rangle $ given by (\ref{eq:p_r}) due to the correlation. As discussed
above, through (\ref{eq:p_r}), (\ref{eq:p_0}) and (\ref{eq:p_compare}), it
is reasonable to expect the existence of a connection between the
correlation and the scattering potential. With non-vanishing
micro-correlation the wave-packets experience the effect of a repulsive
potential; the magnitude of the wave vectors (or momenta) decreases relative
to the corresponding non-correlated value. One may rewrite (\ref{eq:cot}), (%
\ref{eq:ka}) and (\ref{eq:kb}) as%
\begin{equation}
k_{r}\cot \left( k_{r}L\right) =k_{\mathrm{o}}\cot \left( k_{\mathrm{o}%
}L+\theta _{\mathrm{o}}\right) ,  \label{eq:cot_r0}
\end{equation}%
with%
\begin{eqnarray}
k_{r} &=&\frac{\sqrt{2\mu \left( \mathcal{E}-V\right) }}{\hbar },\;0<x<L,
\label{eq:k_r} \\
k_{\mathrm{o}} &=&\frac{\sqrt{2\mu \mathcal{E}}}{\hbar },\;x>L,
\label{eq:k_0}
\end{eqnarray}%
where $\theta _{\mathrm{o}}\equiv \theta \left( k_{\mathrm{o}}\right)
\approx -k_{\mathrm{o}}a_{\mathrm{s}}=-p_{\mathrm{o}}a_{\mathrm{s}}/\hbar $
denotes the $s$-wave scattering phase shift, and $k_{r}$ and $k_{\mathrm{o}}$
represent the wave vectors with and without the correlation, respectively.
The connection between the correlation and the scattering potential can be
established by combining (\ref{eq:k_r}) and (\ref{eq:k_0}).

From (\ref{eq:p_compare}) one finds that the correlation renders 
\begin{equation}
k_{\mathrm{o}}\text{ }\longrightarrow \text{ }k_{r}\equiv \sqrt{1-r}k_{%
\mathrm{o}}.  \label{eq:k_reduce}
\end{equation}%
Then using (\ref{eq:k_r}), (\ref{eq:k_0}) and (\ref{eq:k_reduce}), we
determine the scattering potential, 
\begin{equation}
V=r\mathcal{E}=r\frac{\hbar ^{2}k_{\mathrm{o}}^{2}}{2\mu }=r\frac{p_{\mathrm{%
o}}^{2}}{2\mu }.  \label{eq:potential}
\end{equation}%
Equation (\ref{eq:potential}) clearly establishes a connection between the
correlation coefficient and the scattering potential: the correlation
coefficient is the ratio of the scattering potential to the initial relative
kinetic energy of the system. From (\ref{eq:potential}) it is evident that
our interaction potential is repulsive, i.e. $V>0$ since we consider
non-negative micro-correlations, $r\in \lbrack 0,1)$.

With the potential determined, one can determine the scattering phase shift
by combining equations (\ref{eq:cot_r0}), (\ref{eq:k_r}), (\ref{eq:k_0}) and
(\ref{eq:potential}). By solving (\ref{eq:cot_r0}) for $\theta _{\mathrm{o}}$%
, we find 
\begin{equation}
\tan \theta _{\mathrm{o}}=\frac{k_{\mathrm{o}}\tan \left( k_{r}L\right)
-k_{r}\tan \left( k_{\mathrm{o}}L\right) }{k_{r}+k_{\mathrm{o}}\tan \left(
k_{\mathrm{o}}L\right) \tan \left( k_{r}L\right) }.  \label{eq:theta}
\end{equation}%
Substituting (\ref{eq:k_reduce}) into (\ref{eq:theta}) and expanding the
expression in $k_{\mathrm{o}}L$ and $r$ at the same time, one obtains%
\begin{equation}
\tan \theta _{\mathrm{o}}\approx \left[ -\frac{1}{3}\left( k_{\mathrm{o}%
}L\right) ^{3}+\frac{1}{15}\left( k_{\mathrm{o}}L\right) ^{5}+\mathcal{O}%
\left[ \left( k_{\mathrm{o}}L\right) ^{7}\right] \right] r+\left[ \frac{2}{15%
}\left( k_{\mathrm{o}}L\right) ^{5}+\mathcal{O}\left[ \left( k_{\mathrm{o}%
}L\right) ^{7}\right] \right] r^{2}+\mathcal{O}\left( r^{6}\right) .
\label{eq:phase_shift}
\end{equation}%
For low energy $s$-wave scattering, $k_{\mathrm{o}}L=p_{\mathrm{o}}L/\hbar
\ll 1$, one may reduce (\ref{eq:phase_shift}) to%
\begin{equation}
\tan \theta _{\mathrm{o}}\approx \theta _{\mathrm{o}}\approx -\frac{r\left(
k_{\mathrm{o}}L\right) ^{3}}{3}.  \label{eq:phase_shift2}
\end{equation}%
By means of (\ref{eq:potential}) and (\ref{eq:phase_shift2}) we can express
the scattering phase shift in terms of the scattering potential 
\begin{equation}
\theta _{\mathrm{o}}\approx -\frac{2\mu Vk_{\mathrm{o}}L^{3}}{3\hbar ^{2}}=-%
\frac{2\mu Vp_{\mathrm{o}}L^{3}}{3\hbar ^{3}},  \label{eq:phase_shift_V}
\end{equation}%
which is in agreement with \cite{Mishima}.

As the scattering potential has been determined, so too can the scattering
amplitude be determined. To this end, we write 
\begin{equation}
f\left( k_{\mathrm{o}}\right) =\frac{e^{i\theta _{\mathrm{o}}}\sin \theta _{%
\mathrm{o}}}{k_{\mathrm{o}}}\approx \frac{\theta _{\mathrm{o}}}{k_{\mathrm{o}%
}}\approx -a_{\mathrm{s}}  \label{eq:f01}
\end{equation}%
for low energy $s$-wave scattering, $k_{\mathrm{o}}L=p_{\mathrm{o}}L/\hbar
\ll 1$. Then the squared modulus of (\ref{eq:f01}), by means of (\ref%
{eq:phase_shift_V}), reads 
\begin{equation}
\left\vert f\left( k_{\mathrm{o}}\right) \right\vert ^{2}\approx \frac{%
\theta _{\mathrm{o}}^{2}}{k_{\mathrm{o}}^{2}}\approx \frac{r^{2}k_{\mathrm{o}%
}^{4}L^{6}}{9}=\frac{4\mu ^{2}V^{2}L^{6}}{9\hbar ^{4}}\approx a_{\mathrm{s}%
}^{2}.  \label{eq:f02}
\end{equation}%
Thus, we finally obtain the scattering cross section: 
\begin{equation}
\Sigma =4\pi \left\vert f\left( k_{\mathrm{o}}\right) \right\vert
^{2}\approx \frac{4\pi r^{2}k_{\mathrm{o}}^{4}L^{6}}{9}=\frac{16\pi \mu
^{2}V^{2}L^{6}}{9\hbar ^{4}}\approx 4\pi a_{\mathrm{s}}^{2}.
\label{eq:cross-section}
\end{equation}%
Equations (\ref{eq:P1}) and (\ref{eq:P2}) above demonstrate how the
entanglement can be measured from the loss of purity by use of the
scattering length or cross section. By combining (\ref{eq:P1}) and (\ref%
{eq:f02}) we find the purity%
\begin{equation}
\mathcal{P}\approx 1-\frac{8rk_{\mathrm{o}}^{2}\left( 2k_{\mathrm{o}%
}^{2}+\sigma _{k\mathrm{o}}^{2}\right) R_{\mathrm{o}}L^{3}}{3}=1-\frac{16\mu
V\left( 2k_{\mathrm{o}}^{2}+\sigma _{k\mathrm{o}}^{2}\right) R_{\mathrm{o}%
}L^{3}}{3\hbar ^{2}}.  \label{eq:purity}
\end{equation}%
The correlation coefficient $r$ can now be expressed in terms of the
physical quantities such as the scattering potential, the scattering cross
section and the purity. Solving equations (\ref{eq:potential}), (\ref%
{eq:cross-section}) and (\ref{eq:purity}) for $r$, we obtain%
\begin{eqnarray}
r &=&\frac{V}{\mathcal{E}}=\frac{2\mu V}{\hbar ^{2}k_{\mathrm{o}}^{2}}=\frac{%
2\mu V}{p_{\mathrm{o}}^{2}},  \label{eq:corr_r1} \\
&\approx &\frac{3\sqrt{\Sigma }}{2\sqrt{\pi }k_{\mathrm{o}}^{2}L^{3}},
\label{eq:corr_r2} \\
&\approx &\frac{3\left( 1-\mathcal{P}\right) }{8k_{\mathrm{o}}^{2}\left( 2k_{%
\mathrm{o}}^{2}+\sigma _{k\mathrm{o}}^{2}\right) R_{\mathrm{o}}L^{3}}.
\label{eq:corr_r3}
\end{eqnarray}%
In view of (\ref{rho12}), (\ref{rQM}), (\ref{eq:cross-section}) and (\ref%
{eq:corr_r2}), one obtains the following relation: 
\begin{equation}
\frac{V}{L^{3}}=\frac{4\hbar ^{2}k_{\mathrm{o}}^{4}\left( 2k_{\mathrm{o}%
}^{2}+\sigma _{k\mathrm{o}}^{2}\right) R_{\mathrm{o}}}{3\mu },
\label{Vdensity}
\end{equation}%
which indicates that the uniform scattering potential density is solely
determined by the initial conditions of the given system.

From (\ref{eq:p_0}), (\ref{eq:p_r}) and (\ref{eq:p_compare}) it is observed
that for the micro-correlated Gaussian system considered here, more time is
required to attain the same momentum value compared with the non-correlated
Gaussian system. For example, in order to attain the same value as the
initial momentum $p_{\mathrm{o}}$, the non-correlated system and the
micro-correlated system would require time intervals $\tau _{\mathrm{o}}$
and $\tau _{\ast }$, respectively, where 
\begin{eqnarray}
p_{\mathrm{o}} &=&\sqrt{p_{\mathrm{o}}^{2}+2\sigma _{\mathrm{o}}^{2}}\tanh
\left( A_{\mathrm{o}}\tau _{\mathrm{o}}\right) ,  \label{eq:p0_tau0} \\
p_{\mathrm{o}} &=&\sqrt{\left( 1-r\right) \left( p_{\mathrm{o}}^{2}+2\sigma
_{\mathrm{o}}^{2}\right) }\tanh \left( A_{\mathrm{o}}\tau _{\ast }\right) .
\label{eq:p0_tau*}
\end{eqnarray}%
Combining (\ref{eq:p0_tau0}) and (\ref{eq:p0_tau*}), we obtain%
\begin{equation}
\tanh \left( A_{\mathrm{o}}\tau _{\ast }\right) =\left( 1-r\right)
^{-1/2}\tanh \left( A_{\mathrm{o}}\tau _{\mathrm{o}}\right) .
\label{eq: tanh1}
\end{equation}%
Rewriting and expanding both sides of (\ref{eq: tanh1}), we have%
\begin{equation}
1-2e^{-2A_{\mathrm{o}}\tau _{\star }}+\mathcal{O}\left( e^{-4A_{\mathrm{o}%
}\tau _{\star }}\right) =\left( 1-r\right) ^{-1/2}\left[ 1-2e^{-2A_{\mathrm{o%
}}\tau _{\mathrm{o}}}+\mathcal{O}\left( e^{-4A_{\mathrm{o}}\tau _{\mathrm{o}%
}}\right) \right] .  \label{eq:tanh2}
\end{equation}%
Rounding (\ref{eq:tanh2}) off and arranging terms,%
\begin{equation}
e^{-2A_{\mathrm{o}}\left( \tau _{\star }-\tau _{\mathrm{o}}\right) }\approx
\left( 1-r\right) ^{-1/2}-\frac{1}{2}\left[ \left( 1-r\right) ^{-1/2}-1%
\right] e^{2A_{\mathrm{o}}\tau _{\mathrm{o}}}.  \label{eq:tanh4}
\end{equation}%
The first term on the right hand side of (\ref{eq:tanh4}) can be
approximated to $1$ since $\left( 1-r\right) ^{-1/2}=1+\frac{1}{2}r+\mathcal{%
O}\left( r^{2}\right) $ and $r\ll 1$. However, $r$ in the second term should
not be disregarded in the same way because $\left[ \left( 1-r\right)
^{-1/2}-1\right] e^{2A_{\mathrm{o}}\tau _{\mathrm{o}}}=$ $\left[ \frac{1}{2}%
r+\mathcal{O}\left( r^{2}\right) \right] e^{2A_{\mathrm{o}}\tau _{\mathrm{o}%
}}$ is not negligible. Therefore, we may rewrite (\ref{eq:tanh4}) as%
\begin{equation}
e^{-2A_{\mathrm{o}}\Delta }\approx 1-\left[ \left( 1-r\right) ^{-1/2}-1%
\right] \cdot \eta _{\Delta },  \label{eq:Delta_tau0}
\end{equation}%
where $\Delta \equiv \tau _{\ast }-\tau _{\mathrm{o}}$ represents a new
quantity that we term \textquotedblleft prolongation\textquotedblright , and 
$\eta _{\Delta }\equiv \frac{1}{2}e^{2A_{\mathrm{o}}\tau _{\mathrm{o}%
}}=\left( \frac{p_{\mathrm{o}}}{\sigma _{\mathrm{o}}}\right) ^{2}\exp \left[
\left( \frac{\sigma _{\mathrm{o}}}{p_{\mathrm{o}}}\right) ^{2}-\frac{3}{4}%
\left( \frac{\sigma _{\mathrm{o}}}{p_{\mathrm{o}}}\right) ^{4}+\mathcal{O}%
\left[ \left( \frac{\sigma _{\mathrm{o}}}{p_{\mathrm{o}}}\right) ^{6}\right] %
\right] $ for $\frac{\sigma _{\mathrm{o}}}{p_{\mathrm{o}}}\ll 1$ due to (\ref%
{a0}). From (\ref{eq:Delta_tau0}) we find 
\begin{equation}
\Delta \propto \left\vert \ln \left\{ 1-\left[ \left( 1-r\right) ^{-1/2}-1%
\right] \cdot \eta _{\Delta }\right\} \right\vert .  \label{sette}
\end{equation}

At this juncture we emphasize the following points:

\begin{itemize}
\item The upper bound value of $r$ depends on the initial conditions, namely 
$p_{\mathrm{o}}$ and $\sigma _{\mathrm{o}}$ through the right-hand side of (%
\ref{eq:Delta_tau0}). The right-hand side of (\ref{eq:Delta_tau0}) must
always be positive, so that given $r\ll 1$, we require%
\begin{equation}
r<\frac{2}{\eta _{\Delta }}=2\left( \frac{\sigma _{\mathrm{o}}}{p_{\mathrm{o}%
}}\right) ^{2}\exp \left[ -\left( \frac{\sigma _{\mathrm{o}}}{p_{\mathrm{o}}}%
\right) ^{2}+\frac{3}{4}\left( \frac{\sigma _{\mathrm{o}}}{p_{\mathrm{o}}}%
\right) ^{4}+\mathcal{O}\left[ \left( \frac{\sigma _{\mathrm{o}}}{p_{\mathrm{%
o}}}\right) ^{6}\right] \right]  \label{eq:r_bound}
\end{equation}%
\textbf{\ }for $\frac{\sigma _{0}}{p_{0}}\ll 1$.\textbf{\ }For example, with 
$\sigma _{\mathrm{o}}/p_{\mathrm{o}}\sim 10^{-3}$, for the right-hand side
to be positive we must have $r\lesssim 2\times 10^{-6}$. In view of (\ref%
{eq:purity}), equation (\ref{eq:r_bound}) provides a lower bound estimate of
the purity for a system with well-localized wave-packets, i.e. $\frac{\sigma
_{0}}{p_{0}}\ll 1$.

\item With $r$ being close to the upper bound value, $\Delta $ would be
infinitely large due to (\ref{sette}). On the other hand, with $r$
vanishing, i.e. no correlation, $\Delta $ would vanish. This implies that $%
\Delta $ may serve as an indicator of quantum entanglement.

\item With $r$ held fixed, $\Delta $ depends on the initial conditions $p_{%
\mathrm{o}}$ and $\sigma _{\mathrm{o}}$ through (\ref{eq:Delta_tau0}).
\end{itemize}

From the above points one may infer that the prolongation $\Delta $ could
represent the duration of quantum entanglement for a given micro-correlated
system and that further, the duration can be controlled by the initial
conditions $p_{\mathrm{o}}$, $\sigma _{\mathrm{o}}$ and the
micro-correlation coefficient $r$. From (\ref{eq:Delta_tau0}) it is
anticipated that the maximum duration would be obtained when $r$ is the
greatest, i.e. the micro-correlation is the strongest and the ratio $\sigma
_{\mathrm{o}}/p_{\mathrm{o}}$ is the smallest (see Figure \ref{fig3}). 
\FRAME{fhFU}{11.1595cm}{7.0622cm}{0pt}{\Qcb{$\Delta \equiv \protect\tau %
_{\ast }-\protect\tau _{\mathrm{o}}$ for different values of $r$: $r_{1}$%
(the first plot) $<r_{2}$(the second plot)}}{\Qlb{fig3}}{fig3.png}{\special%
{language "Scientific Word";type "GRAPHIC";maintain-aspect-ratio
TRUE;display "USEDEF";valid_file "F";width 11.1595cm;height 7.0622cm;depth
0pt;original-width 18.1892in;original-height 11.4694in;cropleft "0";croptop
"1";cropright "1";cropbottom "0";filename '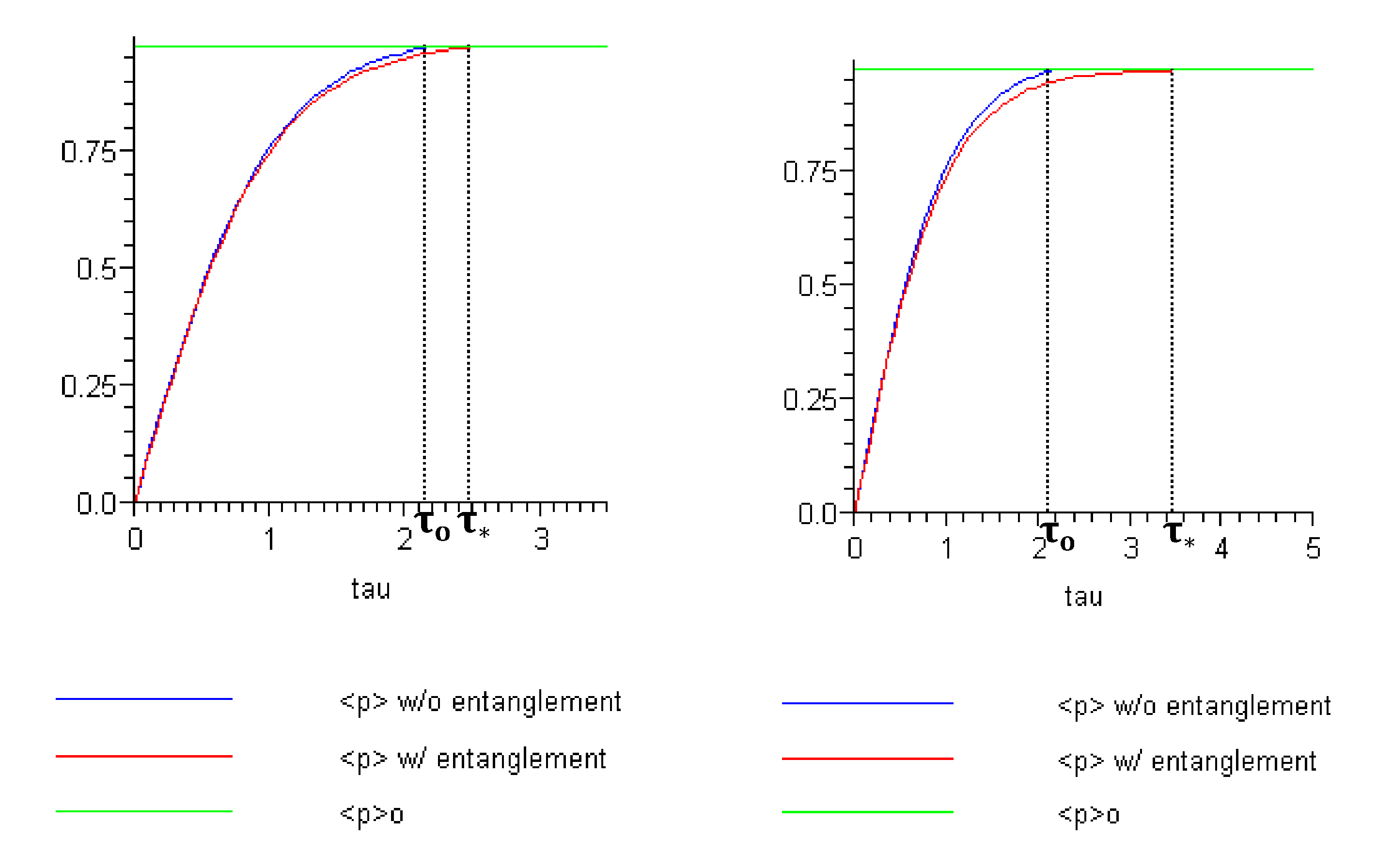';file-properties
"XNPEU";}}We emphasize that the prolongation serves to quantify the time
required by a micro-correlated system - relative to a corresponding
non-correlated one - to attain the same momentum value (relative to the same
initial reference time). The occurrence of a non-vanishing prolongation is
in fact due to the existence of micro-correlations and therefore, due to the
existence of scattering phase shifts. In other words, in the absence of
scattering there is no time difference. This can be stated in yet another
way as follows: \textquotedblleft The prolongation encodes information about
how long it would take an entangled system to overcome the momentum gap
(relative to a corresponding non-entangled system) generated by the
scattering phase shift. The entangled system only attains the full value of
momentum (i.e. the momentum value as seen in the corresponding non-entangled
system) when the scattering phase shift vanishes. For this reason, the
prolongation represents the temporal duration over which the entanglement is
active\textquotedblright .

From (\ref{eq:corr_r1}), (\ref{eq:corr_r2}) and (\ref{eq:corr_r3}) we
observe that the micro-correlation coefficient $r$ is directly associated
with the quantum scattering process, and thus with the quantum entanglement.
For example, the cross term $\left\langle p_{1}p_{2}\right\rangle $ in the
definition of the micro-correlation coefficient $r$ may represent the
average interference between transmitted/reflected modes in the momentum
degrees of freedom of particles $1$ and $2$. This may be viewed from a
different perspective when considering\ the definition of micro-correlations
(\ref{corr-coeff2}). For our statistical system in which $\sigma
_{p_{1}}=\sigma _{p_{2}}=\sigma $, the micro-correlation coefficient reads 
\begin{equation}
r=r\left( p_{1},p_{2}\right) \overset{\text{def}}{=}\frac{\left\langle
p_{1}p_{2}\right\rangle -\left\langle p_{1}\right\rangle \left\langle
p_{2}\right\rangle }{\sigma ^{2}}\text{ with }\sigma =\sqrt{\left\langle
\left( p_{1/2}-\left\langle p_{1/2}\right\rangle \right) ^{2}\right\rangle }.
\label{corr-coeff2}
\end{equation}%
Here the numerator is defined as covariance%
\begin{equation}
\mathrm{Cov}\left( p_{1},p_{2}\right) \overset{\text{def}}{=}\left\langle
p_{1}p_{2}\right\rangle -\left\langle p_{1}\right\rangle \left\langle
p_{2}\right\rangle ,  \label{eq:covariance}
\end{equation}%
and does not vanish if the statistical system is micro-correlated. In other
words, if our statistical system models a quantum scattering process, then
the relevant physical information such as scattering potential and
scattering cross section should be encoded in $\mathrm{Cov}\left(
p_{1},p_{2}\right) $.

With $r\ll 1$, we may split our micro-correlated information geometry (\ref%
{cim}) into two pieces,%
\begin{equation}
g_{ab}=\frac{1}{\sigma ^{2}}\left( 
\begin{array}{ccc}
1+r^{2}+\mathcal{O}\left( r^{3}\right) & -r+\mathcal{O}\left( r^{3}\right) & 
0 \\ 
-r+\mathcal{O}\left( r^{3}\right) & 1+r^{2}+\mathcal{O}\left( r^{3}\right) & 
0 \\ 
0 & 0 & 4%
\end{array}%
\right) =g_{\mathrm{o}\,ab}+h_{ab},  \label{eq:corr_geo}
\end{equation}%
where%
\begin{equation}
g_{\mathrm{o}\,ab}=\frac{1}{\sigma ^{2}}\left( 
\begin{array}{ccc}
1 & 0 & 0 \\ 
0 & 1 & 0 \\ 
0 & 0 & 4%
\end{array}%
\right) ,  \label{eq:non-corr_geo}
\end{equation}%
and%
\begin{equation}
h_{ab}=\frac{1}{\sigma ^{2}}\left( 
\begin{array}{ccc}
r^{2}+\mathcal{O}\left( r^{3}\right) & -r+\mathcal{O}\left( r^{3}\right) & 0
\\ 
-r+\mathcal{O}\left( r^{3}\right) & r^{2}+\mathcal{O}\left( r^{3}\right) & 0
\\ 
0 & 0 & 0%
\end{array}%
\right) .  \label{eq:non-corr_geo1}
\end{equation}%
This decomposition of the micro-correlated information geometry may provide
a different (inherently IG) perspective on the phenomenon of quantum
entanglement. In this view, the non-correlated geometry (\ref%
{eq:non-corr_geo}) is perturbed due to the presence of the quantum
scattering, the information of which is encoded in (\ref{eq:non-corr_geo1}).
Thus, the quantum entanglement manifests as this information geometric
perturbation of the statistical space.\textbf{\ }

\section{ Chaoticity, information geometric complexity and entropy\label%
{sec-complexity}}

In this Section we obtain exact expressions for the information geometric
analogue of standard indicators of chaos such as sectional curvatures,
Jacobi field intensities and Lyapunov exponents. Finally, we present an
analytical estimate of the information geometric entropy (IGE). This will
lead us to uncover connections between quantum entanglement and the
complexity of informational geodesic flows in a quantitative manner.

\subsection{Chaoticity}

\subsubsection{Curvatures of the Statistical Manifold $\mathcal{M}_{\text{%
corr.}}^{3\text{D}}$}

The Riemann curvature tensor $\mathcal{R}_{abcd}$ of the statistical
manifold $\mathcal{M}_{\text{corr.}}^{3\text{D}}$ is defined in the usual
manner as \cite{felice}%
\begin{equation}
\mathcal{R}_{\text{ }bcd}^{a}=\partial _{c}\Gamma _{bd}^{a}-\partial
_{d}\Gamma _{bc}^{a}+\Gamma _{fc}^{a}\Gamma _{bd}^{f}-\Gamma _{fd}^{a}\Gamma
_{bc}^{f},  \label{Riemann1}
\end{equation}%
where the non-vanishing connection coefficients are given in (\ref{c2}). The
non-vanishing components of the Riemann tensor read%
\begin{equation}
\mathcal{R}_{1212}=\frac{1}{4\sigma ^{4}\left( r^{2}-1\right) },\text{ }%
\mathcal{R}_{1313}=\frac{1}{\sigma ^{4}\left( r^{2}-1\right) },\text{ }%
\mathcal{R}_{1323}=-\frac{r}{\sigma ^{4}\left( r^{2}-1\right) },\text{ }%
\mathcal{R}_{2323}=\frac{1}{\sigma ^{4}\left( r^{2}-1\right) }.
\label{Riemann2}
\end{equation}%
The Ricci curvature tensor $\mathcal{R}_{ab}$ of the manifold $\mathcal{M}_{%
\text{corr.}}^{3\text{D}}$ is defined as%
\begin{equation}
\mathcal{R}_{ab}=\partial _{c}\Gamma _{ab}^{c}-\partial _{b}\Gamma
_{ac}^{c}+\Gamma _{ab}^{c}\Gamma _{cd}^{d}-\Gamma _{ac}^{d}\Gamma _{bd}^{c}.
\label{RICCI}
\end{equation}%
The non-vanishing components of the Ricci tensor read%
\begin{equation}
\mathcal{R}_{11}=\frac{1}{2\sigma ^{2}\left( r^{2}-1\right) },\text{ }%
\mathcal{R}_{12}=-\frac{r}{2\sigma ^{2}\left( r^{2}-1\right) }=\mathcal{R}%
_{21},\text{ }\mathcal{R}_{22}=\frac{1}{2\sigma ^{2}\left( r^{2}-1\right) },%
\text{ }\mathcal{R}_{33}=-\frac{2}{\sigma ^{2}}.  \label{blah}
\end{equation}%
Finally, we compute the Ricci scalar curvature $\mathcal{R}$ of the manifold 
$\mathcal{M}_{\text{corr.}}^{3\text{D}}$,%
\begin{equation}
\mathcal{R}=\mathcal{R}_{\text{ }a}^{a}=\mathcal{R}_{ab}g^{ab}=-\frac{3}{2}%
=\sum_{i\neq j}K\left( e_{i},e_{j}\right) .  \label{scalar}
\end{equation}%
That is, the scalar curvature is the sum of all sectional curvatures $%
K\left( e_{i},e_{j}\right) $ of planes spanned by pairs of orthonormal basis
elements $\left\{ e_{a}=\partial _{\vartheta ^{a}(p)}\right\} $ of the
tangent space $T_{p}\mathcal{M}_{\text{corr.}}^{3\text{D}}$ with $p\in 
\mathcal{M}_{\text{corr.}}^{3\text{D}}$ \cite{felice, MTW}, 
\begin{equation}
K_{\mathcal{M}_{\text{corr.}}^{3\text{D}}}\left( u,v\right) =\frac{\mathcal{R%
}_{abcd}u^{a}v^{b}u^{c}v^{d}}{\left( g_{ad}g_{bc}-g_{ac}g_{bd}\right)
u^{a}v^{b}u^{c}v^{d}};\text{ }u\rightarrow h^{i},\text{ }v\rightarrow h^{j}%
\text{ with }i\neq j,  \label{sectionK}
\end{equation}%
where $\left\langle e_{a},h^{b}\right\rangle =\delta _{\text{ }a}^{b}$.
Notice that $K_{\mathcal{M}_{\text{corr.}}^{3\text{D}}}$ completely
determines the curvature tensor. The components of the sectional curvature
are given by%
\begin{equation}
K_{\mu _{1}}=-\frac{1}{4}=K_{-\mu _{1}},\text{ }K_{\mu _{2}}=-\frac{1}{4}%
=K_{-\mu _{2}},\text{ }K_{\sigma }=-\frac{1}{4}=K_{-\sigma }.
\end{equation}

From above, it is worthwhile to note that\ both statistical manifolds $%
\mathcal{M}_{\text{corr.}}^{3\text{D}}$ and $\mathcal{M}_{\text{non-corr.}%
}^{3\text{D}}$ are negatively curved, with the micro-correlation independent
Ricci scalar curvature $\mathcal{R}_{\mathcal{M}_{\text{corr.}}^{3\text{D}%
}}=-\frac{3}{2}=\mathcal{R}_{\mathcal{M}_{\text{non-corr.}}^{3\text{D}}}$
and sectional curvature $K_{\mathcal{M}_{\text{corr.}}^{3\text{D}}}=-\frac{1%
}{4}=K_{\mathcal{M}_{\text{non-corr.}}^{3\text{D}}}$. Moreover, the
constancy of the sectional curvature in all directions imply that both $%
\mathcal{M}_{\text{corr.}}^{3\text{D}}$ and $\mathcal{M}_{\text{non-corr.}%
}^{3\text{D}}$ are isotropic manifolds. Below, this will be verified by the
vanishing of all components of the Weyl projective curvature tensor $%
\mathcal{W}_{abcd}$ defined on each space.

\subsubsection{Anisotropy and the Weyl Projective Tensor}

The anisotropy of the manifold underlying system dynamics plays a crucial
role in the mechanism of instability. In particular, fluctuating sectional
curvatures require also that the manifold be anisotropic. The Weyl
projective tensor quantifies such anisotropy and is defined as%
\begin{equation}
\mathcal{W}_{abcd}\overset{\text{def}}{=}\mathcal{R}_{abcd}-\frac{1}{n-1}%
\left( \mathcal{R}_{bd}g_{ac}-\mathcal{R}_{bc}g_{ad}\right) ,
\label{inter10}
\end{equation}%
where $n$ is the dimension of the manifold on which $\mathcal{W}_{abcd}$ is
defined. By direct computation using (\ref{cim}), (\ref{Riemann2}) and (\ref%
{blah}), we find that all components of (\ref{inter10}) with $n=3$ are
vanishing. The fact that\textbf{\ }$\mathcal{W}_{abcd}=0$\textbf{\ }implies
the manifold $\mathcal{M}_{\text{corr.}}^{3\text{D}}$\textbf{\ }is isotropic.%
\textbf{\ }If the manifold over which a system evolves is maximally
symmetric, then 
\begin{equation}
\mathcal{R}_{ab}=\frac{\mathcal{R}}{n}g_{ab}.  \label{inter11}
\end{equation}%
This is obtained from (\ref{inter10}), using the fact that $\mathcal{W}%
_{abcd}=0$. By inspection of (\ref{blah}), (\ref{cim}) and the fact that $%
\mathcal{R}=-\frac{3}{2}$ and $n=3$, it is evident that (\ref{inter11}) is
indeed valid for our statistical manifold $\mathcal{M}_{\text{corr.}}^{3%
\text{D}}$. Upon substitution of (\ref{inter11}) into (\ref{inter10}) we
obtain%
\begin{equation}
\mathcal{W}_{abcd}=\mathcal{R}_{abcd}-\frac{\mathcal{R}}{n\left( n-1\right) }%
\left( g_{bd}g_{ac}-g_{bc}g_{ad}\right) .  \label{inter12}
\end{equation}%
By the fact that $\mathcal{W}_{abcd}=0$, this\textbf{\ }leads to\textbf{\ }%
\begin{equation}
\mathcal{R}_{abcd}=\frac{\mathcal{R}}{n\left( n-1\right) }\left(
g_{bd}g_{ac}-g_{bc}g_{ad}\right) ,  \label{eq:VII-3}
\end{equation}%
which again proves to be true for our manifold $\mathcal{M}_{\text{corr.}}^{3%
\text{D}}$ by inspecting (\ref{Riemann2}), (\ref{cim}) and the fact that $%
\mathcal{R}=-\frac{3}{2}$ and $n=3$. Contracting the both sides of (\ref%
{inter11}), one finds%
\begin{equation}
\mathcal{\delta }_{~a}^{a}=n,  \label{g}
\end{equation}%
which is also the case for our manifold $\mathcal{M}_{\text{corr.}}^{3\text{D%
}}$, that is, $n=3$.

\subsubsection{Jacobi Fields and Lyapunov Exponents}

For the sake of clarity, consider the behavior of a family of neighboring
geodesics $\left\{ \vartheta _{\mathcal{M}_{\text{corr.}}^{3\text{D}%
}}^{a}\left( \tau ;\vec{\varsigma}\right) \right\} _{\vec{\varsigma}\in 
\mathbb{R}_{+}^{3}}^{a=1,2,3}$ on the statistical manifold $\mathcal{M}_{%
\text{corr.}}^{3\text{D}}$, where $\tau $ and $\vec{\varsigma}=\left(
\varsigma ^{1},\varsigma ^{2},\varsigma ^{3}\right) $ are affine parameters.
The geodesics $\vartheta _{\mathcal{M}_{\text{corr.}}^{3\text{D}}}^{a}\left(
\tau ;\vec{\varsigma}\right) $ are solutions of equation (\ref{GE}). The
relative geodesic spread on $\mathcal{M}_{\text{corr.}}^{3\text{D}}$ is
characterized by the Jacobi-Levi-Civita (JLC) equation \cite{MTW, carmo},%
\begin{equation}
\frac{D^{2}J^{a}}{D\tau ^{2}}+\mathcal{R}_{\text{ }bcd}^{a}\frac{\partial
\vartheta ^{b}}{\partial \tau }J^{c}\frac{\partial \vartheta ^{d}}{\partial
\tau }=0,  \label{eq:VII-4}
\end{equation}%
where $a,b,c,d=1,2,3$, and the second order covariant derivatives $\frac{%
D^{2}J^{a}}{D\tau ^{2}}$ are given by \cite{ohanian} 
\begin{equation}
\frac{D^{2}J^{a}}{D\tau ^{2}}=\frac{d^{2}J^{a}}{d\tau ^{2}}+2\Gamma _{bc}^{a}%
\frac{dJ^{b}}{d\tau }\frac{d\vartheta ^{c}}{d\tau }+\Gamma _{bc}^{a}J^{b}%
\frac{d^{2}\vartheta ^{c}}{d\tau ^{2}}+\Gamma _{bc,d}^{a}\frac{d\vartheta
^{d}}{d\tau }\frac{d\vartheta ^{c}}{d\tau }J^{b}+\Gamma _{bc}^{a}\Gamma
_{df}^{b}\frac{d\vartheta ^{f}}{d\tau }\frac{d\vartheta ^{c}}{d\tau }J^{d},
\end{equation}%
and the Jacobi vector field components $J^{a}$ are given by%
\begin{equation}
J^{a}=\delta _{\vec{\varsigma}}\vartheta ^{a}\equiv \left. \frac{\partial
\vartheta ^{a}\left( \tau ;\vec{\varsigma}\right) }{\partial \varsigma ^{b}}%
\right\vert _{\tau }\delta \varsigma ^{b}.  \label{jacobi}
\end{equation}%
$\mathbf{J}=\left\{ J^{a}\right\} _{a=1,2,3}$ represents how geodesics are
separating. The JLC equation of geodesic deviation is a complicated
second-order system of linear ordinary differential equations. It describes
the geodesic spread on curved manifolds of a pair of nearby freely falling
particles traveling on trajectories $\vartheta ^{a}\left( \tau \right) $ and 
$\vartheta ^{\prime a}\left( \tau \right) \overset{\text{def}}{=}\vartheta
^{a}\left( \tau \right) +\delta \vartheta ^{a}\left( \tau \right) $.
Equation (\ref{eq:VII-4}) forms a system of three coupled ordinary
differential equations \emph{linear} in the components of the deviation
vector field (\ref{jacobi}) but\ \emph{nonlinear} in derivatives of the
metric tensor $g_{ab}\left( \Theta \right) $. It describes the linearized
geodesic flow: the linearization ignores the relative velocity of the
geodesics. When the geodesics are neighboring but their relative velocity is
arbitrary, the corresponding geodesic deviation equation is the so-called
generalized Jacobi equation \cite{chicone}. The nonlinearity is due to the
existence of velocity-dependent terms in the system. Neighboring geodesics
accelerate relative to each other with a rate directly measured by the
curvature tensor $\mathcal{R}_{abcd}$.

By means of (\ref{eq:VII-3}) the second term on the light-hand side of
equation (\ref{eq:VII-4}) can be rewritten as 
\begin{equation}
\mathcal{R}_{abcd}\frac{\partial \vartheta ^{b}}{\partial \tau }J^{c}\frac{%
\partial \vartheta ^{d}}{\partial \tau }=\frac{\mathcal{R}\left\Vert \mathbf{%
v}\right\Vert ^{2}}{n(n-1)}P_{ab}J^{b},  \label{eq:VII-5}
\end{equation}%
where $\left\Vert \mathbf{v}\right\Vert \equiv \sqrt{g_{ab}v^{a}v^{b}}$ with 
$v^{a}\equiv \frac{\partial \vartheta ^{a}}{\partial \tau }$, and $%
P_{ab}\equiv g_{ab}-u_{a}u_{b}$ with $u^{a}\equiv v^{a}/\left\Vert \mathbf{v}%
\right\Vert $ and $u^{a}u_{a}=1$. Due to the orthogonality between $u_{a}$
and $J^{a}$, we have $P_{ab}J^{b}=J_{a}$ and thus (\ref{eq:VII-5}) is now
reduced to%
\begin{equation}
\mathcal{R}_{abcd}\frac{\partial \vartheta ^{b}}{\partial \tau }J^{c}\frac{%
\partial \vartheta ^{d}}{\partial \tau }=\frac{\mathcal{R}\left\Vert \mathbf{%
v}\right\Vert ^{2}}{n(n-1)}J_{a}.  \label{eq:VII-6}
\end{equation}%
Using equations (\ref{cim}) and (\ref{eq:3}), the squared modulus of $v^{a}$
is computed as 
\begin{eqnarray}
\left\Vert \mathbf{v}\right\Vert ^{2} &=&g_{ab}\frac{\partial \vartheta ^{a}%
}{\partial \tau }\frac{\partial \vartheta ^{b}}{\partial \tau }  \notag \\
&=&\frac{1}{(1-r^{2})\sigma ^{2}}\left[ \left( \frac{\partial \mu _{1}}{%
\partial \tau }\right) ^{2}+\left( \frac{\partial \mu _{2}}{\partial \tau }%
\right) ^{2}\right] -\frac{r}{(1-r^{2})\sigma ^{2}}\left( \frac{\partial \mu
_{1}}{\partial \tau }\right) \left( \frac{\partial \mu _{2}}{\partial \tau }%
\right) +\frac{4}{\sigma ^{2}}\left( \frac{\partial \sigma }{\partial \tau }%
\right) ^{2}  \notag \\
&=&-\frac{4}{\sigma }\left[ \frac{\partial ^{2}\sigma }{\partial \tau ^{2}}-%
\frac{2}{\sigma }\left( \frac{\partial \sigma }{\partial \tau }\right) ^{2}%
\right] .  \label{b-norm}
\end{eqnarray}%
Upon substitution of (\ref{eq:SIGMArf}) into (\ref{b-norm}) we find%
\begin{equation}
\left\Vert \mathbf{v}\right\Vert ^{2}=4A_{\mathrm{o}}^{2},  \label{b}
\end{equation}%
where $A_{\mathrm{o}}$ is given by (\ref{a0}). By combining (\ref{eq:VII-4}%
), (\ref{eq:VII-6}) and (\ref{b}) we finally simplify the JLC equation to
the following form: 
\begin{equation}
\frac{D^{2}J^{a}}{D\tau ^{2}}+QJ^{a}=0,  \label{eq:VII-7}
\end{equation}%
where 
\begin{equation}
Q\equiv \frac{\mathcal{R}\left\Vert \mathbf{v}\right\Vert ^{2}}{n(n-1)}=-A_{%
\mathrm{o}}^{2}<0,  \label{eq:Q_A2}
\end{equation}%
which has been computed with $n=3$, $\mathcal{R}=-\frac{3}{2}$ and $%
\left\Vert \mathbf{v}\right\Vert ^{2}$ from (\ref{b}).

The Jacobi vector field intensity is given by%
\begin{equation}
\mathcal{J}_{\mathcal{M}_{\text{corr.}}^{3\text{D}}}\equiv \left\Vert 
\mathbf{J}\right\Vert =\left( g_{ab}J^{a}J^{b}\right) ^{\frac{1}{2}}=\left(
J^{a}J_{a}\right) ^{\frac{1}{2}}.  \label{eq:VII-9}
\end{equation}%
For applications of the asymptotic temporal behavior of $\mathcal{J}_{%
\mathcal{M}_{\text{corr.}}^{3\text{D}}}\left( \tau \right) $ as a reliable
indicator of chaoticity, we refer to our previous articles in references 
\cite{ca-2, cafaroPD, cafaroPA}. Defining the operator $\hat{\Omega}\equiv
D^{2}/D\tau ^{2}$ and observing its action on $\mathcal{J}_{\mathcal{M}_{%
\text{corr.}}^{3\text{D}}}^{2}$ leads to conclude%
\begin{equation}
\hat{\Omega}\mathcal{J}_{\mathcal{M}_{\text{corr.}}^{3\text{D}}}^{2}=\left( 
\hat{\Omega}J^{a}\right) J_{a}+J^{a}\left( \hat{\Omega}J_{a}\right)
=-2QJ^{a}J_{a}=-2Q\mathcal{J}_{\mathcal{M}_{\text{corr.}}^{3\text{D}}}^{2}.
\label{eq:VII-10}
\end{equation}%
Equation (\ref{eq:VII-10}) follows from (\ref{eq:VII-7}) and (\ref{eq:VII-9}%
). We may however, write $\hat{\Omega}\mathcal{J}_{\mathcal{M}_{\text{corr.}%
}^{3\text{D}}}^{2}=2\mathcal{J}_{\mathcal{M}_{\text{corr.}}^{3\text{D}%
}}\left( \hat{\Omega}\mathcal{J}_{\mathcal{M}_{\text{corr.}}^{3\text{D}%
}}\right) $. This fact together with (\ref{eq:VII-10}) enables the further
reduction of (\ref{eq:VII-7}) to a scalar form 
\begin{equation}
\frac{D^{2}\mathcal{J}_{\mathcal{M}_{\text{corr.}}^{3\text{D}}}}{D\tau ^{2}}%
+Q\mathcal{J}_{\mathcal{M}_{\text{corr.}}^{3\text{D}}}=0.  \label{eq:VII-11}
\end{equation}%
Since $Q<0$, the solutions of equation (\ref{eq:VII-11}) assume the form%
\begin{equation}
\mathcal{J}_{\mathcal{M}_{\text{corr.}}^{3\text{D}}}\left( \tau \right) =%
\frac{1}{\sqrt{-Q}}\omega \left( 0\right) \sinh \left( \sqrt{-Q}\tau \right)
,  \label{jac}
\end{equation}%
where $\omega \left( 0\right) \equiv \left. \frac{d\mathcal{J}_{\mathcal{M}_{%
\text{corr.}}^{3\text{D}}}\left( \tau \right) }{d\tau }\right\vert _{\tau
=0} $.

Recalling the definition of the hyperbolic sine function, $\sinh x=\frac{1}{2%
}\left( e^{x}-e^{-x}\right) $, it is clear that the geodesic deviation on $%
\mathcal{M}_{\text{corr.}}^{3\text{D}}$ is described by means of an
exponentially divergent Jacobi vector field intensity $\mathcal{J}_{\mathcal{%
M}_{\text{corr.}}^{3\text{D}}}$, a classical feature of chaos. In this
Riemannian geometric approach, the quantity $\lambda _{\mathcal{M}_{\text{%
corr.}}^{3\text{D}}}$ defined as \cite{casetti}%
\begin{equation}
\lambda _{\mathcal{M}_{\text{corr.}}^{3\text{D}}}\overset{\text{def}}{=}%
\lim_{\tau \rightarrow \infty }\frac{1}{\tau }\ln \left[ \frac{\left\vert 
\mathcal{J}_{\mathcal{M}_{\text{corr.}}^{3\text{D}}}\left( \tau \right)
\right\vert ^{2}+\left\vert \frac{d\mathcal{J}_{\mathcal{M}_{\text{corr.}}^{3%
\text{D}}}\left( \tau \right) }{d\tau }\right\vert ^{2}}{\left\vert \mathcal{%
J}_{\mathcal{M}_{\text{corr.}}^{3\text{D}}}\left( 0\right) \right\vert
^{2}+\left\vert \left. \frac{d\mathcal{J}_{\mathcal{M}_{\text{corr.}}^{3%
\text{D}}}\left( \tau \right) }{d\tau }\right\vert _{\tau =0}\right\vert ^{2}%
}\right] ,  \label{lap}
\end{equation}%
would play the role of the conventional Lyapunov exponents. In order to
evaluate (\ref{lap}) we use (\ref{jac}) to find $\left\vert \mathcal{J}_{%
\mathcal{M}_{\text{corr.}}^{3\text{D}}}\left( \tau \right) \right\vert ^{2}=%
\frac{\omega ^{2}\left( 0\right) }{-Q}\sinh ^{2}\left( \sqrt{-Q}\tau \right) 
$ and $\left\vert d\mathcal{J}_{\mathcal{M}_{\text{corr.}}^{3\text{D}%
}}\left( \tau \right) /d\tau \right\vert ^{2}=\omega ^{2}\left( 0\right)
\cosh ^{2}\left( \sqrt{-Q}\tau \right) $. Thus, for the case being
considered,%
\begin{eqnarray}
\lambda _{\mathcal{M}_{\text{corr.}}^{3\text{D}}} &=&\lim_{\tau \rightarrow
\infty }\frac{1}{\tau }\ln \left[ -\frac{1}{Q}\sinh ^{2}\left( \sqrt{-Q}\tau
\right) +\cosh ^{2}\left( \sqrt{-Q}\tau \right) \right]  \notag \\
&=&\lim_{\tau \rightarrow \infty }\frac{1}{\tau }\ln \left[ \frac{\left(
1-Q\right) }{4}e^{2\sqrt{-Q}\tau }\right] =2\sqrt{-Q}.
\end{eqnarray}%
Therefore, it follows that%
\begin{equation}
\lambda _{\mathcal{M}_{\text{corr.}}^{3\text{D}}}\overset{\tau \rightarrow
\infty }{=}2\sqrt{-Q}=2A_{\mathrm{o}}>0,  \label{due}
\end{equation}%
which is due to (\ref{eq:Q_A2}). From (\ref{due}) we observe the following
points: the information about chaoticity encoded in the positive Lyapunov
exponent does not depend on the statistical correlation, i.e. $\lambda _{%
\mathcal{M}_{\text{corr.}}^{3\text{D}}}=\lambda _{\mathcal{M}_{\text{%
non-corr.}}^{3\text{D}}}\equiv \lambda _{\mathcal{M}^{3\text{D}}}=2A_{%
\mathrm{o}}$, and the\textbf{\ }Lyapunov exponents can be determined solely
from the initial conditions (see equation (\ref{a0})).

\subsection{Information Geometric Complexity and Entropy}

We recall that a suitable indicator of temporal complexity within the IGAC
framework is provided by the \emph{information geometric entropy} (IGE) $%
\mathcal{S}_{\mathcal{M}_{\text{corr.}}^{3\text{D}}}\left( \tau \right) $ 
\cite{carlo-tesi, carlo-CSF},%
\begin{equation}
\mathcal{S}_{\mathcal{M}_{\text{corr.}}^{3\text{D}}}\left( \tau \right) 
\overset{\text{def}}{=}\lim_{\tau \rightarrow \infty }\ln \mathcal{V}_{%
\mathcal{M}_{\text{corr.}}^{3\text{D}}}\left[ \mathcal{D}_{\Theta }^{\left( 
\text{geodesic}\right) }\left( \tau \right) \right] .  \label{ige}
\end{equation}%
The \emph{information geometric complexity} (IGC) is defined as the temporal
average of the dynamical statistical volume,%
\begin{equation}
\mathcal{V}_{\mathcal{M}_{\text{corr.}}^{3\text{D}}}\left[ \mathcal{D}%
_{\Theta }^{\left( \text{geodesic}\right) }\left( \tau \right) \right] 
\overset{\text{def}}{=}\lim_{\tau \rightarrow \infty }\left( \frac{1}{\tau }%
\int_{0}^{\tau }d\tau ^{\prime }\emph{vol}\left[ \mathcal{D}_{\Theta
}^{\left( \text{geodesic}\right) }\left( \tau ^{\prime }\right) \right]
\right) .  \label{inter8}
\end{equation}%
The extended volume $\emph{vol}\left[ \mathcal{D}_{\Theta }^{\left( \text{%
geodesic}\right) }\left( \tau ^{\prime }\right) \right] $ of the effective
parameter space explored by the system at time $\tau ^{\prime }$ is given by%
\begin{equation}
\emph{vol}\left[ \mathcal{D}_{\Theta }^{\left( \text{geodesic}\right)
}\left( \tau ^{\prime }\right) \right] \overset{\text{def}}{=}\int \int
\int_{\mathcal{D}_{\Theta }^{\left( \text{geodesic}\right) }\left( \tau
^{\prime }\right) }\rho _{\left( \mathcal{M}_{\text{corr.}}^{3\text{D}%
},g\right) }\left( \vartheta ^{1},\vartheta ^{2},\vartheta ^{3}\right)
d\vartheta ^{1}d\vartheta ^{2}d\vartheta ^{3},  \label{v}
\end{equation}%
where $\rho _{\left( \mathcal{M}_{\text{corr.}}^{3\text{D}},g\right) }\left(
\vartheta ^{1},\vartheta ^{2},\vartheta ^{3}\right) $ is the so-called
Fisher density and equals the square root of the determinant $g=\left\vert
\det \left( g_{ab}\right) \right\vert $ of the metric tensor $g\left(
\vartheta ^{1},\vartheta ^{2},\vartheta ^{3}\right) $,%
\begin{equation}
\rho _{\left( \mathcal{M}_{\text{corr.}}^{3\text{D}},g\right) }\left(
\vartheta ^{1},\vartheta ^{2},\vartheta ^{3}\right) \overset{\text{def}}{=}%
\sqrt{g\left( \vartheta ^{1},\vartheta ^{2},\vartheta ^{3}\right) }.
\end{equation}%
The set $\mathcal{D}_{\Theta }^{\left( \text{geodesic}\right) }$\textbf{\ }%
represents a subspace of the whole (permitted) parameter space $\mathcal{D}%
_{\Theta }^{\left( \text{total}\right) }$ in (\ref{is}),%
\begin{equation}
\mathcal{D}_{\Theta }^{\left( \text{geodesic}\right) }\left( \tau ^{\prime
}\right) =\left\{ \Theta \equiv \left( \vartheta ^{1},\vartheta
^{2},\vartheta ^{3}\right) :\vartheta ^{a}\left( 0\right) \leq \vartheta
^{a}\leq \vartheta ^{a}\left( \tau ^{\prime }\right) \right\} ,
\end{equation}%
where $a=1,2,3$, and $\vartheta ^{a}\equiv \vartheta ^{a}\left( s\right) $
with $0\leq s\leq \tau ^{\prime }$ such that $\vartheta ^{a}\left( s\right) $
satisfies (\ref{GE}). The elements of $\mathcal{D}_{\Theta }^{\left( \text{%
geodesic}\right) }\left( \tau ^{\prime }\right) $ are the macrovariables $%
\left\{ \Theta \right\} $ whose components $\vartheta ^{a}$ are bounded by
specified limits of integration $\vartheta ^{a}\left( 0\right) $ and $%
\vartheta ^{a}\left( \tau ^{\prime }\right) $. The limits of integration are
obtained via integration of the set of coupled nonlinear second order
ordinary differential equations characterizing the geodesic equations. In
the case of the statistical manifold of three-dimensional Gaussian
probability distributions parametrized in terms of $\Theta =\left( \mu
_{1},\mu _{2},\sigma \right) $, the integration space $\mathcal{D}_{\Theta
}^{\left( \text{geodesic}\right) }\left( \tau ^{\prime }\right) $ in (\ref{v}%
) is the direct product of the parameter subspaces $\mathcal{I}_{\mu _{1}}$, 
$\mathcal{I}_{\mu _{2}}$ and $\mathcal{I}_{\sigma }$\textbf{,} where in the
Gaussian case, $\mathcal{I}_{\mu _{1}}=\left( -\infty ,+\infty \right) _{\mu
_{1}}$, $\mathcal{I}_{\mu _{2}}=\left( -\infty ,+\infty \right) _{\mu _{2}}$
and $\mathcal{I}_{\sigma }=\left( 0,+\infty \right) _{\sigma }$\textbf{\ }%
such that\textbf{\ }%
\begin{equation}
\mathcal{D}_{\Theta }^{\left( \text{geodesic}\right) }=\mathcal{I}_{\mu
_{1}}\otimes \mathcal{I}_{\mu _{2}}\otimes \mathcal{I}_{\sigma }=\left[
\left( -\infty ,+\infty \right) \otimes \left( -\infty ,+\infty \right)
\otimes \left( 0,+\infty \right) \right] .
\end{equation}%
In the IGAC, we are interested in a probabilistic description of the
evolution of a given system in terms of its corresponding probability
distribution on $\mathcal{M}_{\text{corr.}}^{3\text{D}}$ which is
homeomorphic to $\mathcal{D}_{\Theta }^{\left( \text{geodesic}\right) }$. We
are interested in the evolution of the system from $\tau _{\text{initial}}=0$
to $\tau _{\text{final}}=\tau $. Within the probabilistic description,
investigating the evolution of the system from\textbf{\ }$\tau _{\text{%
initial}}=0$\textbf{\ }to\textbf{\ }$\tau _{\text{final}}=\tau $\textbf{\ }%
is equivalent to studying the shortest path (or, in terms of the ME method 
\cite{caticha2, caticha(REII), caticha-giffin, adom}, the maximally probable
path) leading from $\Theta \left( 0\right) $ to $\Theta \left( \tau \right) $%
.

Formally, the IGE $\mathcal{S}_{\mathcal{M}_{\text{corr.}}^{3\text{D}%
}}\left( \tau \right) $ is defined in terms of an averaged parametric $3$%
-fold integral ($\tau $ is the parameter) over the three-dimensional
geodesic paths connecting $\Theta \left( 0\right) $ to $\Theta \left( \tau
\right) $. In the present IG approach, the IGC represents a statistical
measure of complexity of the macroscopic path $\Theta \overset{\text{def}}{=}%
\Theta \left( \tau \right) $ on $\mathcal{M}_{\text{corr.}}^{3\text{D}}$
connecting initial and final macrostates $\Theta _{\text{I}}$ and $\Theta _{%
\text{F}}$, respectively. The path $\Theta \left( \tau \right) $ is obtained
via integration of the geodesic equation on $\mathcal{M}_{\text{corr.}}^{3%
\text{D}}$ generated by the universal ME updating method. At a discrete
level, the path $\Theta \left( \tau \right) $ can be described in terms of
an infinite continuous sequence of intermediate macroscopic states, $\Theta
\left( \tau \right) =\left[ \Theta _{\text{I}},...,\Theta _{\bar{k}%
-1},\Theta _{\bar{k}},\Theta _{\bar{k}+1},...,\Theta _{\text{F}}\right] $
with $\Theta _{\bar{j}}=\Theta \left( \tau _{\bar{j}}\right) $, determined
via the logarithmic relative entropy maximization procedure subjected to
appropriately-specified normalization and information constraints. The
nature of such constraints defines the (correlational) structure of the
underlying probability distribution on the particular curved statistical
manifold $\mathcal{M}_{\text{corr.}}^{3\text{D}}$. In other words, the
correlational structure that emerges in our IG statistical models originates
in the information pertaining to the microscopic degrees of freedom of the
actual physical systems. It is finally quantified in terms of the intuitive
notion of volume growth via the IGC or alternatively in entropic terms by
the IGE. The IGC is then interpreted as the temporally averaged volume of
the statistical macrospace explored by the system, in the asymptotic limit,
in its evolution from $\Theta _{\text{I}}$ to $\Theta _{\text{F}}$.
Otherwise, upon a suitable normalization procedure that makes the IGC an
adimensional quantity, it represents the number of accessible macrostates
(with coordinates living in the accessible parameter space $\mathcal{D}%
_{\Theta }^{\left( \text{geodesic}\right) }\left( \tau \right) $) explored
by the system in its evolution from $\Theta _{\text{I}}$ to $\Theta _{\text{F%
}}$.

The temporal average in (\ref{inter8}) has been introduced in order to smear
out the possibly very complex fine details of the entropic dynamical
description of the system on $\mathcal{M}_{\text{corr.}}^{3\text{D}}$. Thus,
we provide a coarse-grained-like inferential description of the system's
chaotic dynamics. The long-term asymptotic temporal behavior is adopted in
order to properly characterize dynamical indicators of chaoticity (for
instance, Lyapunov exponents, entropies, etc.) eliminating transient effects
which enters the computation of the expected value of (\ref{v}). In chaotic
transients, one observes that typical initial conditions behave in an
apparently chaotic manner for a possibly long time, but then asymptotically
approach a non-chaotic attractor in a rapid fashion.

In the case under investigation, (\ref{inter8}) is given by%
\begin{equation}
\mathcal{V}_{\mathcal{M}_{\text{corr.}}^{3\text{D}}}\left[ \mathcal{D}%
_{\Theta }^{\left( \text{geodesic}\right) }\left( \tau ;r\right) \right]
=\lim_{\tau \rightarrow \infty }\frac{1}{\tau }\int\limits_{0}^{\tau }d\tau
^{\prime }\int\limits_{\mu _{1}\left( 0\right) }^{\mu _{1}\left( \tau
^{\prime }\right) }\int\limits_{\mu _{2}\left( 0\right) }^{\mu _{2}\left(
\tau ^{\prime }\right) }\int\limits_{\sigma \left( 0\right) }^{\sigma \left(
\tau ^{\prime }\right) }\sqrt{g}d\mu _{1}d\mu _{2}d\sigma ,  \label{inter9}
\end{equation}%
where the geodesic paths $\Theta \left( \tau \right) =\left( \mu _{1}\left(
\tau ;r\right) ,\mu _{2}\left( \tau ;r\right) ,\sigma \left( \tau ;r\right)
\right) $ are given in (\ref{eq:MUr1f}), (\ref{eq:MUr2f}) and (\ref%
{eq:SIGMArf}) and the determinant of the information metric reads%
\begin{equation}
g=\det \left( g_{ab}\right) =\frac{4}{\left( 1-r^{2}\right) \sigma ^{6}}.
\label{det-g}
\end{equation}%
Substituting (\ref{det-g}) into\ (\ref{inter9}),\ and evaluating the
integral by means of (\ref{eq:MUr1f}), (\ref{eq:MUr2f}) and (\ref{eq:SIGMArf}%
), we obtain the IGC for the correlated Gaussian statistical models: 
\begin{eqnarray}
\mathcal{V}_{\mathcal{M}_{\text{corr.}}^{3\text{D}}}\left[ \mathcal{D}%
_{\Theta }^{\left( \text{geodesic}\right) }\left( \tau ;r\right) \right] &=&-%
\frac{1}{\sqrt{1-r^{2}}}\lim_{\tau \rightarrow \infty }\frac{1}{\tau }%
\int\limits_{0}^{\tau }d\tau ^{\prime }\left[ \left. \mu _{1}\left( \tau
^{\prime \prime }\right) \right\vert _{\tau ^{\prime \prime }=0}^{\tau
^{\prime \prime }=\tau ^{\prime }}\left. \mu _{2}\left( \tau ^{\prime \prime
}\right) \right\vert _{\tau ^{\prime \prime }=0}^{\tau ^{\prime \prime
}=\tau ^{\prime }}\left. \frac{1}{\sigma ^{2}\left( \tau ^{\prime \prime
}\right) }\right\vert _{\tau ^{\prime \prime }=0}^{\tau ^{\prime \prime
}=\tau ^{\prime }}\right]  \notag \\
&=&\frac{8\sqrt{\frac{1-r}{1+r}}}{\lambda _{\mathcal{M}^{3\text{D}}}}\left[ -%
\frac{3}{4}\lambda _{\mathcal{M}^{3\text{D}}}+\frac{1}{4}\frac{\sinh \left(
\lambda _{\mathcal{M}^{3\text{D}}}\tau \right) }{\tau }+\frac{\tanh \left( 
\frac{1}{2}\lambda _{\mathcal{M}^{3\text{D}}}\tau \right) }{\tau }\right] ,
\label{eq:vol}
\end{eqnarray}%
where $A_{\mathrm{o}}$ has been replaced with $\frac{1}{2}\lambda _{\mathcal{%
M}^{3\text{D}}}$ due to (\ref{due}). For non-correlated Gaussian statistical
models the IGC becomes%
\begin{equation}
\mathcal{V}_{\mathcal{M}_{\text{non-corr.}}^{3\text{D}}}\left[ \mathcal{D}%
_{\Theta }^{\left( \text{geodesic}\right) }\left( \tau ;0\right) \right] =%
\frac{8}{\lambda _{\mathcal{M}^{3\text{D}}}}\left[ -\frac{3}{4}\lambda _{%
\mathcal{M}^{3\text{D}}}+\frac{1}{4}\frac{\sinh \left( \lambda _{\mathcal{M}%
^{3\text{D}}}\tau \right) }{\tau }+\frac{\tanh \left( \frac{1}{2}\lambda _{%
\mathcal{M}^{3\text{D}}}\tau \right) }{\tau }\right] .  \label{vol0}
\end{equation}

Inserting (\ref{eq:vol}) into (\ref{ige}) and working through some
calculations, we obtain the IGE for the correlated Gaussian statistical
models:%
\begin{equation}
\mathcal{S}_{\mathcal{M}_{\text{corr.}}^{3\text{D}}}\left( \tau ;r\right) 
\overset{\tau \rightarrow \infty }{=}\lambda _{\mathcal{M}^{3\text{D}}}\tau
-\ln \left( \lambda _{\mathcal{M}^{3\text{D}}}\tau \right) +\frac{1}{2}\ln
\left( \frac{1-r}{1+r}\right) .  \label{eq:entrp_r}
\end{equation}%
For non-correlated Gaussian statistical models the IGE becomes%
\begin{equation}
\mathcal{S}_{\mathcal{M}_{\text{non-corr.}}^{3\text{D}}}\left( \tau
;0\right) \overset{\tau \rightarrow \infty }{=}\lambda _{\mathcal{M}^{3\text{%
D}}}\tau -\ln \left( \lambda _{\mathcal{M}^{3\text{D}}}\tau \right) .
\label{eq:entrp_0}
\end{equation}

By means of (\ref{eq:vol}) and (\ref{vol0}) we compare the asymptotic
(long-time limit) expressions of the IGCs in the presence and absence of
micro-correlations, respectively, to obtain%
\begin{equation}
\frac{\mathcal{V}_{\mathcal{M}_{\text{corr.}}^{3\text{D}}}\left[ \mathcal{D}%
_{\Theta }^{\left( \text{geodesic}\right) }\left( \tau ;r\right) \right] }{%
\mathcal{V}_{\mathcal{M}_{\text{non-corr.}}^{3\text{D}}}\left[ \mathcal{D}%
_{\Theta }^{\left( \text{geodesic}\right) }\left( \tau ;0\right) \right] }=%
\sqrt{\frac{1-r}{1+r}}.  \label{eq:v-rel}
\end{equation}%
From (\ref{eq:entrp_r}) and (\ref{eq:entrp_0}) we also find 
\begin{equation}
\mathcal{S}_{\mathcal{M}_{\text{corr.}}^{3\text{D}}}\left( \tau ;r\right) -%
\mathcal{S}_{\mathcal{M}_{\text{non-corr.}}^{3\text{D}}}\left( \tau
;0\right) =\frac{1}{2}\ln \left( \frac{1-r}{1+r}\right) .  \label{s-rel}
\end{equation}%
From (\ref{eq:v-rel}) and (\ref{s-rel}) we find that both the IGC and the
IGE decrease in presence of micro-correlations.\ In particular, the IGC
decreases by the factor $\sqrt{\frac{1-r}{1+r}}<1$ for $r>0$ whereas\textbf{%
\ }the IGE decreases by $\frac{1}{2}\ln \left( \frac{1-r}{1+r}\right) <0$
for $r>0$.

It is evident from (\ref{s-rel}) that in presence of micro-correlations the
IGE is attenuated in a correlation-dependent manner: $\mathcal{S}_{\mathcal{M%
}_{\text{corr.}}^{3\text{D}}}\left( \tau ;r\right) $ decreases as the
magnitude of the correlation increases. It is important to observe that this
has no relation to the asymptotic (long-time limit) feature of the IGE. The
correlated IGE is reduced by $\frac{1}{2}\ln \left( \frac{1-r}{1+r}\right)
<0 $ for $r>0$, which is independent of the evolution of the system (see
equation (\ref{s-rel})). When the micro-correlations vanish (i.e. $r=0$), we
obtain the expected result $\mathcal{S}_{\mathcal{M}_{\text{corr.}}^{3\text{D%
}}}\left( \tau ;r\right) =\mathcal{S}_{\mathcal{M}_{\text{non-corr.}}^{3%
\text{D}}}\left( \tau ;0\right) $.

With $\mathcal{V}_{\mathcal{M}_{\text{corr.}}^{3\text{D}}}\left[ \mathcal{D}%
_{\Theta }^{\left( \text{geodesic}\right) }\left( \tau ;r\right) \right] $
in hand, we make the following observations. From (\ref{eq:v-rel}) we find%
\begin{equation}
r=\frac{\Delta \mathcal{C}^{2}}{\mathcal{C}_{\text{total}}^{2}},
\label{due2}
\end{equation}%
where 
\begin{equation}
\Delta \mathcal{C}^{2}\equiv \left\{ \mathcal{V}_{\mathcal{M}_{\text{%
non-corr.}}^{3\text{D}}}\left[ \mathcal{D}_{\Theta }^{\left( \text{geodesic}%
\right) }\left( \tau ;0\right) \right] \right\} ^{2}-\left\{ \mathcal{V}_{%
\mathcal{M}_{\text{corr.}}^{3\text{D}}}\left[ \mathcal{D}_{\Theta }^{\left( 
\text{geodesic}\right) }\left( \tau ;r\right) \right] \right\} ^{2}
\end{equation}%
and 
\begin{equation}
\mathcal{C}_{\text{total}}^{2}\equiv \left\{ \mathcal{V}_{\mathcal{M}_{\text{%
non-corr.}}^{3\text{D}}}\left[ \mathcal{D}_{\Theta }^{\left( \text{geodesic}%
\right) }\left( \tau ;0\right) \right] \right\} ^{2}+\left\{ \mathcal{V}_{%
\mathcal{M}_{\text{corr.}}^{3\text{D}}}\left[ \mathcal{D}_{\Theta }^{\left( 
\text{geodesic}\right) }\left( \tau ;r\right) \right] \right\} ^{2}.
\end{equation}%
Combining (\ref{eq:purity}) and (\ref{due2}), we obtain%
\begin{equation}
\mathcal{P}=1-\eta _{\mathcal{C}}\cdot \frac{\Delta \mathcal{C}^{2}}{%
\mathcal{C}_{\text{total}}^{2}},  \label{nnnn}
\end{equation}%
where the dimensionless coefficient $\eta _{\mathcal{C}}\equiv \frac{8}{3}k_{%
\mathrm{o}}^{2}\left( 2k_{\mathrm{o}}^{2}+\sigma _{k\mathrm{o}}^{2}\right)
R_{\mathrm{o}}L^{3}$. From (\ref{nnnn}) it is evident that quantum
entanglement and the information geometric complexity are connected. It
turns out that when purity goes to unity, the difference between the
correlated and non-correlated information geometric complexities approaches
zero.

\section{Final Remarks\label{sec-final}}

In this article, micro-correlated and non-correlated Gaussian statistical
models were used to model the entanglement of a quantum mechanical system
generated by an $s$-wave scattering event. The IGAC was used to analyze our
specific two-variable micro-correlated Gaussian statistical model. The
manifolds $\mathcal{M}_{\text{corr.}}^{3\text{D}}$ and $\mathcal{M}_{\text{%
non-corr.}}^{3\text{D}}$ were used to model the quantum entanglement induced
by head-on elastic scattering of two spinless, structureless,
non-relativistic particles, each represented by minimum uncertainty
wave-packets. The degree of entanglement was quantified by the purity of the
system. The purity $\mathcal{P}$ for $s$-wave scattering was found in terms
of the micro-correlation coefficient $r$, the interaction potential range $L$%
, the initial separation $R_{\mathrm{o}}$ between particles, the initial
momentum $p_{\mathrm{o}}=\hbar k_{\mathrm{o}}$ and initial momentum spread $%
\sigma _{\mathrm{o}}=\hbar \sigma _{k\mathrm{o}}$. The scattering phase
shift $\theta $ as well as the scattering cross section $\Sigma $ were both
found to be defined in terms of $r$, $L$ and $p_{\mathrm{o}}$. For $r=0$, $%
\theta $, $\Sigma $ and $V$ (interaction potential height) are each zero
while $\mathcal{P}=1$ (indicating that the system is not entangled). The
micro-correlation coefficient $r$, a quantity that parameterizes the
correlated microscopic degrees of freedom of the system, can be understood
as the ratio of the potential to kinetic energy of the system. When $r\neq 0$
the wave-packets experience the effect of a repulsive potential; the
magnitude of the wave vectors (momenta) decreases relative to their
corresponding non-correlated value. The upper bound value of $r$ depends on $%
p_{\mathrm{o}}$ and $\sigma _{\mathrm{o}}$ in such a manner that $r$
increases as $p_{\mathrm{o}}$ decreases. This result constitutes a
significant, explicit connection between micro-correlations (the correlation
coefficient $r$) and physical observables (the macrovariable $p_{\mathrm{o}}$%
). The role played by $r$ in the quantities $\mathcal{P}$, $\Sigma $, $%
\theta $, and $V$ suggests that information about quantum scattering and
therefore about quantum entanglement is encoded in the statistical
micro-correlation, specifically in the covariance term $\mathrm{Cov}\left(
p_{1},p_{2}\right) \overset{\text{def}}{=}\left\langle
p_{1}p_{2}\right\rangle -\left\langle p_{1}\right\rangle \left\langle
p_{2}\right\rangle $ appearing in the definition (\ref{corr-coeff2}) of $r$.

In summary, we proposed that the emergence of scattering-induced quantum
entanglement can be understood by considering pre and post-collisional
quantum dynamical scenarios as macroscopic manifestations emerging from
appropriately chosen statistical microstructures. In this view, the
information geometry associated with the post-collisional statistical
microstructure can be modelled in terms of a weak perturbation of the
information geometry relative to the pre-collisional microstructure. In
particular, quantum entanglement may be interpreted as a perturbation of
statistical space geometry: the non-correlated geometry (\ref%
{eq:non-corr_geo}) is perturbed due to the presence of the quantum
scattering, the information of which is encoded in the statistical
micro-correlation terms present in (\ref{eq:non-corr_geo1}). Indeed, in the
case where $r=0$, the perturbation matrix (\ref{eq:non-corr_geo1}) is null.
Thus, the quantum entanglement manifests as a geometric perturbation of the
statistical space in analogy to the interpretation of a static gravitational
field as a perturbation of flat space. The perturbation of statistical
geometry occurs in the $2$D\ momentum subspace spanned by basis vectors%
\textbf{\ }$e_{1}=\partial _{\mu _{1}}$\textbf{\ }and $e_{2}=\partial _{\mu
_{2}}$. In particular,\textbf{\ }after scattering the two particles maintain
a correlation among their microscopic momentum degrees of freedom regardless
of the extent of their separation in statistical space.\textbf{\ }This fact,
together with the time-independence of the statistical geometry [i.e. the
information metric is Riemannian (rather than pseudo-Riemannian) since its
signature is positive definite (rather than positive semi-definite)] leads
to a notion of statistical non-locality.\textbf{\ }The perturbation of
statistical geometry is associated with the scattering phase shift in the
statistical momentum space.

The prolongation, denoted $\Delta $, was defined as the time required for
the observed momentum difference between a correlated and corresponding
non-correlated system to vanish. The prolongation encodes information about
how long it would take an entangled system to overcome the momentum gap
generated by the scattering phase shift. The entangled system only attains
the full value of momentum\textbf{\ }(i.e. the momentum value as seen in the
corresponding non-correlated system) when the scattering phase shift
vanishes.\textbf{\ }For this reason, the prolongation represents the
temporal duration over which the entanglement is active.\textbf{\ }It was
found that for $r$ values close to its upper bound, the prolongation $\Delta 
$ becomes infinitely large. On the other hand, with $r$ vanishing (i.e., no
micro-correlation) $\Delta $ is identically zero. With $r$ fixed however,
the prolongation $\Delta $ depends on $p_{\mathrm{o}}$ and $\sigma _{\mathrm{%
o}}$. Thus, the prolongation $\Delta $ may be taken to represent the
duration of quantum entanglement for a given correlated system where the
entanglement duration can be controlled by the initial conditions $p_{%
\mathrm{o}}$ and $\sigma _{\mathrm{o}}$ as well as $r$. Maximal prolongation
occurs when $r$ is greatest and the ratio $\sigma _{\mathrm{o}}/p_{\mathrm{o}%
}$ is smallest. For small initial $r$ and $p_{\mathrm{o}}$, $\Delta $ would
be correspondingly small, suggesting that for such scenarios quantum
entanglement is transient.

It was determined that\ both statistical manifolds $\mathcal{M}_{\text{corr.}%
}^{3\text{D}}$ and $\mathcal{M}_{\text{non-corr.}}^{3\text{D}}$ are
negatively curved, with a micro-correlation independent Ricci scalar
curvature $\mathcal{R}_{\mathcal{M}_{\text{corr.}}^{3\text{D}}}=-\frac{3}{2}=%
\mathcal{R}_{\mathcal{M}_{\text{non-corr.}}^{3\text{D}}}$. Moreover, the
sectional curvature throughout both manifolds was determined to be constant, 
$K_{\mathcal{M}_{\text{corr.}}^{3\text{D}}}=-\frac{1}{4}=K_{\mathcal{M}_{%
\text{non-corr.}}^{3\text{D}}}$. The constancy of the sectional curvature in
all directions imply that both $\mathcal{M}_{\text{corr.}}^{3\text{D}}$ and $%
\mathcal{M}_{\text{non-corr.}}^{3\text{D}}$ are isotropic manifolds. This
was verified by the vanishing of all components of the Weyl projective
curvature tensor $\mathcal{W}_{abcd}$ defined on each space. The complexity
of geodesic paths on $\mathcal{M}_{\text{corr.}}^{3\text{D}}$ and $\mathcal{M%
}_{\text{non-corr.}}^{3\text{D}}$ was characterized through the asymptotic
computation of the IGE and the\textbf{\ }Lyapunov exponents on each
manifold. The Lyapunov exponents in both cases were found to be constant and
positive definite, i.e. $\lambda _{\mathcal{M}_{\text{corr.}}^{3\text{D}%
}}=\lambda _{\mathcal{M}_{\text{non-corr.}}^{3\text{D}}}=2A_{\mathrm{o}}>0$.
The IGE $\mathcal{S}_{\mathcal{M}_{\text{corr.}}^{3\text{D}}}\left( \tau
;r\right) $ in presence of micro-correlations assumes smaller values
relative to the non-correlated case $\mathcal{S}_{\mathcal{M}_{\text{%
non-corr.}}^{3\text{D}}}\left( \tau ;0\right) $ while the growth
characteristics of both correlated and non-correlated IGEs were found to be
the same. Specifically, the larger the micro-correlation (i.e. the closer $r$
is to $1$) the lower the values of the IGE $\mathcal{S}_{\mathcal{M}_{\text{%
corr.}}^{3\text{D}}}\left( \tau ;r\right) $. Thus, the stronger the
micro-correlation, the larger the gap between $\mathcal{S}_{\mathcal{M}_{%
\text{corr.}}^{3\text{D}}}\left( \tau ;r\right) $\ and $\mathcal{S}_{%
\mathcal{M}_{\text{non-corr.}}^{3\text{D}}}\left( \tau ;0\right) $. This
implies that $\mathcal{S}_{\mathcal{M}_{\text{corr.}}^{3\text{D}}}\left(
\tau ;r\right) <\mathcal{S}_{\mathcal{M}_{\text{non-corr.}}^{3\text{D}%
}}\left( \tau ;0\right) $. When micro-correlations vanish (i.e. when $r=0$),
we obtain the expected result, $\mathcal{S}_{\mathcal{M}_{\text{corr.}}^{3%
\text{D}}}\left( \tau ;0\right) =\mathcal{S}_{\mathcal{M}_{\text{non-corr.}%
}^{3\text{D}}}\left( \tau ;0\right) $. In the model investigated in this
work, the appearance of\textbf{\ }micro-correlation terms in the elements in
the Fisher-Rao information metric leads to the compression of $\mathcal{V}_{%
\mathcal{M}_{\text{corr.}}^{3\text{D}}}\left[ \mathcal{D}_{\Theta }^{\left( 
\text{geodesic}\right) }\left( \tau ;r\right) \right] $ by the fraction $%
\sqrt{\frac{1-r}{1+r}}$ and thus, to a reduction of the complexity of the
path leading from initial macrostate $\Theta _{\text{I}}$ to final
macrostate $\Theta _{\text{F}}$.

Information Geometry and Maximum Relative Entropy methods hold great promise
for solving computational problems in classical and quantum physics. Our
theoretical formalism allows for the analysis of physical problems by means
of statistical inference and information geometric techniques, that is,
Riemannian (differential) geometric techniques applied to probability
theory. The macroscopic behavior of an arbitrary complex system is a
consequence of the underlying statistical structure of the microscopic
degrees of freedom of the system. We are confident that the present work
represents significant progress toward the goal of understanding the
relationship between statistical micro-correlations and quantum entanglement
on the one hand and the effect of micro-correlations on the dynamical
complexity of informational geodesic flows on the other. It is our hope to
build upon the techniques employed in this work to ultimately establish a
sound information geometric interpretation of quantum entanglement.

\begin{acknowledgments}
This work was partially supported by WCU (World Class University) program of
NRF/MEST (R32-2009-000-10130-0) and by the European Community's Seventh
Framework Program FP7/2007-2013 under grant agreement 213681 (CORNER
Project).
\end{acknowledgments}

\appendix

\section{Integration of the Geodesic Equations\label{app-geod}}

The coupled ODEs (\ref{eq:1}), (\ref{eq:2}) and (\ref{eq:3}) can be solved
via the following strategy. First, (\ref{eq:1}) and (\ref{eq:2}) can be
rewritten as%
\begin{equation}
\frac{\mu _{1/2}^{\prime \prime }}{\mu _{1/2}^{\prime }}=\frac{2\sigma
^{\prime }}{\sigma },  \label{eq:4}
\end{equation}%
where \textquotedblleft $^{\prime }$\textquotedblright\ denotes a
differentiation with respect to $\tau $. (\ref{eq:4}) can be recasted as%
\begin{equation}
\frac{x^{\prime }}{x}=\frac{2\sigma ^{\prime }}{\sigma },\text{ where }x=\mu
_{1/2}^{\prime }\text{ and }x^{\prime }=\mu _{1/2}^{\prime \prime }.
\end{equation}%
Moreover, since 
\begin{equation}
\frac{x^{\prime }}{x}=\frac{2\sigma ^{\prime }}{\sigma }\Rightarrow \frac{d}{%
d\tau }\ln \left\vert x\right\vert =2\frac{d}{d\tau }\ln \sigma ,
\end{equation}%
we find%
\begin{equation}
\int \left( \frac{d}{d\tau }\ln \left\vert x\right\vert \right) d\tau =2\int
\left( \frac{d}{d\tau }\ln \sigma \right) d\tau \Rightarrow \ln \left\vert
x\right\vert +k=2\ln \sigma ,\;\text{where}\;k=\text{const}.
\end{equation}%
Exponentiating both sides of the above equation leads to%
\begin{equation}
\exp \left( \ln \left\vert x\right\vert +k\right) =\exp (2\ln \sigma
)\Rightarrow e^{k}\left\vert x\right\vert =\sigma ^{2}.
\end{equation}%
Thus,%
\begin{equation}
\left\vert \mu _{1/2}^{\prime }\right\vert =\left\vert C_{1/2}\right\vert
\sigma ^{2},  \label{eq:5}
\end{equation}%
where $C_{1/2}$ are the integration constants corresponding to $\mu
_{1/2}^{\prime }$. In order for our Gaussian statistical model to have
smooth and natural evolution, $\sigma (\tau )$ must be positive definite and
well-behaved (continuous and differentiable) over the entire domain of $\tau 
$; $\tau \in \left( -\infty ,+\infty \right) $. Then from (\ref{eq:5}) $\mu
_{1/2}^{\prime }$ must be either positive definite or negative definite over
the entire domain of $\tau $ and the sign of $C_{1/2}$ must be associated
with the sign of $\mu _{1/2}^{\prime }$ so that $\sigma $ is positive
definite and free from nodes. We can rewrite (\ref{eq:5}) as 
\begin{equation}
\mu _{1/2}^{\prime }=C_{1/2}\sigma ^{2};\;\text{with}\;\frac{\mu
_{1/2}^{\prime }}{C_{1/2}}>0.  \label{eq:5-re}
\end{equation}%
Substituting (\ref{eq:5}) into (\ref{eq:3}), we obtain%
\begin{equation}
\sigma ^{\prime \prime }-\frac{\sigma ^{\prime 2}}{\sigma }+\frac{1}{4\left(
r^{2}-1\right) }\left( 2rC_{1}C_{2}-C_{1}^{2}-C_{2}^{2}\right) \sigma ^{3}=0.
\label{eq:6}
\end{equation}%
Dividing both sides of (\ref{eq:6}) by $\sigma $ yields 
\begin{equation}
\frac{\sigma ^{\prime \prime }\sigma -\sigma ^{\prime 2}}{\sigma ^{2}}+\frac{%
1}{4\left( r^{2}-1\right) }\left( 2rC_{1}C_{2}-C_{1}^{2}-C_{2}^{2}\right)
\sigma ^{2}=0.  \label{eq:7}
\end{equation}%
The first term of (\ref{eq:7}) can be rewritten as a complete differential
by means of the following identity: 
\begin{equation}
\frac{d}{d\tau }\left( \frac{\sigma ^{\prime }}{\sigma }\right) =\frac{%
\sigma ^{\prime \prime }\sigma -\sigma ^{\prime 2}}{\sigma ^{2}}.
\label{eq:8}
\end{equation}%
The second term of (\ref{eq:7}) is also a complete differential form due to (%
\ref{eq:5}). Then, for $\mu _{1}$ and $\mu _{2}$, respectively, we may
rewrite (\ref{eq:7}) as 
\begin{eqnarray}
\frac{d}{d\tau }\left( \frac{\sigma ^{\prime }}{\sigma }\right) +\frac{C_{1}%
}{4\left( r^{2}-1\right) }\left[ \frac{C_{2}}{C_{1}}\left( 2r-\frac{C_{2}}{%
C_{1}}\right) -1\right] \mu _{1}^{\prime } &=&0,  \label{eq:9} \\
\frac{d}{d\tau }\left( \frac{\sigma ^{\prime }}{\sigma }\right) +\frac{C_{2}%
}{4\left( r^{2}-1\right) }\left[ \frac{C_{1}}{C_{2}}\left( 2r-\frac{C_{1}}{%
C_{2}}\right) -1\right] \mu _{2}^{\prime } &=&0.  \label{eq:10}
\end{eqnarray}%
Integrating both sides with respect to $\tau $, these become%
\begin{eqnarray}
\frac{\sigma ^{\prime }}{\sigma }+\frac{C_{1}}{4\left( r^{2}-1\right) }\left[
\frac{C_{2}}{C_{1}}\left( 2r-\frac{C_{2}}{C_{1}}\right) -1\right] \mu
_{1}+D_{1} &=&0,  \label{eq:11} \\
\frac{\sigma ^{\prime }}{\sigma }+\frac{C_{2}}{4\left( r^{2}-1\right) }\left[
\frac{C_{1}}{C_{2}}\left( 2r-\frac{C_{1}}{C_{2}}\right) -1\right] \mu
_{2}+D_{2} &=&0,  \label{eq:12}
\end{eqnarray}%
where $D_{1}$ and $D_{2}$ are integration constants.

Substituting (\ref{eq:11}) and (\ref{eq:12}) into (\ref{eq:4}) leads to 
\begin{eqnarray}
\mu _{1}^{\prime \prime }+\frac{C_{1}}{2\left( r^{2}-1\right) }\left[ \frac{%
C_{2}}{C_{1}}\left( 2r-\frac{C_{2}}{C_{1}}\right) -1\right] \mu _{1}\mu
_{1}^{\prime }+2D_{1}\mu _{1}^{\prime } &=&0,  \label{eq:15} \\
\mu _{2}^{\prime \prime }+\frac{C_{2}}{2\left( r^{2}-1\right) }\left[ \frac{%
C_{1}}{C_{2}}\left( 2r-\frac{C_{1}}{C_{2}}\right) -1\right] \mu _{2}\mu
_{2}^{\prime }+2D_{2}\mu _{2}^{\prime } &=&0.  \label{eq:16}
\end{eqnarray}%
Then integration of both sides of (\ref{eq:15}) and (\ref{eq:16}) with
respect to $\tau $ yields%
\begin{eqnarray}
\mu _{1}^{\prime }+\frac{C_{1}}{4\left( r^{2}-1\right) }\left[ \frac{C_{2}}{%
C_{1}}\left( 2r-\frac{C_{2}}{C_{1}}\right) -1\right] \mu _{1}^{2}+2D_{1}\mu
_{1}+E_{1} &=&0,  \label{eq:17} \\
\mu _{2}^{\prime }+\frac{C_{2}}{4\left( r^{2}-1\right) }\left[ \frac{C_{1}}{%
C_{2}}\left( 2r-\frac{C_{1}}{C_{2}}\right) -1\right] \mu _{2}^{2}+2D_{2}\mu
_{2}+E_{2} &=&0.  \label{eq:18}
\end{eqnarray}%
Equations (\ref{eq:17}) and (\ref{eq:18}) can now be represented by the
general form:%
\begin{equation}
\mu ^{\prime }+a\mu ^{2}+b\mu +c=0,  \label{eq:19}
\end{equation}%
which is known as the {}\textquotedblleft Riccati
equation\textquotedblright\ \cite{Gradshtyne}. Due to the fact that $a$, $b$
and $c$ are all constants in our problem, (\ref{eq:19}) may be modified to a
more tractable form:%
\begin{equation}
\nu ^{\prime }+A\nu ^{2}+B=0,  \label{eq:20}
\end{equation}%
where 
\begin{eqnarray}
\nu &=&\mu +\frac{b}{2a},  \label{eq:21} \\
A &=&a,  \label{eq:22} \\
B &=&-\frac{b^{2}}{4a}+c.  \label{eq:23}
\end{eqnarray}%
The solution of (\ref{eq:20}) is given by the form:%
\begin{equation}
\nu =\frac{1}{A}\frac{\alpha u^{\prime }+\beta v^{\prime }}{\alpha u+\beta v}%
,  \label{eq:24}
\end{equation}%
with $\alpha $ and $\beta $ being arbitrary constants, not both zero, while $%
u$ and $v$ are linearly independent solutions of 
\begin{equation}
\frac{d}{d\tau }\left( \frac{1}{A}\frac{dz}{d\tau }\right)
+Bz=0\;\;\;\Rightarrow \;\;\;\frac{d^{2}z}{d\tau ^{2}}+AB\,z=0.
\label{eq:25}
\end{equation}%
One finds easily 
\begin{eqnarray}
u &=&e^{\gamma \tau },  \label{eq:26} \\
v &=&e^{-\gamma \tau },  \label{eq:27}
\end{eqnarray}%
where%
\begin{equation}
\gamma =\pm \sqrt{-AB}=\pm \frac{\sqrt{b^{2}-4ac}}{2}.  \label{eq:28}
\end{equation}%
Then by means of equations (\ref{eq:21}), (\ref{eq:22}), (\ref{eq:23}), (\ref%
{eq:24}), (\ref{eq:26}), (\ref{eq:27}) and (\ref{eq:28}), we find%
\begin{eqnarray}
\mu \left( \tau \right) &=&\frac{\gamma }{a}\frac{\alpha e^{\gamma \tau
}-\beta e^{-\gamma \tau }}{\alpha e^{\gamma \tau }+\beta e^{-\gamma \tau }}-%
\frac{b}{2a}  \notag \\
&=&\pm \frac{\sqrt{b^{2}-4ac}}{2a}\frac{\alpha \exp \left[ \pm \frac{\sqrt{%
b^{2}-4ac}}{2}\tau \right] -\beta \exp \left[ \mp \frac{\sqrt{b^{2}-4ac}}{2}%
\tau \right] }{\alpha \exp \left[ \pm \frac{\sqrt{b^{2}-4ac}}{2}\tau \right]
+\beta \exp \left[ \mp \frac{\sqrt{b^{2}-4ac}}{2}\tau \right] }-\frac{b}{2a}.
\label{eq:29}
\end{eqnarray}%
Finally, we may identify equations (\ref{eq:17}) and (\ref{eq:18}) with (\ref%
{eq:19}) to find the solutions $\mu _{1}$ and $\mu _{2}$ via (\ref{eq:29}):%
\begin{equation}
\mu _{1/2}\left( \tau \right) =\frac{\gamma _{1/2}}{a_{1/2}}\frac{\alpha
_{1/2}e^{\gamma _{1/2}\tau }-\beta _{1/2}e^{-\gamma _{1/2}\tau }}{\alpha
_{1/2}e^{\gamma _{1/2}\tau }+\beta _{1/2}e^{-\gamma _{1/2}\tau }}-\frac{%
b_{1/2}}{2a_{1/2}},  \label{eq:30}
\end{equation}%
where for $\mu _{1}$%
\begin{equation}
\gamma _{1}\equiv \pm \frac{\sqrt{b_{1}^{2}-4a_{1}c_{1}}}{2},
\label{eq:gamma1}
\end{equation}%
with 
\begin{eqnarray}
a_{1} &=&\frac{C_{1}}{4\left( r^{2}-1\right) }\left[ \frac{C_{2}}{C_{1}}%
\left( 2r-\frac{C_{2}}{C_{1}}\right) -1\right] ,  \label{eq:31} \\
b_{1} &=&2D_{1},  \label{eq:32} \\
c_{1} &=&E_{1},  \label{eq:33}
\end{eqnarray}%
and for $\mu _{2}$%
\begin{equation}
\gamma _{2}\equiv \pm \frac{\sqrt{b_{2}^{2}-4a_{2}c_{2}}}{2},
\label{eq:gamma2}
\end{equation}%
with%
\begin{eqnarray}
a_{2} &=&\frac{C_{2}}{4\left( r^{2}-1\right) }\left[ \frac{C_{1}}{C_{2}}%
\left( 2r-\frac{C_{1}}{C_{2}}\right) -1\right] ,  \label{eq:34} \\
b_{2} &=&2D_{2},  \label{eq:35} \\
c_{2} &=&E_{2}.  \label{eq:36}
\end{eqnarray}%
In order for our system to have non-oscillatory and non-constant evolution, $%
\gamma _{1/2}$ must be real, thus the quantities $%
b_{1/2}^{2}-4a_{1/2}c_{1/2} $ must be positive definite. Later, we will find
the conditions for this (see (\ref{eq:real_gamma})).

We may rewrite (\ref{eq:30}) as%
\begin{equation}
\mu _{1/2}\left( \tau \right) =\frac{\gamma _{1/2}}{a_{1/2}}\frac{\delta
_{1/2}e^{\gamma _{1/2}\tau }-e^{-\gamma _{1/2}\tau }}{\delta _{1/2}e^{\gamma
_{1/2}\tau }+e^{-\gamma _{1/2}\tau }}-\frac{b_{1/2}}{2a_{1/2}},
\label{eq:mu_def}
\end{equation}%
where 
\begin{equation}
\delta _{1/2}\equiv \frac{\alpha _{1/2}}{\beta _{1/2}}.  \label{eq:delta}
\end{equation}%
By means of (\ref{eq:5-re}) and (\ref{eq:mu_def}) we find%
\begin{equation}
\sigma \left( \tau \right) =\sqrt{\frac{\mu _{1/2}^{\prime }\left( \tau
\right) }{C_{1/2}}}=2\sqrt{\left\vert \frac{\delta _{1/2}}{a_{1/2}C_{1/2}}%
\right\vert }\left\vert \frac{\gamma _{1/2}}{\delta _{1/2}e^{\gamma
_{1/2}\tau }+e^{-\gamma _{1/2}\tau }}\right\vert ,  \label{eq:37}
\end{equation}%
where $a_{1/2}$, $\gamma _{1/2}$, $\delta _{1/2}$ are given by (\ref{eq:31}%
), (\ref{eq:34}), (\ref{eq:gamma1}), (\ref{eq:gamma2}) and (\ref{eq:delta}).
However, our $\sigma $ obtained either via $\mu _{1}$ or via $\mu _{2}$ must
be identical. This yields the following equality: 
\begin{equation}
\frac{\mu _{1}^{\prime }}{C_{1}}=\frac{\mu _{2}^{\prime }}{C_{2}}%
\;\;\;\Leftrightarrow \;\;\;\frac{4\delta _{1}\gamma _{1}^{2}}{%
a_{1}C_{1}\left( \delta _{1}e^{\gamma _{1}\tau }+e^{-\gamma _{1}\tau
}\right) ^{2}}=\frac{4\delta _{2}\gamma _{2}^{2}}{a_{2}C_{2}\left( \delta
_{2}e^{\gamma _{2}\tau }+e^{-\gamma _{2}\tau }\right) ^{2}}.
\label{eq:identical}
\end{equation}%
In order for (\ref{eq:identical}) to be generally true, the following
conditions must be satisfied:%
\begin{eqnarray}
\left\vert \gamma _{1}\right\vert &=&\left\vert \gamma _{2}\right\vert ,
\label{eq:cond1} \\
\delta _{1} &=&\delta _{2}=1,  \label{eq:cond2} \\
a_{1}C_{1} &=&a_{2}C_{2}.  \label{eq:cond3}
\end{eqnarray}%
From (\ref{eq:31}) and (\ref{eq:34}) one finds that (\ref{eq:cond3}) holds
true by itself. In order for (\ref{eq:cond1}) to hold true, we require%
\begin{equation}
D_{1}^{2}-\frac{C_{1}E_{1}}{4\left( r^{2}-1\right) }\left[ \frac{C_{2}}{C_{1}%
}\left( 2r-\frac{C_{2}}{C_{1}}\right) -1\right] =D_{2}^{2}-\frac{C_{2}E_{2}}{%
4\left( r^{2}-1\right) }\left[ \frac{C_{1}}{C_{2}}\left( 2r-\frac{C_{1}}{%
C_{2}}\right) -1\right] .  \label{eq:cond1-2}
\end{equation}%
Substituting the conditions (\ref{eq:cond1}) and (\ref{eq:cond2}) into (\ref%
{eq:mu_def}), we obtain%
\begin{equation}
\mu _{1/2}=\frac{\gamma }{a_{1/2}}\tanh \left( \gamma \tau \right) -\frac{%
D_{1/2}}{a_{1/2}},  \label{eq:mu1/2}
\end{equation}%
where $\gamma =\left\vert \gamma _{1}\right\vert =\left\vert \gamma
_{2}\right\vert $. Adding $\mu _{1}$ and $\mu _{2}$, we find 
\begin{equation}
\mu _{1}+\mu _{2}=\left( \frac{1}{a_{1}}+\frac{1}{a_{2}}\right) \gamma \tanh
\left( \gamma \tau \right) -\left( \frac{D_{1}}{a_{1}}+\frac{D_{2}}{a_{2}}%
\right) .  \label{eq:mu1+mu2}
\end{equation}%
We make use of this Gaussian system to model a head-on collision between two
Gaussian packets in momentum space, where each particle carries the average
momentum, $\left\langle p_{1}\right\rangle =\mu _{1}$ and $\left\langle
p_{2}\right\rangle =\mu _{2}$, respectively. Thus, the total momentum of the
two-particle system represented by (\ref{eq:mu1+mu2}) must be conserved.
This requires 
\begin{equation}
\frac{1}{a_{1}}+\frac{1}{a_{2}}=0.  \label{eq:a1a2}
\end{equation}%
For convenience we require both $\mu _{1}\left( \tau \right) $ and $\mu
_{2}\left( \tau \right) $ cross $0$ at $\tau =0$. From (\ref{eq:mu1/2}) we
find that this condition implies 
\begin{equation}
D_{1}=D_{2}=0.  \label{eq:D12=0}
\end{equation}%
From (\ref{eq:a1a2}) one finds 
\begin{equation}
C_{1}=-C_{2}\equiv C\neq 0.  \label{eq:cond9}
\end{equation}%
Then due to (\ref{eq:cond3}), (\ref{eq:D12=0}) and (\ref{eq:cond9}), (\ref%
{eq:cond1-2}) is reduced to 
\begin{equation}
E_{1}=-E_{2}\equiv E\neq 0.  \label{eq:cond10}
\end{equation}%
Substituting (\ref{eq:D12=0}), (\ref{eq:cond9}) and (\ref{eq:cond10}) into (%
\ref{eq:cond1-2}), we obtain the above mentioned reality condition for $%
\gamma _{1/2}$, namely%
\begin{equation}
CE<0.  \label{eq:real_gamma}
\end{equation}%
From (\ref{eq:gamma1}) together with (\ref{eq:31}), (\ref{eq:32}), (\ref%
{eq:33}), (\ref{eq:cond1}), (\ref{eq:D12=0}), (\ref{eq:cond9}), (\ref%
{eq:cond10}), we find 
\begin{equation}
\gamma =\left\vert \gamma _{1}\right\vert =\left\vert \gamma _{2}\right\vert
=\sqrt{\frac{CE}{2\left( r-1\right) }}.  \label{eq:cond12}
\end{equation}%
Then substituting (\ref{eq:D12=0}), (\ref{eq:cond9}) and (\ref{eq:cond12})
into (\ref{eq:mu1/2}), and using (\ref{eq:37}), we finally obtain 
\begin{eqnarray}
\mu _{1}\left( \tau ;r\right) &=&-\sqrt{\frac{2E\left( r-1\right) }{C}}\tanh
\left( \sqrt{\frac{CE}{2\left( r-1\right) }}\tau \right) ,
\label{eq:mu1_final} \\
\mu _{2}\left( \tau ;r\right) &=&\sqrt{\frac{2E\left( r-1\right) }{C}}\tanh
\left( \sqrt{\frac{CE}{2\left( r-1\right) }}\tau \right) ,
\label{eq:mu2_final} \\
\sigma \left( \tau ;r\right) &=&\sqrt{-\frac{E}{C}}\frac{1}{\cosh \left( 
\sqrt{\frac{CE}{2\left( r-1\right) }}\tau \right) },  \label{eq:sigma_final}
\end{eqnarray}%
where we have set $C<0$ and $E>0$. In our probabilistic macroscopic approach
to dynamics, these geodesic trajectories represent the maximum probability
paths on $\mathcal{M}_{\text{corr.}}^{3\text{D}}$.

For the non-correlated Gaussian system, we set $r=0$ in (\ref{eq:mu1_final}%
), (\ref{eq:mu2_final}) and (\ref{eq:sigma_final}) to obtain 
\begin{eqnarray}
\mu _{1}\left( \tau ;0\right) &=&-\sqrt{-\frac{2E}{C}}\tanh \left( \sqrt{-%
\frac{CE}{2}}\tau \right) ,  \label{eq:MU01} \\
\mu _{2}\left( \tau ;0\right) &=&\sqrt{-\frac{2E}{C}}\tanh \left( \sqrt{-%
\frac{CE}{2}}\tau \right) ,  \label{eq:MU02} \\
\sigma \left( \tau ;0\right) &=&\sqrt{-\frac{E}{C}}\frac{1}{\cosh \left( 
\sqrt{-\frac{CE}{2}}\tau \right) }.  \label{eq:SIGMA0}
\end{eqnarray}%
Distinguishing the constants $C$ and $E$ for the correlated Gaussian system
from those for the non-correlated Gaussian system, we rewrite (\ref%
{eq:mu1_final}), (\ref{eq:mu2_final}) and (\ref{eq:sigma_final}) as%
\begin{eqnarray}
\mu _{1}\left( \tau ;r\right) &=&-\sqrt{\frac{2E_{r}\left( r-1\right) }{C_{r}%
}}\tanh \left( \sqrt{\frac{C_{r}E_{r}}{2\left( r-1\right) }}\tau \right) ,
\label{eq:MUr1} \\
\mu _{2}\left( \tau ;r\right) &=&\sqrt{\frac{2E_{r}\left( r-1\right) }{C_{r}}%
}\tanh \left( \sqrt{\frac{C_{r}E_{r}}{2\left( r-1\right) }}\tau \right) ,
\label{eq:MUr2} \\
\sigma \left( \tau ;r\right) &=&\sqrt{-\frac{E_{r}}{C_{r}}}\frac{1}{\cosh
\left( \sqrt{\frac{C_{r}E_{r}}{2\left( r-1\right) }}\tau \right) },
\label{eq:SIGMAr}
\end{eqnarray}%
where the subscript \textquotedblleft $_{r}$\textquotedblright\ in $C_{r}$
and $E_{r}$ implies that the constants are dependent upon the correlation
coefficient $r$ of the given statistical manifold.

\section{Refining the Geodesic Trajectories\label{app-refn}}

In this Appendix we join two different charts of Gaussian statistical
manifolds, one without correlation (before collision) and the other with
correlation (after collision). The set of geodesic curves for each model is
represented by equations (\ref{eq:MU01}), (\ref{eq:MU02}), (\ref{eq:SIGMA0})
(for the non-correlated model) and by equations (\ref{eq:MUr1}), (\ref%
{eq:MUr2}), (\ref{eq:SIGMAr}) (for the correlated model). The two sets are
joined at the junction, $\tau =0$: $\tau <0$ (before collision) for the
non-correlated model and $\tau \geq 0$ (after collision) for the correlated
model.

The constants, $C$ and $E$ in (\ref{eq:MU01}), (\ref{eq:MU02}) and (\ref%
{eq:SIGMA0}) can be determined via the conditions at the initial affine
time, $-\tau _{\mathrm{o}}$. We assign the initial momenta and the
dispersion of the wave-packets as%
\begin{eqnarray}
\mu _{1}\left( -\tau _{\mathrm{o}};0\right) &=&-\sqrt{-\frac{2E}{C}}\tanh
\left( -\sqrt{-\frac{CE}{2}}\tau _{\mathrm{o}}\right) \equiv p_{\mathrm{o}},
\label{eq:MU01tau} \\
\mu _{2}\left( -\tau _{\mathrm{o}};0\right) &=&\sqrt{-\frac{2E}{C}}\tanh
\left( -\sqrt{-\frac{CE}{2}}\tau _{\mathrm{o}}\right) \equiv -p_{\mathrm{o}},
\label{eq:MU02tau} \\
\sigma \left( -\tau _{\mathrm{o}};0\right) &=&\sqrt{-\frac{E}{C}}\frac{1}{%
\cosh \left( -\sqrt{-\frac{CE}{2}}\tau _{\mathrm{o}}\right) }\equiv \sigma _{%
\mathrm{o}}.  \label{eq:SIGMA0tau}
\end{eqnarray}%
Combining (\ref{eq:MU01tau}) and (\ref{eq:SIGMA0tau}), one obtains%
\begin{equation}
-\frac{E}{C}=\frac{1}{2}p_{\mathrm{o}}^{2}+\sigma _{\mathrm{o}}^{2}.
\label{eq:E_over_C}
\end{equation}%
Also, taking the ratio between $\sigma _{\mathrm{o}}$ and $p_{\mathrm{o}}$
via (\ref{eq:MU01tau}) and (\ref{eq:SIGMA0tau}) yields,%
\begin{equation}
\frac{\sigma _{\mathrm{o}}}{p_{\mathrm{o}}}=\frac{1}{\sqrt{2}\sinh \left( 
\sqrt{-\frac{CE}{2}}\tau _{\mathrm{o}}\right) }.  \label{eq:sigma_over_p}
\end{equation}%
Upon considering large $\tau _{\mathrm{o}}$ in (\ref{eq:sigma_over_p}), we
find%
\begin{eqnarray}
\sqrt{-\frac{CE}{2}} &=&\frac{1}{\tau _{\mathrm{o}}}\sinh ^{-1}\left( \frac{%
p_{\mathrm{o}}}{\sqrt{2}\sigma _{\mathrm{o}}}\right)  \notag \\
&\overset{\frac{\sigma _{\mathrm{o}}}{p_{\mathrm{o}}}\ll 1}{=}&\frac{1}{\tau
_{\mathrm{o}}}\left\{ \ln \left( \frac{\sqrt{2}p_{\mathrm{o}}}{\sigma _{%
\mathrm{o}}}\right) +\frac{1}{2}\left( \frac{\sigma _{\mathrm{o}}}{p_{%
\mathrm{o}}}\right) ^{2}-\frac{3}{8}\left( \frac{\sigma _{\mathrm{o}}}{p_{%
\mathrm{o}}}\right) ^{4}+\mathcal{O}\left[ \left( \frac{\sigma _{\mathrm{o}}%
}{p_{\mathrm{o}}}\right) ^{6}\right] \right\} .  \label{eq:sqrt_CE}
\end{eqnarray}%
Equation (\ref{eq:sqrt_CE}) implies that $\tau _{\mathrm{o}}$ should be
chosen sufficiently large so that the ratio $\sigma _{\mathrm{o}}/p_{\mathrm{%
o}}$ will be very small, while $\sqrt{-\frac{CE}{2}}$ remains finite. The
constants $C$ and $E$ can be individually determined by simultaneously
solving (\ref{eq:E_over_C}) and (\ref{eq:sqrt_CE}).

In a similar manner, the constants $C_{r}$ and $E_{r}$ in (\ref{eq:MUr1}), (%
\ref{eq:MUr2}) and (\ref{eq:SIGMAr}) can be determined via the conditions at
the reversal time $\tau _{\mathrm{o}}$. We assign the momenta and dispersion
of the wave-packets according to 
\begin{eqnarray}
\mu _{1}\left( \tau _{\mathrm{o}};r\right) &=&-\sqrt{\frac{2E_{r}\left(
r-1\right) }{C_{r}}}\tanh \left( \sqrt{\frac{C_{r}E_{r}}{2\left( r-1\right) }%
}\tau _{\mathrm{o}}\right) \equiv -p_{\mathrm{o}}^{\prime },
\label{eq:MUr1tau} \\
\mu _{2}\left( \tau _{\mathrm{o}};r\right) &=&\sqrt{\frac{2E_{r}\left(
r-1\right) }{C_{r}}}\tanh \left( \sqrt{\frac{C_{r}E_{r}}{2\left( r-1\right) }%
}\tau _{\mathrm{o}}\right) \equiv p_{\mathrm{o}}^{\prime },
\label{eq:MUr2tau} \\
\sigma \left( \tau _{\mathrm{o}};r\right) &=&\sqrt{-\frac{E_{r}}{C_{r}}}%
\frac{1}{\cosh \left( \sqrt{\frac{C_{r}E_{r}}{2\left( r-1\right) }}\tau _{%
\mathrm{o}}\right) }\equiv \sigma _{\mathrm{o}}^{\prime }.
\label{eq:SIGMArtau}
\end{eqnarray}%
Combination of (\ref{eq:MUr1tau}) with (\ref{eq:SIGMArtau}) leads to 
\begin{equation}
-\frac{E_{r}}{C_{r}}=\frac{p_{\mathrm{o}}^{\prime }{}^{2}}{2\left(
1-r\right) }+\sigma _{\mathrm{o}}^{\prime }{}^{2}.  \label{eq:Er_over_Cr}
\end{equation}%
From (\ref{eq:MUr1tau}) and (\ref{eq:SIGMArtau}) it is found that the ratio
between $\sigma _{\mathrm{o}}^{\prime }$ and $p_{\mathrm{o}}^{\prime }$
reads 
\begin{equation}
\frac{\sigma _{\mathrm{o}}^{\prime }}{p_{\mathrm{o}}^{\prime }}=\frac{1}{%
\sqrt{2\left( 1-r\right) }}\frac{1}{\sinh \left( \sqrt{\frac{C_{r}E_{r}}{%
2\left( r-1\right) }}\tau _{\mathrm{o}}\right) }.  \label{eq:sigmar_over_pr}
\end{equation}%
From (\ref{eq:sigmar_over_pr}) it is found that for large $\tau _{\mathrm{o}%
},$%
\begin{eqnarray}
\sqrt{\frac{C_{r}E_{r}}{2\left( r-1\right) }} &=&\frac{1}{\tau _{\mathrm{o}}}%
\sinh ^{-1}\left( \frac{p_{\mathrm{o}}^{\prime }}{\sqrt{2\left( 1-r\right) }%
\sigma _{\mathrm{o}}^{\prime }}\right)  \notag \\
&\overset{\frac{\sigma _{\mathrm{o}}^{\prime }}{p_{\mathrm{o}}^{\prime }}\ll
1}{=}&\frac{1}{\tau _{\mathrm{o}}}\left\{ \ln \left( \frac{\sqrt{2}p_{%
\mathrm{o}}^{\prime }}{\sqrt{1-r}\sigma _{\mathrm{o}}^{\prime }}\right) +%
\frac{1-r}{2}\left( \frac{\sigma _{\mathrm{o}}^{\prime }}{p_{\mathrm{o}%
}^{\prime }}\right) ^{2}-\frac{3}{8}\left( 1-r\right) ^{2}\left( \frac{%
\sigma _{\mathrm{o}}^{\prime }}{p_{\mathrm{o}}^{\prime }}\right) ^{4}+%
\mathcal{O}\left[ \left( \frac{\sigma _{\mathrm{o}}^{\prime }}{p_{\mathrm{o}%
}^{\prime }}\right) ^{6}\right] \right\} .  \label{eq:sqrt_CrEr}
\end{eqnarray}%
Here again it is implied that $\tau _{\mathrm{o}}$ should be taken
sufficiently large so that the ratio, $\sigma _{\mathrm{o}}^{\prime }/p_{%
\mathrm{o}}^{\prime }$ can be very small while $\sqrt{\frac{C_{r}E_{r}}{%
2\left( r-1\right) }}$ remains finite. Furthermore, the constants $C_{r}$
and $E_{r}$ can be individually determined by simultaneously solving (\ref%
{eq:Er_over_Cr}) and (\ref{eq:sqrt_CrEr}).

The two sets of geodesic curves (with and without correlations) are joined
at the junction $\tau =0.$ The two sets of geodesic curves must be
continuous at the junction $\tau =0$ so as to ensure the collision does not
assume any unphysical irregularity in the momentum dispersion. From (\ref%
{eq:MU01}), (\ref{eq:MU02}), (\ref{eq:SIGMA0}) and (\ref{eq:MUr1}), (\ref%
{eq:MUr2}), (\ref{eq:SIGMAr}) it is found that this continuity condition is
satisfied by 
\begin{equation}
\frac{E}{C}=\frac{E_{r}}{C_{r}}.  \label{eq:E_over_C_cont}
\end{equation}%
Using condition (\ref{eq:E_over_C_cont}) together with (\ref{eq:MU01tau})
and (\ref{eq:MUr1tau}), one may compare $p_{\mathrm{o}}$ with $p_{\mathrm{o}%
}^{\prime }$ as follows,%
\begin{eqnarray}
\frac{p_{\mathrm{o}}^{\prime }}{p_{\mathrm{o}}} &=&\sqrt{1-r}\frac{\tanh
\left( \sqrt{\frac{C_{r}E_{r}}{2\left( r-1\right) }}\tau _{\mathrm{o}%
}\right) }{\tanh \left( \sqrt{-\frac{CE}{2}}\tau _{\mathrm{o}}\right) } 
\notag \\
&=&\sqrt{1-r}\left\{ 1+2\left( \epsilon -\epsilon ^{\prime }\right) +%
\mathcal{O}\left[ \left( \epsilon -\epsilon ^{\prime }\right) ^{2}\right]
\right\} ,  \label{eq:pr_over_p}
\end{eqnarray}%
where $\epsilon \equiv \exp \left( -\sqrt{-2CE}\tau _{\mathrm{o}}\right) $
and $\epsilon ^{\prime }\equiv \exp \left( -\sqrt{\frac{2C_{r}E_{r}}{r-1}}%
\tau _{\mathrm{o}}\right) $. In a similar manner, by way of (\ref%
{eq:SIGMA0tau}) and (\ref{eq:SIGMArtau}) one may also compare $\sigma _{%
\mathrm{o}}$ with $\sigma _{\mathrm{o}}^{\prime }$,%
\begin{eqnarray}
\frac{\sigma _{\mathrm{o}}^{\prime }}{\sigma _{\mathrm{o}}} &=&\frac{\cosh
\left( \sqrt{-\frac{CE}{2}}\tau _{\mathrm{o}}\right) }{\cosh \left( \sqrt{%
\frac{C_{r}E_{r}}{2\left( r-1\right) }}\tau _{\mathrm{o}}\right) }  \notag \\
&=&\sqrt{\frac{\epsilon ^{\prime }}{\epsilon }}\left\{ 1+\left( \epsilon
-\epsilon ^{\prime }\right) +\mathcal{O}\left[ \left( \epsilon -\epsilon
^{\prime }\right) ^{2}\right] \right\} .  \label{eq:sigmar_over_sigma}
\end{eqnarray}%
From (\ref{eq:pr_over_p}) it is observed that for sufficiently large $\tau _{%
\mathrm{o}}$ the ratio $p_{\mathrm{o}}^{\prime }/p_{\mathrm{o}}$ is not
significantly influenced by how the functional arguments $\sqrt{-\frac{CE}{2}%
}\tau _{\mathrm{o}}$ and $\sqrt{\frac{C_{r}E_{r}}{2\left( r-1\right) }}\tau
_{\mathrm{o}}$ compare with each other, since the quantities on the
right-hand side of $\epsilon $ and $\epsilon ^{\prime }$ are very small (as
is the difference $\epsilon -\epsilon ^{\prime }$). From (\ref%
{eq:sigmar_over_sigma}) however, the ratio $\sigma _{\mathrm{o}}^{\prime
}/\sigma _{\mathrm{o}}$ appears to be influenced by how those functional
arguments compare with each other since the leading approximation reads%
\begin{equation}
\sqrt{\frac{\epsilon ^{\prime }}{\epsilon }}=\exp \left[ \left( \sqrt{-\frac{%
CE}{2}}-\sqrt{\frac{C_{r}E_{r}}{2\left( r-1\right) }}\right) \tau _{\mathrm{o%
}}\right] .  \label{eq:sigmar_over_sigma_approx}
\end{equation}%
From (\ref{eq:sigmar_over_sigma_approx}), one observes that the difference $%
\sqrt{-\frac{CE}{2}}-\sqrt{\frac{C_{r}E_{r}}{2\left( r-1\right) }}$ must
vanish in order for $\sqrt{\epsilon ^{\prime }/\epsilon }$ to remain finite
given that $\tau _{\mathrm{o}}$ is sufficiently large; otherwise a
non-vanishing difference could result in a sufficiently large exponent when
multiplied by a large value of $\tau _{\mathrm{o}}$ - this would cause $%
\sqrt{\epsilon ^{\prime }/\epsilon }$ to grow or decay exponentially. A
vanishing value of $\sqrt{-\frac{CE}{2}}-\sqrt{\frac{C_{r}E_{r}}{2\left(
r-1\right) }}$ implies\textbf{\ } 
\begin{equation}
\epsilon =\epsilon ^{\prime }.  \label{eq:epsilons_equal}
\end{equation}%
From (\ref{eq:pr_over_p}), (\ref{eq:sigmar_over_sigma}) and (\ref%
{eq:epsilons_equal}) it follows that%
\begin{eqnarray}
p_{\mathrm{o}}^{\prime } &=&\sqrt{1-r}p_{\mathrm{o}},  \label{eq:pr_p} \\
\sigma _{\mathrm{o}}^{\prime } &=&\sigma _{\mathrm{o}}.
\label{eq:sigmar_sigma}
\end{eqnarray}%
Equations (\ref{eq:pr_p}) and (\ref{eq:sigmar_sigma}) also satisfies the
condition (\ref{eq:E_over_C_cont}) through (\ref{eq:E_over_C}) and (\ref%
{eq:Er_over_Cr}).

Substituting (\ref{eq:E_over_C}), (\ref{eq:sqrt_CE}) and (\ref{eq:Er_over_Cr}%
), (\ref{eq:sqrt_CrEr}) into (\ref{eq:MU01}), (\ref{eq:MU02}), (\ref%
{eq:SIGMA0}) and (\ref{eq:MUr1}), (\ref{eq:MUr2}), (\ref{eq:SIGMAr}),
respectively, and using (\ref{eq:epsilons_equal}), (\ref{eq:pr_p}), (\ref%
{eq:sigmar_sigma}), we may rewrite the geodesic trajectories as follows: for
the non-correlated Gaussian system,\newline
\begin{eqnarray}
\mu _{1}\left( \tau ;0\right) &=&-\sqrt{p_{\mathrm{o}}^{2}+2\sigma _{\mathrm{%
o}}^{2}}\tanh \left( A_{\mathrm{o}}\tau \right) , \\
\mu _{2}\left( \tau ;0\right) &=&\sqrt{p_{\mathrm{o}}^{2}+2\sigma _{\mathrm{o%
}}^{2}}\tanh \left( A_{\mathrm{o}}\tau \right) , \\
\sigma \left( \tau ;0\right) &=&\sqrt{\frac{1}{2}p_{\mathrm{o}}^{2}+\sigma _{%
\mathrm{o}}^{2}}\frac{1}{\cosh \left( A_{\mathrm{o}}\tau \right) },
\end{eqnarray}%
while for the correlated Gaussian system,%
\begin{eqnarray}
\mu _{1}(\tau ;r) &=&-\sqrt{\left( 1-r\right) \left( p_{\mathrm{o}%
}^{2}+2\sigma _{\mathrm{o}}^{2}\right) }\tanh \left( A_{\mathrm{o}}\tau
\right) , \\
\mu _{2}(\tau ;r) &=&\sqrt{\left( 1-r\right) \left( p_{\mathrm{o}%
}^{2}+2\sigma _{\mathrm{o}}^{2}\right) }\tanh \left( A_{\mathrm{o}}\tau
\right) , \\
\sigma (\tau ;r) &=&\sqrt{\frac{1}{2}p_{\mathrm{o}}^{2}+\sigma _{\mathrm{o}%
}^{2}}\frac{1}{\cosh \left( A_{\mathrm{o}}\tau \right) },
\end{eqnarray}%
where%
\begin{eqnarray}
A_{\mathrm{o}} &\equiv &\sqrt{-\frac{CE}{2}}=\frac{1}{\tau _{\mathrm{o}}}%
\sinh ^{-1}\left( \frac{p_{\mathrm{o}}}{\sqrt{2}\sigma _{\mathrm{o}}}\right)
. \\
&\overset{\frac{\sigma _{\mathrm{o}}}{p_{\mathrm{o}}}\ll 1}{=}&\frac{1}{\tau
_{\mathrm{o}}}\left\{ \ln \left( \frac{\sqrt{2}p_{\mathrm{o}}}{\sigma _{%
\mathrm{o}}}\right) +\frac{1}{2}\left( \frac{\sigma _{\mathrm{o}}}{p_{%
\mathrm{o}}}\right) ^{2}-\frac{3}{8}\left( \frac{\sigma _{\mathrm{o}}}{p_{%
\mathrm{o}}}\right) ^{4}+\mathcal{O}\left[ \left( \frac{\sigma _{\mathrm{o}}%
}{p_{\mathrm{o}}}\right) ^{6}\right] \right\} ,  \notag
\end{eqnarray}%
which is defined from (\ref{eq:sqrt_CE}).

\end{document}